\documentclass[11pt,preprint]{article}
\pdfoutput=1
\usepackage{amssymb}
\usepackage{color,xcolor}
\usepackage{amsmath, amsfonts}
\usepackage{graphicx}
\usepackage{multirow, multicol}
\usepackage{rotating}
\usepackage{latexsym}
\usepackage{amsfonts}

\usepackage{cancel}

\usepackage[font=scriptsize]{caption}

\usepackage[colorlinks,bookmarks]{hyperref}
\definecolor{linkblue}{rgb}{0,0,0.8}
\definecolor{linkgreen}{rgb}{0,0.5,0}
\definecolor{vert}{rgb}{0,0.6,0.2}

\hypersetup{pdfpagemode=UseNone, pdfstartview=FitH, linkcolor=linkblue, %
            citecolor=linkgreen, urlcolor=linkblue}

\usepackage{amssymb}
\usepackage{amsmath}
\usepackage{graphicx}
\usepackage{hep}
\usepackage{soul}
\usepackage{latexsym}
\usepackage[latin1]{inputenc}
\usepackage{graphicx}
\usepackage{mathrsfs}
\usepackage{amssymb}
\usepackage{amsmath}
\usepackage{verbatim}
\usepackage{multirow}
\usepackage{booktabs}
\usepackage{subfigure}

\newcommand{\hzero}{\ensuremath{h}} 
\newcommand{\Hzero}{\ensuremath{H}} 
\newcommand{\CP}{\ensuremath{\mathcal{C}\mathcal{P}}}
\newcommand{\lag}{\ensuremath{\mathscr{L}}}

\newcommand{\rself}{\ensuremath{\hat{\Sigma}}}
\newcommand{\retildehat}{\ensuremath{\mbox{Re}\,\hat{\Sigma}}}
\newcommand{\retilde}{\ensuremath{\mbox{Re}\,\Sigma}}

\newcommand{\squark}{\ensuremath{\tilde{q}}}

\newcommand{\msbar}{\ensuremath{\overline{\text{MS}}}}

\newcommand{\GammaLO}{\ensuremath{\Gamma^{\text{LO}}_{\Hhh}}}
\newcommand{\GammaNLO}{\ensuremath{\Gamma^{\text{NLO}}_{\Hhh}}}

\newcommand{\sw}{\ensuremath{s_W}}
\newcommand{\swd}{\ensuremath{s^2_W}}
\newcommand{\cw}{\ensuremath{c_W}}
\newcommand{\cwd}{\ensuremath{c^2_W}}
\newcommand{\GHhh}{\ensuremath{\Gamma_{\Hzero \to \hzero\hzero}}}
\newcommand{\mw}{\ensuremath{m_W}}
\newcommand{\mz}{\ensuremath{m_Z}}
\newcommand{\pstar}{\ensuremath{p_*}}
\newcommand{\mHHd}{\ensuremath{m^2_{\Hzero}}}
\newcommand{\mHH}{\ensuremath{m_{\Hzero}}}
\newcommand{\mhd}{\ensuremath{m^2_{\hzero}}}
\newcommand{\mh}{\ensuremath{m_{\hzero}}}

\newcommand{\htb}[1]{{\color{blue} #1}}

\newcommand{\eqn}{equation}
\newcommand{\al}{\alpha}
\newcommand{\lb}{\left(}
\newcommand{\rb}{\right)}
\newcommand{\be}{\beta}

\newcommand{\aem}{\ensuremath{\alpha_{\rm{em}}}}
\newcommand{\gf}{\ensuremath{G_F}}
\newcommand{\Hhh}{\ensuremath{\Hzero \to \hzero\hzero}}

\newcommand{\mhHd}{\ensuremath{m^2_{\hzero\Hzero}}}

\newcommand{\gzero}{\ensuremath{\text{G}^0}}
\newcommand{\gp}{\ensuremath{\text{G}^+}}
\newcommand{\gm}{\ensuremath{\text{G}^-}}

\newcommand{\br}{\ensuremath{\text{BR}}}

\newcommand{\smrenorm}{Sirlin:1980nh,Bohm:1986rj,Hollik:1988ii,Denner:1991kt,Baro:2009gn}
\newcommand{\singlet}{Silveira:1985rk,Schabinger:2005ei,Patt:2006fw}
\newcommand{\singletours}{Pruna:2013bma,Lopez-Val:2014jva,Robens:2015gla}

\newcommand{\squarkos}{Guasch:1998as,Eberl:1999cy,Guasch:2001kz}
\newcommand{\higgs}{Higgs:1964pj,Higgs:1964ia}
\newcommand{\sbt}{s_{2\beta}}
\newcommand{\cbt}{c_{2\beta}}

\newcommand{\anlg}{\tilde{\alpha}}
\newcommand{\bnlg}{\tilde{\beta}} 
\newcommand{\dnlg}{\tilde{\delta}}
\newcommand{\enlg}{\tilde{\epsilon}}

\newcommand{\knlg}{\tilde{\kappa}}

\def\l{\lambda}

\def\b{\beta}

\def\G{\Gamma}
\newcommand{\Lag}{\mathscr{L}}

\def\ba{\begin{array}}
\def\ea{\end{array}}
\def\bea{\begin{eqnarray}}
\def\eea{\end{eqnarray}}
\def\noi{\noindent}

\newcommand{\HS}{\texttt{HiggsSignals}}
\newcommand{\HB}{\texttt{HiggsBounds}}

\newcommand{\ra}{\rightarrow}
\def\lsim{\;\raise0.3ex\hbox{$<$\kern-0.75em\raise-1.1ex\hbox{$\sim$}}\;}
\def\gsim{\;\raise0.3ex\hbox{$>$\kern-0.75em\raise-1.1ex\hbox{$\sim$}}\;}
\def\noi{\noindent}
\def\non{\nonumber}

\newcommand{\lam}{\lambda}

\begin{document}

\bibliographystyle{JHEP}
\author{F. Bojarski$^{a}$, G. Chalons$^{a}$, D. L\'opez-Val$^{b}$, T. Robens$^{c}$}
\title {
\begin{flushright}
\normalsize LPSC15335
\end{flushright}
\vskip0.2cm
\textbf{Heavy to light Higgs boson decays at NLO in the 
    Singlet Extension of the Standard Model }}

\maketitle

\begin{center}

\textit{$^{a}$
Laboratoire de Physique Subatomique et de Cosmologie \\  Universit\'e Grenoble-Alpes, CNRS/IN2P3
\\ 53 Rue des Martyrs \\
F-38026 Grenoble Cedex, France \\ \vskip5mm
}

\textit{$^{b}$ Center for Cosmology, Particle Physics and Phenomenology CP3 \\
Universit\'e catholique de Louvain \\ 
Chemin du Cyclotron 2,  
B-1348 Louvain-la-Neuve,  Belgium}\\

\vskip5mm

\textit{$^{c}$ IKTP, Technische Universit\"at Dresden \\ 
Zellescher Weg 19, D-01069 Dresden, Germany}\\

\vskip5mm

{E-mails: francois.bojarski@ens-lyon.fr, chalons@lpsc.in2p3.fr, david.lopezval@uclouvain.be, tania.robens@tu-dresden.de}
\vskip1mm

\end{center}

\vspace{.2cm}

\hrule \vspace{0.3cm}
{\small  \noindent \textbf{Abstract} \\[0.3cm]
\noindent  
We study the decay of a heavy Higgs boson into a light Higgs pair at one loop
in the singlet extension of the Standard Model.
To this purpose, we construct several
renormalization schemes for the extended
Higgs sector of the model.
We apply these schemes 
to {calculate} the heavy-to-light
Higgs decay width $\Gamma_{\Hhh}$ at next-to-leading order electroweak accuracy,
and demonstrate that certain  prescriptions
lead to gauge-dependent results.
We comprehensively examine how the NLO predictions depend 
on the relevant singlet model parameters,
with emphasis on the trademark behavior of 
the quantum effects, and how these change under
different renormalization
schemes and a variable renormalization scale. 
Once all present constraints on the model are included, 
we find mild NLO corrections, typically of few percent, and with small
theoretical uncertainties. 
\vspace{0.15cm}
\hrule

 \vspace{0.15cm}
 }

\section{Introduction}

With the discovery of the Higgs boson~\cite{Aad:2012tfa,Chatrchyan:2012xdj}, 
the LHC
has finally reached the very frontiers of the Standard Model (SM).
Dedicated  analyses based on Run I data have so far shown
excellent agreement between the observed $125$ GeV bosonic resonance
and the scalar particle originally postulated by 
Higgs~\cite{\higgs}, Englert and Brout~\cite{Englert:1964et},
and Guralnik, Hagen, and Kibble~\cite{Guralnik:1964eu}. 
Notwithstanding, it is an ongoing task to decipher whether such a  state
corresponds indeed to the SM agent 
of electroweak (EW) symmetry breaking~\cite{Spira:1997qz,Djouadi:2005gi,Plehn:2009nd}, or if alternatively 
the LHC has unveiled just one Higgs-like state
among many others, a composite state, or the
overlap of multiple resonances, just to mention few possibilities. 
Moreover, the state-of-the-art precision in the Higgs coupling extraction lies
within the $10-20\%$ level \cite{Englert:2014uua}, right at the ballpark of the deviations
predicted by popular new physics models. The overall picture
strengthens the belief that perhaps 
the Higgs discovery is in fact our first glimpse at
a more fundamental {UV complete structure.}

\medskip{}
The arguably most simple, renormalizable extension of the Higgs sector, is constructed
by expanding 
the SM Lagrangian with one additional spinless real electroweak singlet \cite{\singlet}. 
This adds up one extra
Higgs companion to the physical spectrum of the model, 
providing an excellent framework to guide collider searches for exotic scalars, either
via direct production or through off-shell effects.
{Moreover}, the coupling between the doublet and the
singlet
fields mixes the two neutral Higgs states,
leading to rescaled Higgs couplings to the SM particles. 
A chief prediction for collider phenomenology are 
the universally suppressed cross sections and partial amplitudes
in all Higgs production modes and decay channels.

\medskip{}
Another paramount signature of a second Higgs is the possibility 
of the heavy-to-light 
Higgs decay mode $\Hzero \to \hzero\hzero$~\cite{Schabinger:2005ei,Bowen:2007ia}. The process
is governed by the Higgs self-coupling
$\lambda_{Hhh}$, and as such it constitutes a direct
probe of the scalar potential of the model.
If the new Higgs boson 
is lighter than the SM Higgs, the novel $\Hhh$ decay
distorts all Higgs branching ratios,
and thereby its signal strengths, from the SM expectations.
Alternatively, if the extra scalar is identified
with a heavier Higgs companion and $m_{\Hzero} > 2 m_{\hzero}$,
$\Hzero \to \hzero\hzero$ can significantly contribute to the heavy Higgs {width and} lineshape
and modify its decay pattern.

\medskip{}
In this paper we present a 
thorough study of the heavy-to-light Higgs decay mode 
$\Hzero \to \hzero\hzero$ in the singlet extension of the SM, including the complete
set of radiative corrections at one loop.
Aside from
being relevant on its own, 
we use this process as a physics
case to construct and compare different renormalization schemes for the 
extended Higgs sector of the model. 
{Using} the {\sc Sloops} \cite{Boudjema:2005hb,Baro:2007em,Baro:2008bg,Baro:2009gn} 
general non-linear
gauge fixing setup, we illustrate how certain prescriptions still exhibit gauge dependence for physical quantities.
Our task carries to completion
the steps initiated in our previous publication ~\cite{Lopez-Val:2014jva}, 
and sharpen up 
all theoretical tools necessary to completely characterize the singlet model phenomenology at next-to-leading order (NLO) accuracy.

\medskip{}
The remainder of the paper is organized as follows. In Section~\ref{sec:model} we 
provide a brief description of the model setup and constraints. In Section~\ref{sec:renormalization}
we discuss in full detail our renormalization setup. 
We devote Section~\ref{sec:decay} to characterize the general
aspects of heavy-to-light Higgs decays at leading order and at one loop,
while in Section~\ref{sec:numerical} we present a detailed phenomenological analysis. 
We summarize and conclude in Section~\ref{sec:conclusions}.
Additional analytical details are provided in the Appendix.

\section{Model setup at the classical level}
\label{sec:model}
We construct the singlet extension of the SM by adding
one colorless,
real scalar field, which transforms
as a singlet under the $SU(2)_L\otimes\,U(1)_Y$ gauge charges \cite{\singlet,\singletours}.
Such a simple renormalizable extension 
can be viewed as a
simplified model approach to the low-energy Higgs sector of a more fundamental UV completion,
for instance the decoupling
limit of the Next-to-Minimal Supersymmetric Standard Model (NMSSM) 
\cite{Ellwanger:2009dp}, some realizations of GUTs 
\cite{Hetzel:2015bla}, models with additional
gauge sectors \cite{Basso:2010jm} or hidden valleys \cite{Strassler:2006im}. The implications
of this model were addressed for the first time
in Refs.~\cite{Silveira:1985rk,Schabinger:2005ei,Patt:2006fw}, and it has been
the object of dedicated
investigation for the past two
decades, displaying a very attractive phenomenology, 
especially in the context of collider physics, ~see e.g. 
\cite{O'Connell:2006wi,BahatTreidel:2006kx,Barger:2007im, Bhattacharyya:2007pb, 
Gonderinger:2009jp, Dawson:2009yx, Bock:2010nz,Fox:2011qc, 
Englert:2011yb,Englert:2011us,Batell:2011pz, Englert:2011aa, Gupta:2011gd, Dolan:2012ac, 
Bertolini:2012gu,Batell:2012mj,Bazzocchi:2012pp,Lopez-Val:2013yba,Heinemeyer:2013tqa,
Chivukula:2013xka,Englert:2013tya,Cooper:2013kia,Caillol:2013gqa,Coimbra:2013qq,
Eichhorn:2014qka,Martin-Lozano:2015dja,Falkowski:2015iwa,Dawson:2015haa,Buttazzo:2015bka,Banerjee:2015hoa,Englert:2015zra,Chen:2014ask, Englert:2014ffa}.
It has also been subject to many {dedicated} searches by the LHC experiments, 
cf. e.g. \cite{ATLAS:2014rxa,ATLAS:2014kua,Aad:2014ioa,Khachatryan:2015cwa,Aad:2015pla,Aad:2015xja} for {recent studies.}  

\subsection{Classical Lagrangian}
\medskip{}
The singlet scalar extension of the SM (denoted as $xSM$) is defined by 
the Lagrangian
\begin{equation}
\mathscr{L}_{xSM}=\mathscr{L}_{\rm gauge}+ \mathscr{L}_{\rm fermions} + \mathscr{L}_{\rm 
Yukawa} + \mathscr{L}_{\rm scalar} + \mathscr{L}_{GF} + \mathscr{L}_{ghost}
 \label{eq:mainlag}
\end{equation}\noi
where the gauge boson and fermionic kinetic parts $\mathscr{L}_{\rm gauge, fermions}$, as well as the
Yukawa interaction $\mathscr{L}_{\rm Yukawa}$, are given by the respective SM contributions.
The gauge-fixing and ghost Lagrangians $\mathscr{L}_{GF, ghost}$ will be defined in more 
detail below. The scalar sector is given by  
\begin{equation}
\mathscr{L}_{\rm scalar}= \left( \mathcal{D}^{\mu} \Phi \right) ^{\dagger} \mathcal{D}_{\mu} \Phi + 
 \partial^{\mu} S \partial_{\mu} S - \mathcal{V}(\Phi,S ) \, , \label{eq:singlet-lagrangian}
\end{equation}
\noindent where $\mathcal{D}_\mu$ is the covariant derivative and
$\mathcal{V}(\Phi,S )$ the scalar potential
\begin{alignat}{5}
\mathcal{V}(\Phi,S) = 
  \mu^2\,\Phi^\dagger\,\Phi 
+ \lambda_1\,|\Phi^{\dagger}\Phi|^2 
+ \mu^2_{s}\,S^2 
+ \lambda_2\,S^4 
+ \lambda_3\,\Phi^{\dagger}\,\Phi S^2 \; .
\label{eq:singlet-potential}
\end{alignat}
\noindent The latter corresponds to the most general,
$SU_L(2) \otimes U(1)_Y$-invariant, renormalizable Lagrangian 
involving the Higgs doublet $\Phi$ and the singlet $S$ fields, {and compatible
with an additional discrete $\mathcal{Z}_2$ symmetry, that precludes other terms with odd (e.g. cubic) field powers in the potential.}
By assuming all parameters in Eq.~\eqref{eq:singlet-potential} to be real,
we disregard additional sources of \CP~violation. 

\subsection{Mass spectrum}

The doublet and singlet
fields are expanded as
\begin{equation}
\Phi = \begin{pmatrix} G^+ \\[1mm] \dfrac{v + \phi_h + i G^0}{\sqrt{2}}
         \end{pmatrix} 
\qquad \qquad S = \cfrac{v_s +\phi_s}{\sqrt{2}} \; ,
\label{eq:fields-def}
\end{equation}
where $v \equiv \sqrt{2}\braket{\Phi} =  (\sqrt{2}\,G_F)^{-1/2} \simeq 246$~GeV
and $v_s \equiv \sqrt{2}\braket{S}$ stand for their respective vacuum expectation values (vevs).
The fields $G^\pm, G^0$ denote the charged and neutral Goldstone bosons. {
Since the singlet transforms trivially under the electroweak gauge group,}
only the doublet vev $v$ takes
part in electroweak symmetry breaking, which proceeds exactly as in the SM case.

\smallskip{}
The linear terms in the fields $\phi_h$ and $\phi_s$ from Eq.~\eqref{eq:singlet-potential} lead to the tadpole relations
\begin{align}
 T_{\phi_h}&= \mu^{2}v + v^{3}\lambda_1 + \frac{1}{2}v v_{s}^{2} \lambda_3
 \label{Tadpole_h} \\
T_{\phi_s}&= \mu_{s}^{2}v_s + v_{s}^{3}\lambda_2 + \frac{1}{2}v_{s} v^{2} \lambda_3
 \label{Tadpole_s},
\end{align}\noi 
\noindent by which we can express the minimization condition of the Higgs potential \eqref{eq:singlet-potential} as $T_{\phi_h,\phi_s} = 0$.

\medskip{}
In turn, the
quadratic terms in the Higgs fields may be arranged into 
a $2\times 2$ squared mass matrix $\mathcal{M}_{hs}^2$. In the gauge basis
$\left(\phi_h,\phi_s\right)$ 
these take the form
\begin{equation}
\label{eq:sqmassmatrix}
 {\cal M}^2_{hs}= {\cal T}_{\phi_h,\phi_s} +  {\cal M}^2_{\phi_h,\phi_s},
\end{equation}\noi 
where the tadpole component ${\cal T}_{\phi_h,\phi_s}$ and the squared mass matrix ${\cal M}^2_{\phi_h,\phi_s}$ are defined 
by
\begin{equation}
{\cal T}_{\phi_h,\phi_s}= 
\begin{pmatrix}
\frac{T_{\phi_h}}{v} & 0 \\
0 &  \frac{T_{\phi_s}}{v_{s}}
\end{pmatrix}, \qquad\qquad
\mathcal{M}^2_{\phi_h,\phi_s} =  \left( \begin{array}{cc}  
      2\lambda_1\,v^2 & \lambda_3\,v\,v_s \\
      \lambda_3\,v\,v_s & 2\lambda_2\,v_s^2  
     \end{array} \right) 
\label{eq:massmatrix}\; .
\end{equation}\noi 
Requiring this matrix to be positively defined leads
to the stability conditions
\begin{alignat}{5}
 \lambda_1, \lambda_2 > 0; \qquad 4\lambda_1\lambda_2 - \lambda_3^2 > 0 \label{eq:stability}\; .
\end{alignat}

Once we impose the tadpoles $T_{\phi_{h,s}}$ to vanish, Eq.~\eqref{eq:massmatrix} can 
be readily transformed into
the (tree-level) Higgs mass basis through the rotation
$U(\al)\cdot\mathcal{M}^2_{\phi_h,\phi_s}\cdot U^{-1}(\al) = \mathcal{M}^2_{hH} =
\text{diag}(m_{\hzero}^2 \;, m^2_{\Hzero})$, the physical 
Higgs masses reading
\begin{alignat}{5}
 m^2_{\hzero,\Hzero} &= \lambda_1\,v^2 + 
 \lambda_2\,v_s^2 \mp |\lambda_1\,v^2 - 
 \lambda_2\,v_s^2|\,\sqrt{1+\tan^2(2\al)}
\label{eq:masseigen}.
\end{alignat}
We identify the corresponding mass-eigenstates as a light [$\hzero$] and a heavy
[$\Hzero$] $\mathcal{CP}$-even Higgs boson, which are related 
to the gauge eigenstates through
\begin{equation}
\label{eq:singlet-rotation}
 \begin{pmatrix}
          h\\
          H \\
  \end{pmatrix}
  = U (\alpha)  \begin{pmatrix}
          \phi_h\\
          \phi_s \\
  \end{pmatrix} = 
  \begin{pmatrix}
   \cos \alpha & - \sin \alpha \\
   \sin \alpha & \cos \alpha \\
  \end{pmatrix}
   \begin{pmatrix}
          \phi_h\\
          \phi_s\\
  \end{pmatrix},
\end{equation}
\noindent where 
the rotation angle $\alpha$ is defined in the range $-\pi/2 \leq \alpha \leq \pi/2$ by 
\begin{equation}
 \sin 2 \alpha = \frac{\l_3 v v_s}{\sqrt{\left(\l_1 v^2-\l_2v_s^2\right)^2+(\l_3 v v_s)^2}}, \quad 
 \cos 2 \alpha = \frac{\l_2v_s^2-\l_1 v^2}{\sqrt{\left(\l_1 v^2-\l_2v_s^2\right)^2+(\l_3 v v_s)^2}}
 \label{angle}.
\end{equation}
Likewise the tadpoles in the gauge basis [$T_{\phi_h},T_{\phi_s}$] may be rotated into the mass basis [$T_h$,$T_H$] through $U(\alpha)$:
\begin{equation}
\label{tadpolephys}
 \begin{pmatrix}
          T_h\\
          T_H \\
  \end{pmatrix} = 
  \begin{pmatrix}
   \cos \alpha & - \sin \alpha \\
   \sin \alpha & \cos \alpha \\
  \end{pmatrix}
   \begin{pmatrix}
          T_{\phi_h}\\
          T_{\phi_s}\\
  \end{pmatrix}
\end{equation}
The above equations are of service to recast the quartic couplings in the Higgs 
potential~\eqref{eq:singlet-potential} as given by the physical Higgs boson masses $m^2_{\hzero,\Hzero}$
and the mixing angle $\alpha$,
\begin{align}
 \l_1 &=\frac{m_{h}^{2}}{2 v ^{2}} \cos^{2}\alpha + \frac{m_{H}^{2}}{2 v ^{2}} \sin^{2}\alpha 
 - \frac{\cos \alpha T_{h} + \sin \alpha T_{H}}{2 v^{3}}+ \frac{m^2_{hH}}{2 v^2}\sin 2 \al 
 \label{l1_fctparamphy}\\
 \l_2 &=\frac{m_{h}^{2}}{2 v_{s} ^{2}} \sin^{2}\alpha + \frac{m_{H}^{2}}{2 v_{s} ^{2}} \cos^{2}\alpha 
 - \frac{\cos \alpha T_{H} - \sin \alpha T_{h}}{2 v_{s}^{3}}- \frac{m^2_{hH}}{2 v_s^2}\sin 2 \al 
 \label{l2_fctparamphy} \\
 \l_3&=\frac{m_{H}^{2}-m_{h}^{2}}{2 v v_{s}}\sin 2\alpha+ \frac{m^2_{hH}}{v v_s}\cos 2 \al
 \label{l3_fctparamphy}\, .
\end{align}\noi
The mixed mass term $m^2_{hH}$ denotes the (symmetric) off-diagonal element of the squared
mass matrix $\mathcal{M}^2_{hH}$, defined in the mass-eigenstate basis.
While at tree-level we have $T_{h,H} = 0$ and $m^2_{hH}=0$, keeping these dependencies  
explicit in Eqs.~\eqref{l1_fctparamphy}-~\eqref{l3_fctparamphy} will be useful
to link the Lagrangian parameter counterterms to the corresponding mass counterterms, as we discuss in Section~\ref{sec:renormalization}. 
Similarly,
it is practical to rephrase 
the bilinear mass terms $\mu$ and $\mu_s$ in Eq.\eqref{eq:singlet-potential} as 

\begin{align}
 \mu^2 &= -\frac{1}{2}m_h^2 \cos^2 \alpha - \frac{1}{2}m_H^2 \sin^2 \alpha 
- \frac{(m_H^2-m_h^2)v_s}{4 v}\sin 2\alpha+\frac{3(\cos\alpha T_h+\sin\alpha 
T_H)}{2v} -\frac{m^2_{hH}v_s }{2 v}\cos 2 \al
 \label{eq:muphys}.\\
 \mu_{s}^2 &= -\frac{1}{2}m_h^2 \sin^2 \alpha - \frac{1}{2}m_H^2 \cos^2 
\alpha - \frac{(m_H^2-m_h^2)v}{4 v_s}\sin 2\alpha-\frac{3(\sin\alpha T_h-\cos\alpha 
T_H)}{2v_s} -\frac{m^2_{hH}v }{2 v_s}\cos 2 \al
 \label{eq:musphys}.
\end{align}

\subsection{Input parameters\label{sec:countparam}}

The Higgs sector of the model
is determined at tree-level by
i) the doublet vev, bilinear mass term and quartic self-coupling; ii) their counterparts for the singlet field;
and iii) the portal coupling $\lambda_3$ between both. 
 The singlet vev is traded as customary via the parameter $\tan\beta \equiv\,\frac{v_s}{v}$. 
\footnote{{Note the different conventions
for $\tan\be$ employed in the literature. The definition we adopt herewith 
is the inverse of that from 
Refs.~\cite{Pruna:2013bma,Lopez-Val:2014jva,Robens:2015gla}.}}
With the help of the above relations~\eqref{Tadpole_h}-\eqref{eq:musphys}, we can conveniently recast them
in terms of the following five independent parameters: 
\begin{equation}
m_{\hzero},\,m_{\Hzero}, \,\sin\al, \,{v}, \,\tan\beta\, \label{eq:inputs}.
\end{equation}
\noindent Two of them are readily fixed in terms of experimental data:  
the doublet vev is linked to the Fermi constant through $v^2 = (\sqrt{2}\,G_F)^{-1}$, 
while one of the Higgs masses is given by the LHC value $125.09\,\GeV$ \cite{Aad:2015zhl}.
Overall, we are left with three quantities which 
define the relevant directions in the new physics parameter space. This is also 
helpful to identify which physical parameters can be used 
to fix the  
renormalization conditions.

\subsection{Gauge-Fixing Lagrangian}

Gauge invariance will play an important role when 
discussing the renormalization of the singlet Higgs sector, in particular in defining
a gauge-independent mixed mass counterterm, as we discuss
in detail in Sections~\ref{sec:OS} and~\ref{sec:gauge-numerical}.
Such non-linear gauges have proven 
useful to check the {gauge independence} of {higher order calculations} within the SM 
\cite{Belanger:2003sd,Boudjema:1995cb,Boudjema:2009pw}, and its supersymmetric extensions 
\cite{Boudjema:2005hb,Baro:2007em,Baro:2008bg,Baro:2009gn,Baro:2009na,Chalons:2011ia,
Chalons:2012qe, Chalons:2012xf,Belanger:2014roa}. 
The gauge-fixing Lagrangian can be written in general as
\begin{equation}
\label{gaugefixing}
{\mathscr L}_{GF} = -\frac{1}{\xi_W} F^+ F^- - \frac{1}{2  {{ \xi_Z}}  }|
F^Z|^2 - \frac{1}{2 {{ \xi_A}} } | F^A|^2\, ,
\end{equation}\noi 
where the functions $F$ depend non-linearly on the Higgs and gauge fields,
\begin{eqnarray}
F^\pm & = & \bigg(\partial_\mu \mp ie {\tilde{\alpha}}  A_\mu \mp ig\cos
\theta_W 
{\tilde{\beta}} Z_\mu\bigg) W^{\mu \, +}\non \\
&  &\pm i{{ \xi_W}}  \frac{g}{2}\bigg(v +  {\tilde{\delta}}_1
h + {\tilde{\delta}_2} H \pm i{\tilde{\kappa}} G^0\bigg)G^+\\
F^Z & = & \partial_\mu Z^\mu + {{ \xi_Z}}  \frac{g}{2\cos \theta_W}\bigg(v +
    {\tilde{\epsilon}}_1  h +  {\tilde{\epsilon}}_2 H   \bigg)G^0 \\
F^A & = & \partial_\mu A^\mu\,.
\end{eqnarray}\noi 
{In the above equations} $e$ is the electromagnetic coupling constant, $g$ the $SU(2)_L$
coupling constant and $\theta_W$ denotes the weak mixing angle.
The quantities $\{ \tilde \alpha,\tilde \beta \cdots \tilde \epsilon_2\}$ correspond to
the generalized gauge-fixing parameters. Setting these parameters to zero leads to the
standard linear $R_{\xi}$ gauge fixing, with $\xi_i\,=\,1$ defining
the familiar 't~Hooft--Feynman gauge. In our renormalization setup we
will take 
these gauge-fixing terms already as \textit{renormalized} quantities -- in such
a way that no additional counterterms ${\delta}{\mathscr L}_{GF}$ are introduced
for this part of the Lagrangian. 

\smallskip{} In turn,
the ghost Lagrangian $\mathscr{L}_{ghost}$ is derived by requiring the complete
Lagrangian at the quantum level to be invariant under BRST transformations. 
This means $\delta_{\mbox{\tiny BRST}} \Lag_{xSM} = 0$ and hence
$\delta_{\mbox{\tiny BRST}}\Lag_{GF} = - \delta_{\mbox{\tiny BRST}}\Lag_{ghost}$.
This follows from the fact that by construction
the gauge, fermionic, Yukawa and scalar components of $\Lag_{xSM}$ are invariant under 
BRST transformations, as {the latter} are equivalent to gauge transformations.  
We consider both ${\mathscr L}_{GF}$ and $\mathscr{L}_{ghost}$ to be written in terms of 
renormalized quantities. The BRST transformations specific to the singlet extension are 
given by
\begin{align}
 \delta_{\mbox{\tiny BRST}}G^0 &= + \frac{g}{2}[G^- c^+ + 
G^+ c^- ] - \frac{e}{2c_W s_W} c^Z [ v + c_\alpha h + s_\alpha H]\\
 \delta_{\mbox{\tiny BRST}} G^\pm & = \mp \frac{ig}{2}c^\pm [ v +
 c_\alpha h + s_\alpha H \mp i G^0 ]\non
\mp i e \left(c^A - \frac{s_W^2 - c_W^2}{2 s_W c_W} \right) G^\pm \\
\delta_{\mbox{\tiny BRST}} h &=  c_\alpha \left( \frac{ig}{2}[G^- c^+ -  G^+
c^- ] +\frac{e}{2c_W s_W} c^Z G^0\right) \\
\delta_{\mbox{\tiny BRST}} H &=  s_\alpha \left( \frac{ig}{2}[G^- c^+ -  G^+
c^- ] +\frac{e}{2c_W s_W} c^Z G^0\right),
\end{align}\noi 
where $c_\theta,s_\theta$ are shorthand notations for $\cos \theta, \sin
\theta$,  while 
$c^Z, c^\pm, c^A$ stand for the Faddeev-Popov ghost fields associated to the
$Z^0$, $W^\pm$ and the photon field respectively. 
Within this particular gauge fixing, 
we set in practice $\xi_{W,Z,A} = 1$. This is convenient {since the gauge boson propagators take a very 
simple form}, while we still keep the possibility to check the gauge {independence}
of the final result, at the expense of adding new {gauge parameter-dependent} vertices 
to the model \cite{Belanger:2003sd}. 
The gauge independence of a calculation can be examined numerically by varying 
the non-linear gauge parameters $\tilde \alpha,\tilde \beta \cdots \tilde 
\epsilon_2$. Technically,  
we perform our implementation of the singlet model with non-linear gauge fixing  
using {\sc LanHEP} \cite{Semenov:1996es,Semenov:2014rea} and {\sc Sloops} 
\cite{Boudjema:2005hb,Baro:2007em,Baro:2008bg,Baro:2009gn}. 

\subsection{Interactions}

The key theoretical structure in the model is the 
doublet-singlet portal coupling 
$\lag_{xSM} \supset \lambda_3(\Phi^\dagger\,\Phi)\,S^2$, which is responsible
for the Higgs mass eigenstates
to be admixtures of the doublet $\phi_h$ and the singlet $\phi_s$ neutral
components. One main consequence of this mixing
is the universal depletion of all Higgs boson couplings to the 
SM particles as
\begin{alignat}{5}
g_{xxy} = g_{xxy}^{\text{SM}}(1 + \Delta_{xy}) \qquad \text{with} \qquad 1+\Delta_{xy} = 
 \begin{cases}\cos\alpha& y\,=\,\hzero\\ \sin\al&y=\Hzero \end{cases} \label{eq:coupling}.
\end{alignat}
\noindent This global rescaling is ultimately due to the fact that, owing to electroweak gauge invariance, only the doublet component
can couple to the fermions (via ordinary Yukawa interactions) and the gauge bosons
(via the gauge covariant derivative). 
The limits
of $\sin\alpha = 0$ (resp. $\cos\alpha =0$) correspond to the decoupling
scenarios for the light (resp. heavy) Higgs bosons, in which all
couplings of the additional scalar to SM fields identically vanish.
The Higgs self-interactions do not obey
such a plain rescaling pattern. Instead, they depend non-trivially
on the cross-talk between the singlet and the doublet fields. 
Analytic expressions for their Feynman rules can be found in the Appendix.

\subsection{Constraints}\label{sec:constraints}

As discussed above, the singlet model specified by the {Lagrangian~\eqref{eq:mainlag} }
is subject to numerous constraints, which have been explicitly discussed
in \cite{Pruna:2013bma,Lopez-Val:2014jva,Robens:2015gla}. 
Although {our primary focus in this paper
is the structure of the higher order corrections in this model irrespectively} of the parameter constraints, 
we briefly remind the reader which ranges are still feasible from {the} theoretical {and} experimental sides
-- and include {all of them} in our phenomenological analysis of Section~\ref{sec:numerical}. We consider
\paragraph{Theoretical constraints}
\begin{itemize}
\item{}perturbative unitarity at tree level \cite{Lee:1977eg,Luscher:1988gc}, 
which {is} taken into account by diagonalizing the five-dimensional $X\,\rightarrow\,Y$
scattering basis, with $X,Y\,\in\,\left\{ hh,\,hH,\,HH\,W^+ W^-,\,ZZ   \right\}$\footnote{{Electro}weak {gauge} bosons are replaced by Goldstone scalars according to the Goldstone equivalence theorem \cite{Chanowitz:1985hj}.}.
\item{}perturbativity of the {self-}couplings in the scalar potential, i.e. $|\lam_i \lb \mu_\text{run} \rb|\,\leq\,4\,\pi$, 
where the couplings {are evolved} through standard renormalization group equations \cite{Lerner:2009xg}
and {evaluated} at {a reference high-energy scale} 
$\mu_\text{run}\,\sim \,4\,\times\,10^{10}\,\GeV$\footnote{{This scale has been chosen such that the model still guarantees vacuum stability 
at a scale slightly larger than the SM breakdown scale for which $\lam_1\,\leq\,0$ in the SM limit ($\sin\al\,=\,0$). Requiring validity up to
higher scales leads to stronger constraints, cf. the discussion in \cite{Pruna:2013bma}.}}
for the high-mass and at $\mu_\text{run}= v = 246$ GeV for the low-mass scenario (see
\cite{Pruna:2013bma,Robens:2015gla} for a more detailed discussion). The high-mass (resp. low-mass) regions
correspond to $m_{H} > 2m_h$, with $m_{h} = 125.09$ GeV (resp. $m_{H} > 2m_h$, with $m_{H} = 125.09$ GeV); 
\item{}{vacuum stability} (cf. Eqn. (\ref{eq:stability})) {up to} the same {high-energy} 
scale.
\end{itemize}
\paragraph{Experimental constraints}
\begin{itemize}
\item{} electroweak parameters $S,T,U$ \cite{Altarelli:1990zd, Peskin:1990zt, Peskin:1991sw, Maksymyk:1993zm}  {in agreement with the  $95 \%$ C.L best-fit values from} \cite{Baak:2014ora};
\item{}similarly, agreement with the measured value of the $W$-mass at $95 \%$ C.L. (see \cite{Lopez-Val:2014jva} for more details);
\item{}agreement with collider searches from LEP, Tevatron, and the LHC, as implemented in HiggsBounds \cite{Bechtle:2008jh, Bechtle:2011sb,Bechtle:2013wla};
\item{}agreement with the Higgs signal strength measurements {at $95 \%$ C.L.},
as implemented in HiggsSignals \cite{Bechtle:2013xfa}. 
In addition, we have applied the {constraints from the} combined signal strength fit, presented in \cite{ATLAS-CONF-2015-044}, which lead to $|\sin\al|\,\leq\,0.36$ for $m_H\,\geq\,152\,\GeV$
\footnote{A detailed discussion of the determination of limits from the Higgs signal strength can be found in \cite{Robens:2015gla}. 
For $m_H\,\leq\,152\,\GeV$, we test a two-scalar hypothesis versus the LHC data, leading to an $m_H$-dependence for the respective $\chi^2$. Results in \cite{ATLAS-CONF-2015-044}, 
however, are derived under an SM assumption. For $m_H\,\geq\,152\,\GeV$, the $\chi^2$ is independent of the second resonance mass 
{and in this range we therefore adopt the improved combined experimental limit.}}.
\end{itemize}
It is interesting to observe the interplay of these different constraints on the overall parameter space. 
We here only summarize {the  main features}\footnote{See also \cite{sbm}.} -  
a dedicated discussion can be found in \cite{Robens:2015gla}. 
\begin{itemize}
\item{\bf in the high mass region} the {leading} constraints 
stem from i) direct searches (for $m_H\,\lesssim\,300\,\GeV$)\footnote{Note that {the most recent} experimental searches published in 2015 have not been included. 
These {potentially} influence the allowed regions for $m_H\,\lesssim\,300\,\GeV$. Indeed, preliminary studies show that results from \cite{Khachatryan:2015cwa} especially modify constraints for $m_H\leq\,250\,\GeV$ \cite{Robens:2016xkb}.}; 
{ii) the difference between the experimental W-mass measurement and its theoretical prediction~\cite{Lopez-Val:2014jva}} (in the 
intermediate range $M_H\,\in\,[300\,\GeV;\;800\,\GeV]$);
and iii) perturbativity of the {self-}couplings in the scalar potential (for $m_H\,\geq\,800\,\GeV$). 
All these features are summarized in Figure~\ref{fig:sinamw} {and Table~\ref{tab:highm}};
\begin{figure}
\centering
\includegraphics[width=0.85\textwidth]{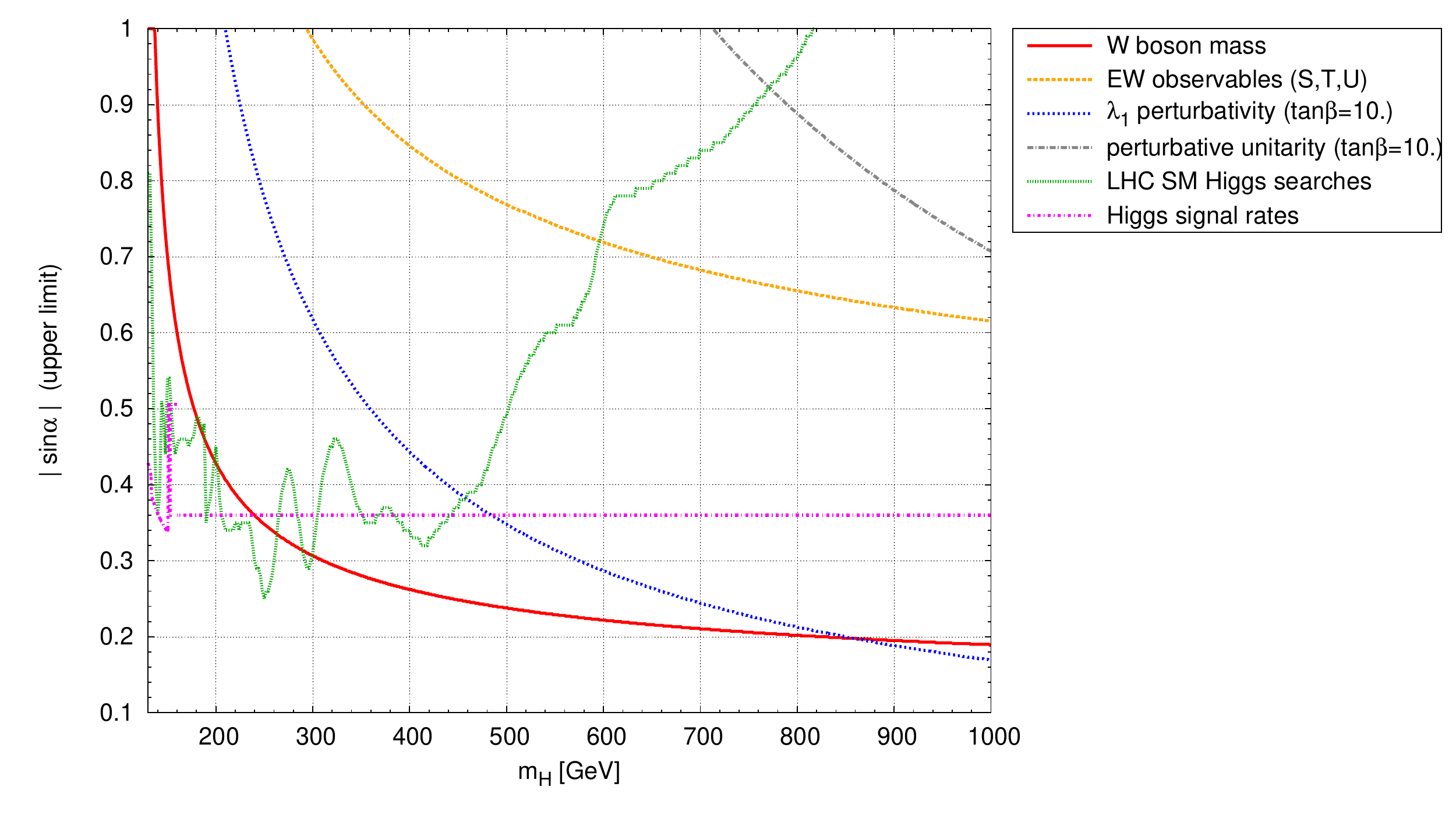}
\caption{\label{fig:sinamw} Maximal allowed values for $| \sin\al |$ in the high mass region, 
{for a heavy Higgs boson mass in the range} $m_H\in [130, 1000]\,\GeV$, from the {following constraints:} 
i) $W$ boson mass {measurement} (\emph{red, solid})~\cite{Lopez-Val:2014jva}; ii) electroweak precision observables (EWPOs) 
tested via the oblique parameters $S$, $T$ and $U$ (\emph{orange, dashed}); iii) perturbativity,
of the RG-evolved coupling $\lam_1$ (\emph{blue, dotted}), evaluated {for an exemplary choice} $\tan\be\,=\,10$, 
iv) perturbative unitarity (\emph{grey, dash-dotted}), v) direct LHC searches (\emph{green, dashed}), and vi) Higgs signal strength measurement (\emph{magenta, dash-dotted}).
For Higgs masses $m_H\,\in\,[300\,\GeV;800\,\GeV]$ the $W$ boson mass {measurement} 
{yields the strongest constraint} \cite{Lopez-Val:2014jva}. {The present plot} corresponds to an update of figure 8 from \cite{Robens:2015gla}, 
where the latest constraints from the combined signal strength \cite{ATLAS-CONF-2015-044} have been taken into account.}
\end{figure}
\item{\bf in the low mass region} where $m_H\,\sim\,125\,\GeV$, the parameter space is extremely 
constrained, especially {from demanding} agreement with the {LHC} Higgs signal strength measurement and the LEP constraints. 
{In Table~\ref{tab:lowscale} }
we summarize these constraints. 
Note that {in this regime} the {SM limit} corresponds to $|\sin\al|\,=\,1$. 
\end{itemize}

\begin{center}
\begin{table}[t]
\begin{center}
\begin{tabular}{| c || c | c | c |}
\toprule
$m~[\GeV]$ & $|\sin\al|$& source upper limit &$(\tan\be)_\text{min}$\\
\hline
{$1000$} &{$[0.018, 0.17]$}&{$\lam_1$ perturbativity}&$ {4.34}$\\
{$900$}&{$[0.022, 0.19]$}&{$\lam_1$ perturbativity}&${3.85}$ \\
{$800$}&{$[0.027, 0.21]$}&{$m_W$ at NLO}& ${3.45}$\\
{$700$}&{$[0.031, 0.21]$}&{$m_W$ at NLO}& ${3.03}$ \\
$600$ &  $[{0.038}, {0.23}]$	&	$m_W$ at NLO		&${2.56}$\\
$500$  &  $[{0.046}, {0.24}]$	& 	$m_W$ at NLO			&${2.13}$\\
$400$   &  $[{0.055}, {0.27}]$	&	$m_W$ at NLO				&${1.69}$ \\
$300$   &  $[0.067, {0.31}]$		& 	$m_W$ at NLO					  &${1.28}$\\
$200$  &  $[{0.090}, {0.36}]$	&signal rates   &${0.85}$\\
$180$ &  $[0.10, {0.36}]$		&      signal rates				    &$0.77$\\
$160$ &  $[{0.11}, {0.36}]$		&      signal rates				     &${0.68}$\\
$140$  &  $[{0.16}, 0.31]$	       & {signal rates}& ${0.60}$\\
\hline
\end{tabular}
\caption{\label{tab:highm} Table II from \cite{Robens:2015gla}, with adjusted conventions for $\tan\be$, 
and updated constraints on the maximally allowed mixing angle from {the combined Higgs signal strength fit} \cite{ATLAS-CONF-2015-044}. {It presents} allowed ranges 
for $\sin\alpha$ and $\tan\beta$ in the high mass region for fixed Higgs masses $m$. The allowed interval of $\sin\alpha$ 
was determined {fixing $(\tan\be)^{-1}=0.15$}.
The $95\%~\mathrm{C.L.}$ limits on $\sin\al$ from the Higgs signal rates are derived 
from one-dimensional fits and taken at $\Delta\chi^2 = 4$. The lower limit on $\sin\al$ always stems from vacuum stability, 
and the upper limit on $\tan\be$ always from perturbativity of $\lam_2$, evaluated at $\sin\al\,=\,0.1$. 
The source of the most stringent upper limit on $\sin\alpha$ is named in the third column. 
We fixed $m_h={125.1}~\GeV$
and the stability and perturbativity were tested at the reference scale $\mu_\text{run}\sim\,4\,\times\,10^{10}\,\GeV$.
}
\end{center}
\end{table}
\end{center}

Tables~\ref{tab:highm} and \ref{tab:lowscale} show the current constraints for the maximal ({minimal}) allowed values of $\sin\al$ and $\tan\be$, following the analysis 
presented in \cite{Robens:2015gla}. Note that the minimal $\tan\be$ values shown here were taken at 
a fixed value of $\sin\al$, so results from a {generic} scan might slightly differ. 
{All the constraints mentioned above have been taken into account} when considering {viable parameter space regions of the model} {for our numerical analysis in} Section \ref{sec:numerical}.
{Also} the results from the combined ATLAS and CMS signal strength fit have been included when applicable. 
We expect the results from the {most} recent LHC searches to influence the {global} picture {in the mass region} $m_H\lesssim\,350\,-\,400\,\GeV$, {while} 
for higher values the $W$ boson mass still poses the strongest constraint on the mixing angle.

\begin{center}
\begin{table}
\begin{center}
\begin{tabular}{| c || c | c | c | c|}
\toprule
$m_h~[\GeV]$& $|\sin\al|_\text{min, {HB}}$ & $|\sin\al|_\text{min, {HS}}$ &$(\tan\be)_\text{min}$&$(\tan\be)_{\text{no}~H\to hh} $\\
\hline
{$120$} & {0.410}  & {0.918}    & {0.12}&-- \\
$110$&${0.819}$&${0.932}$&${0.11}$&--\\
$100$&${0.852}$&${0.933}$&$0.10$&--\\
$90$&${0.901}$& -- &$0.09$&-- \\
$80$&${0.974}$&--&$0.08$&--\\
$70$&${0.985} $&--&$0.07$&--\\
$60$&${0.978}$&${0.996}$&$0.06$&{4.76}\\
$50$&${0.981}$&${0.998}$&$0.05$&{5.00}\\
$40$&${0.984}$&${0.998}$&$0.04$&{5.56}\\
{$30$} &{0.988}&{0.998}& {0.03}&{6.25} \\
{$20$} &{0.993}&{0.998}&{0.02}&{8.33} \\
{$10$} &{0.997}&{0.998}&{0.01}&{12.5} \\
\hline
\end{tabular}
\caption{\label{tab:lowscale} Table III from \cite{Robens:2015gla}, 
with adjusted definition for $\tan\be$  and
updated constraints on the {minimally} allowed mixing angle from {the combined Higgs signal strength fit} \cite{ATLAS-CONF-2015-044}.
{It presents} limits on $\sin\al$ and $\tan\be$ in the low mass scenario
{for various light Higgs masses $m_h$}. The limits on $\sin\al$ have been determined at $\tan\be=1$. 
{The lower limit on $\sin\al$ stemming from exclusion limits from LEP or LHC Higgs searches {is obtained} using 
\HB\ \cite{Bechtle:2008jh,Bechtle:2011sb,Bechtle:2013gu,Bechtle:2013wla} and given in the second column.
If the lower limit on $\sin\al$ obtained from the test against the Higgs signal rates using \HS\ \cite{Bechtle:2013xfa} results in stricter limits, we display them in the third column.}
The upper limit on $\tan\be$ in the fourth column stems from perturbative unitarity for the complete decoupling case ($|\sin\al|\,=\,1$).
In the fifth column we give {the $\tan\be$ value for which $\Gamma_{H\rightarrow hh}=0$ is obtained,
given the maximal mixing angle {allowed by the Higgs exclusion limits (second column)}. {At 
this $\tan\be$ value, the $|\sin\alpha|$ limit obtained from the Higgs signal rates (third column) is abrogated.}}}
\end{center}
\end{table}
\end{center}

\section{Renormalization}
\label{sec:renormalization}

\subsection{Setup}

The renormalization program we present here sticks close 
to the general strategy followed in
multidoublet Higgs extensions such as the MSSM \cite{Heinemeyer:2004ms,Baro:2008bg}
and the Two-Higgs-Doublet Model \cite{LopezVal:2009qy}.
We generate the required counterterms by introducing multiplicative renormalization constants 
to the weak coupling constant, fields, tadpoles,
masses and vevs. 
These are then
fixed by as many renormalization conditions as independent
parameters are present in the model \cite{Hollik:2002mv}. 
We adopt on-shell conditions 
to renormalize the electroweak gauge parameters \cite{\smrenorm} and the diagonal terms of the Higgs boson mass matrices. Using an on-shell scheme, as customary in this context, provides 
an unambiguous interpretation of the bare parameters in the classical Lagrangian
in terms of physically measurable quantities.
We also recall that field renormalization constants 
are not needed if we only require the observables derived from S-matrices to be
finite, but not each of the Greens' functions individually. They are nonetheless
convenient from the technical viewpoint, as they {account for} loop corrections
to the external legs and less Feynman diagrams have to be explicitly included.

We proceed as customary by splitting the
bare Lagrangian of the model \eqref{eq:mainlag}  
into the renormalized and the counterterm pieces as $\lag^0(\{X^0\}) \to
\lag(\{ X \}) + \delta\lag(\{ \delta X \})$.
Accordingly, we rewrite
each of the bare parameters
$X^0$ as a renormalized part $X$ and its counterterm $\delta X$.
For the purpose of this work we only need to deal with the scalar and gauge 
sectors
$\lag^0_{\rm scalar, gauge}$, as the other sectors
do not feature for the remainder of our discussion. We also recall that the gauge-fixing Lagrangian ~$\lag_{GF}$
does not contribute to $\delta\lag$,
since we choose to write it already in terms of \textit{renormalized fields and
parameters}~\cite{Belanger:2003sd,Baro:2008bg}.
The physical parameters of the
gauge sector are the electromagnetic coupling constant $e$ and the gauge boson
masses $m_W,m_Z$ that we split as \cite{Denner:1991kt},
\begin{equation}
 e^0 \ra e + \delta Z_e, \qquad (m^0_W)^{2} \ra m^2_W + \delta m_W^2, \qquad
(m^0_Z)^{2} \ra m^2_Z + \delta m_Z^2 \; . 
\end{equation}\noi 
Also the bare parameters appearing in $\lag^0_{\rm scalar}$ in
the gauge-eigenstate basis are decomposed as
\begin{align} \label{eq:shiftpotential}
 \l_{i}^{0} & \ra \l_i + \delta \l_i \quad [i=1\cdots 3], \qquad 
  v^0  \ra v + \delta v, \qquad v_{s}^{0} \ra v_s + \delta v_s,\nonumber\\
  \mu^0 & \ra \mu +\delta \mu, \qquad \mu_{s}^{0} \ra \mu_s + \delta \mu_s\; .
\end{align}\noi 
A similar splitting is introduced for the Higgs tadpoles $T^{0}_{\phi} \to T_{\phi} + 
\delta T_{\phi}$ ($[\phi = \phi_h,\phi_s]$),
which feature explicitly for calculations beyond the leading order.
Equivalent expressions
can be written trading some of the
above bare parameters for more physical ones through the relations given by 
Eqs.~\eqref{l1_fctparamphy}-\eqref{eq:musphys}.

\medskip{}
In our setup we choose not to renormalize the mixing angle $\alpha$.
Instead, we promote the
relation between the Higgs eigenstates in the gauge $(\phi_h,\phi_s)$ and mass
basis $(h,H)$ to be valid to all orders, 
\begin{equation}
 \begin{pmatrix}
          h\\
          H \\
  \end{pmatrix}^0
  = U (\alpha)  \begin{pmatrix}
          \phi_h\\
          \phi_s \\
  \end{pmatrix}^0\quad \mbox{and equivalently} \quad 
\begin{pmatrix}
          h\\
          H \\
  \end{pmatrix}
  = U (\alpha)  \begin{pmatrix}
          \phi_h\\
          \phi_s \\
  \end{pmatrix}.
\label{eq:nomixingCT}
\end{equation}\noi Doing so, the bare and the {physical} mixing angle
coincide and we need no additional {mixing angle} counterterm.

\medskip{}
In turn, field renormalization constants for the physical
Higgs states are introduced by shifting the bare Higgs
fields in the mass basis as
\begin{equation}
 \left( \begin{array}{c} \hzero \\ \Hzero \end{array} \right)^0
 \rightarrow
 \left(\begin{array}{cc} 
 1 + \cfrac{1}{2}\,\delta Z_h & \frac{1}{2} \delta Z_{hH} \\
\frac{1}{2} \delta Z_{Hh} & 1 + \cfrac{1}{2}\,\delta Z_H 
 \end{array}\right)\,\left( \begin{array}{c} \hzero \\ \Hzero \end{array}
\right)\, + \mathcal{O}(\alpha^2_{ew})
\label{eq:rc-massbasis1}\; .
\end{equation}
Finally, we introduce the {(matrix-valued) Higgs mass counterterm} via
\begin{equation} {\cal M}_{\phi}^2 \ra {\cal M}^2_{\phi} + \delta {\cal
M}^2_{\phi}\; , 
\end{equation}
\noindent where the generic index $\phi$ applies to the squared mass matrix in both the gauge and
the mass basis. Their respective matrix counterterms are linked through
\begin{equation} 
 \delta {\cal M}^2_{hH}= U(\alpha)\cdot \delta {\cal M}^2_{\phi_h, \phi_s}
\cdot U(-\alpha) = 
\begin{pmatrix} 
 \delta m_h^2 & \delta m^2_{hH} \\
  \delta m^2_{Hh} & \delta m_H^2
\end{pmatrix}\, , \label{eq:massct-main}
\end{equation}
\noindent where the mixed mass counterterms are symmetric $\delta m_{Hh}^2\,=\,\delta
m_{hH}^2$. 

Thus, aside from the purely SM inputs, the renormalization of the
scalar sector in the singlet model is completely specified
by four renormalization constants for the neutral Higgs fields, the respective
singlet and doublet vev counterterms, and five additional counterterms
linked to the parameters in the Higgs potential~\eqref{eq:singlet-potential}. In the mass eigenstate basis,
these can be traded by:
\begin{equation}
\begin{array}{lll}
\bullet\,\text{tadpoles:} \; \delta T_h, \delta T_{\Hzero} 
& \bullet\,\text{vev:} \;  \delta v, \delta v_s  &
\bullet\,\text{mixing:} \; \delta m^2_{\hzero\Hzero}

\\
\bullet\,\text{Higgs masses:} \; \delta m_{\hzero}^2, \delta
m_{\Hzero}^2 &\bullet\,\text{fields:} \; \delta Z_\hzero, \delta
Z_\Hzero, \delta
Z_{\hzero\Hzero}, \delta Z_{\Hzero\hzero} & \\
\end{array}
\label{eq:param-physical}
\end{equation}
Defining a renormalization
scheme is then tantamount to identifying a set of independent conditions by which to link {the above quantities} to physical inputs. The renormalization conditions by which we fix
these counterterms will rely on the one-point and two-point Greens' functions
of physical fields. Depending on the scheme we choose for Higgs field renormalization,
not all of the above field renormalization constants will be independent from each other.

A complete renormalization scheme fixes
 all the counterterms which are necessary to absorb the UV-divergent contributions from loop-level amplitudes, such that one obtains
 UV finite predictions for physical observables. Another important property of a renormalization scheme is gauge independence.
More precisely, 
maintaining  gauge {independence when defining} a scheme allows to write
 physical predictions as a function
of the input parameters in a way that does not vary when the gauge-fixing is changed. Only in this case one can unambiguously relate physical observables to
Lagrangian parameters. In this work 
we examine different strategies to extend the conventional SM
renormalization to the singlet model case,
and discuss in detail whether these comply with gauge independence. 
{The {\sc Sloops }non-linear gauge-fixing setup (cf. Eq.~\eqref{gaugefixing}) turns out to be instrumental in this task.}

\subsection{Gauge sector}

\medskip{}
We begin by introducing the on-shell definition of the electroweak mixing
angle 
$\sin^2\theta_W\,=\,1-m_W^2/m_Z^2$, {along with the shorthand notations $\swd \equiv \sin^2\theta_W$, $c^2_W \equiv 1-\swd$  \cite{Sirlin:1980nh}.
This relation
fixes the weak mixing angle counterterm $(\swd)^{0} \to \swd + \delta\swd$ as
\begin{alignat}{5}
 \cfrac{\delta \swd}{\swd} &= -\cfrac{\cwd}{\swd}\,\left(\cfrac{\delta \mw^2}{\mw^2} - \cfrac{\delta \mz^2}{\mz^2}\right)
 \label{eq:sw-ct}\, .
\end{alignat}}
The weak gauge boson masses are renormalized in the 
standard on-shell scheme~\cite{\smrenorm}, 
i.e. by requiring the real part of the transverse renormalized weak gauge boson
self-energies to vanish
at the respective
pole masses. The condition
\begin{alignat}{5}
\text{Re}\,\hat{\Sigma}^V_T(p^2)\,& = \text{Re}\,\Sigma^V_T(p^2) + \delta
Z_V\,(p^2-m_V^2) - \delta m_V^2\Bigg{\lvert}_{p^2 = m_V^2} = \,0 \qquad [V =
W^{\pm}, Z]
\label{eq:gaugeself}\; ,
\end{alignat}
\noindent where $\delta Z_V$ stands for the weak gauge boson field renormalization $V\to Z_V^{1/2}\,V = (1+1/2\delta Z_V)\,V + \mathcal{O}(\alpha^2_{\text{ew}})$,
leads to
\begin{equation}
 \delta m^2_W = - \text{Re}\,{\Sigma}^W_T(m_W^2) \qquad \mbox{and} \qquad \delta
m^2_Z =- \text{Re}\,\Sigma^Z_T(m_Z^2) 
 \label{eq:gaugemassct}\; .
\end{equation}
\noindent 
All renormalized self-energies are denoted hereafter by a hat.
The transverse part of the gauge boson self-energies follows from
the vacuum polarization tensor, 
\begin{alignat}{5}
 \Sigma_{VV'}^{\mu\nu}(p^2) & \equiv  \Sigma^{VV'}_T(p^2) + p^\mu\,p^\nu\,\Sigma^{VV'}_L(p^2)
 \label{eq:polarization-tensor}\, .
\end{alignat}
\noindent
The explicit form of the weak gauge boson two-point functions in the singlet model can be found in Ref.~\cite{Lopez-Val:2014jva}.

\medskip{}
To renormalize the electromagnetic coupling constant, we require
the electric charge to be equal to the full $ee\gamma$ vertex in the Thompson
limit. With 
the help of the QED Ward identities, this condition
is given in terms of the photon and mixed $Z-\gamma$ two-point functions 
\begin{alignat}{5}
\cfrac{\delta Z_e}{e} &= \cfrac{1}{2}\Pi_{\gamma}(0) + \cfrac{\sw}{\cw}\,\cfrac{\Sigma^T_{\gamma\,Z}(0)}{\mz^2},
\qquad \text{with} \qquad \Pi_{\gamma} = \cfrac{d^2}{\partial p^2}\,\Sigma_{\gamma\gamma}(p^2)\Bigg{\lvert}_{p^2=0}
 \label{eq:charge-ct1}.
\end{alignat}
To avoid large logarithms from light fermion masses, we rephrase as customary 
the photon vacuum polarization as
\begin{alignat}{5}
 \Pi_\gamma(0) &= \Delta\alpha_{\text{lep}} + \Delta\alpha_{\text{had}} + \cfrac{1}{\mz^2}\,\text{Re}\,\Sigma_{\gamma}^{\text{light f}}(m_Z^2)
 \label{eq:charge-ct2},
\end{alignat}
\noindent where the superindex indicates that 
only the light fermion contributions (all leptons and quarks, except the top) are included in the photon self-energy,
while the QED-induced shift to the fine structure constant,
\begin{alignat}{5}
 \Delta\alpha = \Delta\alpha_{\text{lep}} + \Delta\alpha_{\text{had}} = -\text{Re}\,\hat{\Pi}^{\text{lep}}_{\gamma}(\mz^2)
 -\text{Re}\,\hat{\Pi}^{\text{had}}_{\gamma}(\mz^2) \label{eq:dalpha1},
\end{alignat}
\noindent  is {known to very good accuracy} \cite{Steinhauser:1998rq,Hagiwara:2003da}.

The improved electric charge counterterm in the Thompson limit is thus given by
\begin{alignat}{5}
 \cfrac{\delta Z_e}{e} &= \cfrac{1}{2}\,\cfrac{d^2}{dp^2}\,\Sigma^{\text{no light f}}_{\gamma}(p^2)\Bigg{\lvert}_{p^2 = 0} + \cfrac{1}{2}\,\Delta\,\alpha
 + \cfrac{1}{{2} \mz^2}\,\text{Re}\,\Sigma^{\text{light f}}_{\gamma}(\mz^2) + \cfrac{\sw}{\cw}\,\cfrac{\Sigma^T_{\gamma\,Z}(0)}{\mz^2}
 \label{eq:charge-ct3}\, ,
\end{alignat}
\noindent in such a way that the 
value of the renormalized electric charge at zero momentum transfer $e(0) = \sqrt{4\pi\,\alpha_{em}(0)}$ can be extracted
from the measured fine-structure constant in this limit: \\ $\alpha_{\text{em}}(0) = 1/137.035999074(44)$~\cite{Agashe:2014kda}.

\medskip{}
On the other hand, the very precise measurement 
of the muon lifetime provides a 
link between the weak gauge boson masses, the fine structure
constant and the Fermi constant. This allows for
different input choices to fix the electroweak sector.
In our numerical analysis we shall use two alternative parametrizations:

\begin{itemize}
 \item The $\alpha_{\text{em}}$-parametrization, in which we select 
   $\alpha_{\rm em}(0)$ and $m_{W,Z}$ as input parameters;

   \item The $\gf$-parametrization, in which we instead replace  the W-boson mass 
   by the Fermi constant $G_F = 1.1663787(6)\times 10^{-5}$ GeV$^{-2}$ \cite{Agashe:2014kda}, 
   the latter being fixed
   by the muon lifetime via \cite{Hollik:1993cg,Hollik:1988ii,Hollik:2003cj,Hollik:2006hd}. 
   \end{itemize}

\noindent These two parametrizations are related via the conventional
parameter $\Delta r$ \cite{Kennedy:1988sn,Hollik:1988ii,Hollik:1993cg,Langacker:1996qb,Hollik:2003cj,Hollik:2006hd} as
\begin{eqnarray}
m_{\PW}^2 \left(1 - \frac{m_{\PW}^2}{m_Z^2}\right)=
\frac{\pi \alpha_{\text{{em}}}}{\sqrt{2} G_F} \left(1 + \Delta r\right)\,
\quad
\label{eq:deltar_def1}
\end{eqnarray}
\noindent  
where $m_{W,Z}$ and $\sw$ are renormalized in the on-shell scheme. 
For a detailed analysis of $\Delta r$ in the singlet model cf. Ref~\cite{Lopez-Val:2014jva}.
Since $\Delta r$ vanishes at leading-order, both parametrizations are 
trivially linked at tree-level as
$ \cfrac{G_F}{\sqrt{2}}= \cfrac{\pi\alpha_{\rm em}}{2\mw^2\,\swd}$, 
while they depart from each other at higher perturbative orders. We will explicitly 
quantify these departures
further down in Section~\ref{sec:numerical}.

\subsection{Extended Higgs sector}

\subsubsection{Tadpole renormalization}

For the tadpole renormalization we proceed as customary \cite{\smrenorm} and impose
\begin{alignat}{5}
 \hat{T}_{\hzero} = T_h + \delta T_h = 0; \qquad \qquad \hat{T}_H = T_{\Hzero} + \delta T_{\Hzero} = 0 
 \label{eq:rtadpole}.
\end{alignat}
\noindent This is equivalent to requiring that $v$ and $v_s$ are the physical 
vacuum expectation values of the doublet and the singlet fields respectively,
so that they define the  
(renormalized) minimum of the Higgs potential. 
In practice, this implies that no Higgs one-point
insertions feature {explicitly} in our calculation.

\subsubsection{Doublet vev renormalization}
The vev $v$ of the scalar doublet $\Phi$ is fixed as in the SM through
its relation to the electroweak on-shell parameters
\begin{alignat}{5}
v = \cfrac{2\,m_W\,s_W}{e} \quad \to \quad \cfrac{\delta v}{v} =
\cfrac{1}{2}\,\cfrac{\delta m_W^2}{m_W^2} + \cfrac{\delta s_W}{s_W} -
\cfrac{\delta Z_e}{e}
 \label{eq:ct-v}.
\end{alignat}\noi 
where all needed counterterms are defined in \eqref{eq:gaugemassct},
\eqref{eq:charge-ct3} and \eqref{eq:sw-ct}.

\subsubsection{Singlet vev renormalization\label{sec:singletvev}}

{The general renormalization transformation of} {a generic scalar field vev \cite{Sperling:2013eva}
can be particularized to the singlet vev case as}
\begin{equation}
 \phi_s + v_s \to Z_S^{1/2} (\phi_s + v_s + \delta \bar{v}_s) \label{eq:renorm-vevs},
\end{equation}\noi 
\noindent where we have introduced for convenience the singlet field renormalization 
in the gauge basis
$ S^{0} \ra Z_S^{1/2} S = \left(1+ \delta Z_S/2\right) S + \mathcal{O}(\aem^2)$. 
The additional counterterm $ \delta \bar{v}_s$ characterizes to what extent the
singlet vev renormalizes 
differently from the singlet {field} $\phi_s$ {itself}. In Ref. \cite{Sperling:2013eva} it was shown that, in an $R_\xi$ gauge,
{a} divergent part {for} $\delta \bar{v}_s$ is forbidden if 
the scalar field obeys a rigid invariance
(see
also Ref. \cite{Gorbahn:2009pp} and references therein). This is precisely the case 
in {the singlet} model, since the singlet field is unlinked from
the gauge sector and hence invariant
under global gauge transformations. 
In addition to that, the singlet field renormalization
constant $\delta Z_S$ is also UV-finite. This can be easily shown
by computing $\delta Z_S$
in the unbroken phase where $\braket{\Phi} = \braket{S} = 0$. Such a scenario is analogous 
to a plain $\lambda \phi^4$-theory, in which the (singlet) scalar field is coupled to a second scalar (doublet) field {only through} 
the gauge-singlet quartic coupling $\lag \supset \lambda_3\,\Phi^\dagger \Phi\,S^2$. In this case, 
all one-loop contributions to the singlet two-point function are momentum-independent, implying 
that $\delta Z_S$ does not get a UV pole (cf. \textit{e.g.} \cite{Itzykson:1980rh}).\footnote{We have numerically verified that $\delta Z_S$ is UV-finite at one loop in all of the different renormalization schemes.}
We thus conclude that the singlet vev counterterm
$\delta v_s = \delta \overline{v}_s + \delta Z_S/2$ defined by Eqs.~\eqref{eq:shiftpotential} and ~\eqref{eq:renorm-vevs}  
gets at most a finite contribution at this order.

\smallskip{} {We finally note that,}
given the condition of vanishing tadpoles~\eqref{eq:rtadpole}, 
the singlet vev $v_s$ corresponds to the physical minimum of the Higgs
potential in the broken phase in the singlet field direction, viz. $\braket{S} = v_s/\sqrt{2}$, at a given
order in perturbation theory. Since $v_s$ does not contribute to the
electroweak symmetry breaking, 
it cannot be fixed in terms of SM observables. Instead,
we must promote it to an independent input parameter (which should eventually be determined
from a future measurement of e.g. the $Hhh$ coupling). So doing, 
any finite shift  $\delta v_s$
can be subsumed into the physical definition
of $v_s$ itself at one-loop. Therefore, in our renormalization setup we can simply fix
$\delta v_s$ in the
$\msbar$ scheme, 
$\delta v_s^{\msbar} = 0$, so that no singlet vev counterterm features in one-loop calculations.
%
\subsubsection{Higgs masses renormalization}

The (matrix-valued) Higgs mass counterterm 
in the gauge basis yields
\begin{equation}
\label{system}
 \delta {\cal M}^2_{\phi_h, \phi_s} =
\begin{pmatrix}
  2(v^2 \delta \l_1+ 2 v \l_1\delta v)+\delta T_{\phi_h} /v & v_s( v\delta \l_3 + \l_3 \delta v) \\
  v_s( v\delta \l_3 + \l_3 \delta v) & 2v_s^2 \delta \l_2 +\delta T_{\phi_s}/v_s
\end{pmatrix},
\end{equation}
\noindent where we have already fixed $\delta v_s =0$, as justified above. This result can be linked as customary to the mass basis 
through Eq.~\eqref{eq:massct-main}. 

To renormalize the physical Higgs masses we impose on-shell conditions on 
the renormalized diagonal Higgs self-energies,
\begin{alignat}{5} \retildehat_\phi(m^2_{\phi}) = 0 \qquad \text{with} \qquad
 \retildehat_{\phi}(p^2) = \retilde_{\phi}(p^2)
 +\delta Z_{\phi}(p^2-m^2_{\phi})- \delta m^2_{\phi}, \qquad [\phi = \hzero,\Hzero]
  \label{eq:twopoint-diagonal},
\end{alignat}
\noindent whereby we obtain
\begin{alignat}{5}  \delta m^2_{\hzero} = \retilde_{\hzero}(p^2)\Big{\lvert}_{p^2 = m^2_{\hzero}} & \qquad \text{and}
  \qquad \delta m^2_{\Hzero} = \retilde_{\Hzero}(p^2)\Big{\lvert}_{p^2 = m^2_{\Hzero}}
  \label{eq:osmasses}.
 \end{alignat}\noi 
The explicit form of the field renormalization constants
$\delta Z_\phi$ in different schemes is discussed below in Section \ref{sec:wvfdiag}.

\bigskip{}
In theories where the gauge eigenstates mix, 
the renormalization of the non-diagonal or mixing terms
must be addressed with 
care 
(cf. \cite{Heinemeyer:2004ms,Baro:2008bg,Baro:2009gn}
for an analogue discussion in the context of the {squark sector in the} MSSM). 
As we have seen in Section~\ref{sec:model},
a \emph{bare} angle $\alpha_0 \equiv \al$ rotates 
the scalar fields from the gauge basis to the mass basis through Eq.~\eqref{eq:singlet-rotation}.
While such diagonal form is valid at leading order, radiative corrections
will in general misalign the (tree-level) mass eigenstates.  
This is reflected in the off-diagonal terms of the loop-corrected
propagators, 

\begin{\eqn*}
\Delta^{-1}_\text{Higgs}\,=\,\lb 
\begin{array}{cc}
p^2-m_{\hzero}^2+\hat{\Sigma}_{\hzero}(p^2)&\hat{\Sigma}_{\hzero \Hzero}(p^2)\\
 \hat{\Sigma}_{\Hzero \hzero}(p^2)&p^2-m_{\Hzero}^2+\hat{\Sigma}_{\Hzero}(p^2)
\end{array}\rb
\end{\eqn*}
\noindent traded by the non-diagonal Higgs two-point function
\begin{alignat}{5}
\retildehat_{\hzero\Hzero}(p^2) = \retilde_{\hzero\Hzero}(p^2)
 +\cfrac{1}{2}\,\delta Z_{\hzero\Hzero}(p^2-m^2_{\hzero})
 +\cfrac{1}{2}\,\delta Z_{\Hzero\hzero}(p^2-m^2_{\Hzero})  - \delta m^2_{\hzero\Hzero} \label{eq:twopoint-nondiagonal}.
\end{alignat}

\medskip{}
One possibility is to 
absorb these additional quantum effects into the renormalization of the mixing angle.  
This is equivalent to
diagonalizing the loop-corrected mass matrix further through an additional
rotation $U(\delta\alpha)$,
where $\delta\alpha$ plays the role of a mixing angle counterterm, 
such that $\alpha^0 \to \alpha + \delta\alpha$. 
Alternatively,
in our approach we take the mixing matrix
$U(\al)$ {as written in terms of the \emph{physical}
mixing angle and hence} valid to all orders.
The two alternative approaches are related through 
\begin{alignat}{5}
 \delta m^2_{hH} = ({m_{\Hzero}^2 - m_{\hzero}^2})\,\delta\alpha\, .
 \label{eq:mixedmass-angle}
\end{alignat}
The residual mixing induced by the off-diagonal terms in the mass matrix is
instead removed
by the non-diagonal field renormalization constants, which we present below.


\subsubsection{Higgs field renormalization: diagonal parts \label{sec:wvfdiag}}

Taking the Higgs boson masses $m_{h,H}$ as experimental inputs, 
we fix the diagonal field renormalization constants
via the on-shell conditions 
\begin{alignat}{5}
 \retildehat_{h}'(m^2_{\hzero}) = 0 \qquad \mbox{and} \qquad 
\retildehat_{H}'(m^2_{\Hzero}) = 0
\end{alignat}
\noindent where $\retildehat_{\phi}(p^2)$ was defined in 
Eq.~\eqref{eq:twopoint-diagonal},
while the familiar shorthand notation $f'(p^2)\equiv df(p^2)/dp^2$
denotes the derivative
with the respect to the momentum squared. 
This leads to
\begin{align}
\delta Z_h &= -\mbox{Re}\Sigma'_{hh}(m_h^2) \qquad \text{and} \qquad
\delta Z_H = -\mbox{Re}\Sigma'_{HH}(m_H^2) \label{eq:fielddiagOS}
\end{align}

\noindent which set the Higgs propagator residues to unity
in the limit $p^2 \to m^2_{\phi}$ ($\phi = \hzero,\Hzero$). 

\subsubsection{Higgs field renormalization: non-diagonal parts}
\label{sec:nondiag}

{Fixing} the non-diagonal {field renormalization} 
is {a crucial step} in setting up a \emph{gauge-invariant} scheme, in which the
renormalized one-loop amplitudes are independent of the gauge-fixing parameters, as discussed above. 
 We first {construct a set of schemes in analogy to the more familiar approaches} in the literature. As we will show, these
lead in general to gauge-dependent predictions for physical observables. {To circumvent this problem, we introduce an additional (dubbed \emph{improved}) scheme, 
which is defined
merely in terms of two-point functions and gives numerically stable results throughout the entire parameter space}. Similar discussions are addressed e.g. when defining  renormalization
schemes for the parameter $\tan\beta$ in the MSSM ~\cite{Freitas:2002um,Baro:2008bg}.

\paragraph{$\bullet$\, \underline{Minimal field:}}
\label{sec:msmin}

As a first setup to fix the non-diagonal Higgs field renormalization $\delta Z_{hH}$ we resort to a 
minimal field renormalization. We attach one single 
renormalization factor per field in the gauge basis,
\begin{alignat}{5}
\Phi & \to  Z^{1/2}_{\Phi}\,\Phi = \left(1+\cfrac{\delta Z_{\Phi}}{2} \right)\,\Phi + \mathcal{O}(\alpha^2_{ew}); 
\qquad \qquad S & \to Z_{S}^{1/2}\, S = \left(1+\cfrac{\delta Z_S}{2}\right)\,S+ \mathcal{O}(\alpha^2_{ew})\, ,
\label{eq:fieldrc1}
\end{alignat} 
\noindent where we have expanded them to first order. 

This procedure is in straight analogy to the conventional renormalization of the 
Higgs sector in multidoublet extensions such as the MSSM \cite{Heinemeyer:2004ms}
and the Two-Higgs-Doublet Model \cite{LopezVal:2009qy}. 
Assuming symmetric off-diagonal components, and using 
the rotation matrix $U(\al)$ in 
Eq.~\eqref{eq:singlet-rotation}, 
we can write the physical Higgs wave function renormalization constants
in terms of the gauge basis ones $\delta Z_{\Phi,S}$ as
\begin{alignat}{5}
 \delta Z_h &= c^2_\alpha\,\delta Z_{\Phi} + s^2_{\alpha}\,\delta Z_S; 
 \quad \delta Z_{H} = s^2_{\alpha}\,\delta Z_{\Phi} + c^2_\alpha\,\delta Z_S;
\quad
  \quad \delta\,Z_{hH} = s_{\alpha}c_{\alpha}(\delta Z_{\Phi} - \delta Z_S) =
\cfrac{1}{2}\,t_{2\alpha}\,
 \left[\delta Z_h - \delta Z_H\right] \label{eq:rc-relations},
\end{alignat}
\noindent with the shorthand notation
$\{s_\alpha,c_\alpha, t_{\alpha}\}$ $=$ $\{\sin \alpha,\cos \alpha, \tan {\alpha}\}$. 
The scheme is dubbed \emph{minimal} as 
the non-diagonal field renormalization $\delta Z_{hH}$ is not independent.
Instead, it  is linked to the
diagonal parts $\delta Z_{h,H}$, which we have already fixed via on-shell conditions
~\eqref{eq:fielddiagOS}. {Additionally}, since at one-loop
we have $\delta Z_S^{\msbar} = 0$ (cf. Section~\ref{sec:singletvev}), we can
further simplify the relations above to get 
\begin{alignat}{5}
 \delta Z_h = c^2_\alpha\,\delta Z_{\Phi} ; 
 \qquad \delta Z_{H} = s^2_{\alpha}\,\delta Z_{\Phi} ; 
\qquad \delta Z_{hH} = s_{\alpha}c_{\alpha}\delta Z_{\Phi}  = 
\cfrac{1}{2}\,s_{2\alpha}\,
 \left[\delta Z_h + \delta Z_H\right] \label{eq:rc-relations-modified}\, .
\end{alignat}
Finally, for the mixed mass counterterm, which enters explicitly in the $Hhh$ vertex 
counterterm, 
we demand the off-diagonal renormalized Higgs self-energy
to vanish at an arbitrary renormalization scale, 
\begin{alignat}{5}
 & \retildehat_{\hzero\Hzero}({p^2}){\Big{\lvert}_{p^2 = \mu_R^2}} = 0; \qquad
 \text{wherefrom} \qquad \delta m^2_{\hzero\Hzero} =  \retilde_{\hzero\Hzero}(p^2)\Big{\lvert}_{p^2 = \mu_R^2} + 
 \delta Z_{\hzero\Hzero}\left(\mu_R^2-\cfrac{m^2_{\hzero} + m^2_{\Hzero} }{2}\right)
  \label{eq:renorm-mixing}.
\end{alignat}
From Eq.~\eqref{eq:renorm-mixing} we see that in this scheme 
all vertices with external Higgs legs receive a
finite wave-function renormalization correction, which absorbs the
residual loop-induced $\hzero-\Hzero$ mixing for an external on-shell Higgs state. These
finite wave-function factors are
given in general by \cite{Heinemeyer:2004ms}
\begin{align}
     & \hat{Z}_{\hzero\Hzero} = -\cfrac{\rself_{hH}(m_{h}^2)}{m_h^2-m_{H}^2 +
\rself_{\Hzero}(m_h^2)} = 
          -\cfrac{\rself_{hH}(m_{h}^2)}{m_h^2-m_{H}^2} + \mathcal{O}(\alpha^2_{\text{ew}});  \notag \\
    & \hat{Z}_{\Hzero\hzero} = -\cfrac{\rself_{Hh}(m_{H}^2)}{m_H^2-m_{h}^2 + \rself_{\Hzero}(m_H^2)}
     =  -\cfrac{\rself_{Hh}(m_{H}^2)}{m_H^2-m_{h}^2} + \mathcal{O}(\alpha^2_{\text{ew}})
 \label{eq:finitewf},
\end{align}
\noindent where $\mathcal{O}(\alpha^2_{\text{ew}})$ denote the contributions
beyond one-loop accuracy. Since the diagonal field renormalization has been fixed via
on-shell conditions~Eq.~(\ref{eq:fielddiagOS}),
the diagonal finite factors at one loop yield $\hat{Z}_{h,H} = 1$ and hence
we do not include them explicitly.

\paragraph{$\bullet\,$\underline{On-shell:}}
\label{sec:OS}

We define a second prescription in close analogy to squark renormalization~\cite{\squarkos} \footnote{{While this work
was being finalized, we learned of the work ~\cite{Kanemura:2015fra}, which presents
a study of the quantum corrections to
the Higgs couplings to fermions and gauge bosons
in a similar {singlet} model {setup}. The renormalization
scheme for the extended Higgs sector used by these authors is equivalent
to the on-shell scheme we discuss here, and which, as we analyse in the following, 
is not gauge-independent.}}.
This time we attach one field renormalization constant $\delta Z_h, \delta Z_H$ per
Higgs field directly in the mass-eigenstate basis ~\eqref{eq:rc-massbasis1}, {in which case} 
the off-diagonal field renormalization constants $\delta Z_{hH}$ and $\delta
Z_{Hh}$ are \emph{not} directly related to the diagonal terms.
The diagonal parts $\delta Z_{h,H}$ are again given by the on-shell relations of Eq.~\eqref{eq:fielddiagOS}.
The non-diagonal field renormalization
constants are set up  by imposing
that no loop-induced $H-h$ transitions occur for external
Higgs states on their mass shell, \textit{i.e.}
\begin{alignat}{5}\label{eq:onshellcond}
  \retildehat_{\hzero\Hzero}(m_{\hzero}^2) = 0; \qquad \text{and} \qquad  \retildehat_{\hzero\Hzero}(m_{\Hzero}^2) = 0\, .
  \end{alignat}
Using Eq.~\eqref{eq:twopoint-nondiagonal} leads to
\begin{align}
\label{dZhH}
\delta Z_{hH} &=  \frac{2}{m_h^2-m_H^2}\left[\mbox{Re}\Sigma_{hH}(m_H^2)-\delta 
m^2_{hH}\right] \\ 
\label{dZHh}
\delta Z_{Hh} &=  \frac{2}{m_H^2-m_h^2}\left[\mbox{Re}\Sigma_{hH}(m_h^2)-\delta 
m^2_{hH}\right]\; .  
\end{align}\noi 
Therefore, to fully fix the non-diagonal renormalization constants one must provide a proper
definition of the mixed mass counterterm $\delta m^2_{hH}$.
One possibility, as inspired from \cite{\squarkos}, is to impose $\delta
Z_{\hzero\Hzero} = \delta Z_{\Hzero\hzero}$, which fixes $\delta m^2_{hH}$
accordingly as
 \begin{alignat}{5}
\delta m^2_{\hzero\Hzero} =
\cfrac{1}{2}\,\left[\retilde_{\hzero\Hzero}(m^2_{\hzero}) 
 +\retilde_{\hzero\Hzero}(m^2_{\Hzero})\right] \qquad \text{and} \qquad \delta Z_{\hzero\Hzero} = \cfrac{\retilde_{\hzero\Hzero}(m^2_{\Hzero}) -
 \retilde_{\hzero\Hzero}(m^2_{\hzero}) }{m_{\hzero}^2 - m_{\Hzero}^2} \label{eq:mixing-scheme2}.
\end{alignat}
The above condition 
removes the loop-induced $H-h$
mixing when either of the two Higgs bosons are on shell, so that
the physical states
propagate independently and do not oscillate. 

\medskip{}
The customary on-shell scheme, as well as the minimal field scheme
discussed above, show indisputable benefits,
e.g. the fact that all counterterms are given in terms
of two-point functions and related to physically measurable quantities.
However, both of them lead to renormalized one-loop 
amplitudes {which, albeit UV finite, {may still have a left-over dependence on the 
parameters of the gauge-fixing Lagrangian~\eqref{gaugefixing}.} 
This is a well known fact for on-shell fermion
\cite{Gambino:1998ec,Kniehl:2000rb,Barroso:2000is,Pilaftsis:2002nc,Denner:2004bm} and 
sfermion mixing in supersymmetric theories 
\cite{Espinosa:2001xu,Espinosa:2002cd,Baro:2009gn}. 
Exploiting the non-linear gauge fixing of Eq.~\eqref{gaugefixing},
we explicitly {verify} this drawback to appear in the singlet model case as well, and illustrate it
numerically in Section~\ref{sec:gauge-numerical}. {In this discussion,}
it is worthwhile recalling that
gauge dependencies may well persist in general in all non-physical building blocks which are involved in 
the renormalization of any gauge theory (e.g.
field renormalization constants). The key test for a given renormalization scheme
is thus
whether it leads to gauge-independent {\sl predictions} for physical observables. 
{In the minimal field and the on-shell schemes,
 renormalized one-loop amplitudes {are proven to}
contain left-over gauge-dependent contributions. 
These can be traced back to the mixed mass 
counterterm $\delta m^2_{hH}$,  
which also enters the non-diagonal field renormalization constants $\delta Z_{hH,Hh}$.}  {The former is 
fixed in these schemes through Eqs.~\eqref{eq:renorm-mixing} and~\eqref{eq:mixing-scheme2} respectively,
and ultimately follows from the $h-H$ mixing self-energy.} 
Using the non-linear
gauge of~\eqref{gaugefixing},  we find
\begin{eqnarray}
\label{mixselfNLG}
 \Sigma_{hH}(p^2) &=& \Sigma_{hH}(p^2)\big|_{\xi_{W}=\xi_Z=1,\tilde\delta_i=0} \non \\
&+& \frac{1}{16 \pi^2}\left\{ \frac{g^2}{2}\left[\tilde\delta_1(m^2_H-p^2)s_\alpha  
+\tilde\delta_2(m^2_h-p^2)c_\alpha\right]B_0\left(p^2,m_W^2,m_W^2\right) \right\} \non \\
&+&\frac{1}{16 \pi^2}\left\{ \frac{g^{'2}}{4 
s_W^2}\left[\tilde\epsilon_1(m^2_H-p^2)s_\alpha  
+\tilde\epsilon_2(m^2_h-p^2)c_\alpha\right]B_0\left(p^2,m_Z^2,m_Z^2\right) \right\}\; , 
\end{eqnarray}\noi 
where $B_0(p^2,m^2,m^2)$ is the two-point Passarino--Veltman scalar integral 
\cite{Passarino:1978jh} and the $\tilde{\delta}_i$ terms are a short-hand notation
for the non-linear gauge parameters in Eq.~\eqref{gaugefixing}. The first line of Eq.~\eqref{mixselfNLG} 
is identical to the {result of the self-energy}
computation in the 't Hooft--Feynman gauge.
The second and third lines correspond to the genuine gauge-dependent 
contributions in the non-linear gauge (we recall that for practical calculations we always set $\xi_{A,W,Z}=1$).
The latter enter the mixed mass counterterm definition through Eqs.~\eqref{eq:renorm-mixing} or ~\eqref{eq:mixing-scheme2}
and are responsible for the uncancelled dependencies
on the gauge-fixing parameters in the renormalized $\Hhh$ one-loop amplitude, which we pin down numerically in Section~\ref{sec:gauge-numerical}.

One first roadway to construct
a gauge-independent definition of $\delta m^2_{hH}$
alternative to Eq.~\eqref{eq:mixing-scheme2} would be to exploit the pole
structure of a process-specific one-loop amplitude (e.g. a Higgs decay)
in the limit $m_h^2 \to
m_H^2$, as suggested by Ref.~\cite{Baro:2009gn}. Such a limit
corresponds though to a vanishing quartic coupling $\l_3$  
~\eqref{l3_fctparamphy} and hence to a vanishing mixing angle $\alpha$~\eqref{angle}.
Therefore, in this no-mixing situation, $\delta m^2_{hH}$ cannot be defined
through the mixed self-energy $\Sigma_{hH}$, because it is identically zero. 
A second possibility would be  
to link
the problematic mixed mass counterterm to a physical observable directly - viz.
using a \emph{per se} gauge-independent quantity such as a decay rate or scattering cross section \cite{Freitas:2002um,Baro:2008bg}.
The price one would pay would be
a process-dependent scheme definition, and sometimes one would have to resort
to quantities out of current experimental reach. A third option is {retaining only} the UV-divergent part of such a quantity via an $\msbar$ prescription}, 
{which we examine next}. {Besides this possibility,}
we also propose an {\sl additional} prescription leading to {a} gauge invariant scheme, which furthermore do{es} not render artificially enhanced contributions in any part of the parameter space.

\paragraph{$\bullet$\,\underline{Mixed $\msbar$/on-shell:}}\label{sec:scheme-mixed}

{{In this case}} we trade $\delta m^2_{hH}$ by one of the Higgs 
self-coupling counterterms $\delta\lambda_i$ from Eq.~\eqref{eq:singlet-potential}, 
and fix it using $\msbar$ conditions. For convenience
we choose $\lambda_2$ and compute the divergent part of the one-loop correction to the singlet field four-point
coupling $\lambda_2\, S^4$. So doing we find
\begin{equation}
 \label{msbardl2}
\delta \l_2^{\msbar} = \frac{-1}{16 \pi^2}\left[\l_3^2 + 9 \l_2^2 \right]\Delta\, ,
\end{equation}\noi 
where $\Delta$ stands for the 
UV divergent part in dimensional regularization  
\begin{\eqn}\label{eq:delta}
\Delta \equiv 1/\epsilon - \gamma_E + \log(4\pi).
\end{\eqn}
This result is manifestly gauge independent, as it should,
and 
agrees
with the beta function for the singlet quartic coupling $\l_2$ given in 
Ref.~\cite{Robens:2015gla}.
The corresponding 
gauge-invariant counterterms for $\lambda_{1,3}$ can now be traded
by $\delta m^2_{h,H},\delta v,\delta T_h, \delta T_H$ and $\delta 
\l_2^{\msbar}$ using the relations from Eqs.~\eqref{l1_fctparamphy}-\eqref{l3_fctparamphy}, 
\begin{eqnarray}
\label{eq:dl1mixed}
 \delta \l_1 &= & \frac{\delta m_{h}^{2}+\delta m_{H}^{2}}{2 v^{2}} +  \frac{v s_\alpha - v_{s} c_\alpha}{2 v^{3}v_{s}}\delta T_h 
 - \frac{v_{s} s_\alpha + v c_\alpha}{2 v^{3}v_{s}}\delta T_{H} -\frac{v_{s}^{2}}{v^{2}}\delta \l_2^{\msbar}- \frac{2 \l_1}{v}\delta v;    \\
\label{eq:dl3mixed}
  \delta \l_3 &=& \frac{ct_\alpha \delta m^2_H-t_\alpha \delta m^2_h}{2 v v_s }+\frac{c_{2\alpha}}{2 v v_s^2}\left[\frac{\delta T_h}{c_\alpha}-\frac{\delta T_H}{s_\alpha}\right]
  - \frac{2}{t_{2\alpha}} \frac{v_s}{v}\delta \l_2^{\msbar}- \frac{\l_3}{v}\delta v. 
\end{eqnarray}\noi 
We are thus left with
\begin{eqnarray}
\label{eq:dMhHSloopS}
  \delta m_{hH}^{2} &= &v^{2} s_{2 \alpha} \delta \l_1 - \delta \l_2^{\msbar}
  v_{s}^{2} s_{2\alpha} + v v_{s} c_{2 \alpha}\delta \l_3 +\frac{s_{2\alpha}}{2}\left[\left(\frac{c_\alpha}{v}+\frac{s_\alpha}{v_s}\right)\delta T_h+\left(\frac{s_\alpha}{v}-\frac{c_\alpha}{v_s}\right) \delta T_H\right] \nonumber\\
& &+ \left(  2 v s_{2\alpha}\l_1 + v_{s} c_{2\alpha}\l_3 \right)  \delta v.
\end{eqnarray}
\noindent Finally, we {use the on-shell relations}~\eqref{dZhH}-\eqref{dZHh} to 
obtain the non-diagonal field renormalization 
constants, which are now fixed in terms of Eq.~\eqref{eq:dMhHSloopS}.

\medskip{}
Since all of the renormalization 
constants within $\delta m^2_{hH}$ are either related to physical observables
and/or correspond to prefactors of gauge invariant operators (e.g. $\lambda_{1,2,3}$)
the mixed mass counterterm $\delta m^2_{hH}$ is by construction gauge-invariant -- and leads
in turn to gauge-{independent} renormalized one-loop amplitudes,
as we prove numerically in Section~\ref{sec:numerical}. This observation,
together with the analytical structure of the mixed self-energy $\Sigma_{hH}(p^2)$ from Eq.~\eqref{mixselfNLG},
{reflects} that the renormalization conditions chosen for $\delta m^2_{hH}$
{(and linked to them, for $\delta Z_{hH,Hh}$)}
are the {ultimate} origin of the uncancelled gauge dependence found in
the minimal field and the on-shell schemes. 

In spite of leading to gauge-{independent} results, this mixed $\msbar$/on-shell scheme
tends to produce overestimated radiative corrections in the phenomenologically interesting 
regions ($s_{\alpha} \to 0, c_{\alpha} \to 0$), as manifest
from the analytic dependencies of the counterterms ~\eqref{eq:dl1mixed}-\eqref{eq:dl3mixed},
which are proportional to inverse powers of small trigonometric factors.
We therefore refrain from using this scheme explicitly in our phenomenological analysis, 
and instead propose an improved gauge-{independent} setup right below.
 
\paragraph{$\bullet$\, \underline{Improved on-shell}} \label{sec:scheme-impr}

A second alternative to sidestep {the gauge-dependent $\delta m^2_{hH}$} definition in the default
on-shell scheme is to isolate the gauge invariant part of the mixed self 
energy of Eq.~\eqref{mixselfNLG}. {In so doing, we can use it to define
the problematic mixed mass counterterm through 
a gauge-independent \emph{improved} self-energy \cite{Binosi:2009qm}.}
This is actually possible if
the mixed scalar self-energy \eqref{mixselfNLG} is computed in 
the linear 't Hooft--Feynman gauge and evaluated
at the average geometrical mass 
$\pstar^2 = (m_h^2+m_H^2)/2$. 
As shown in Ref.~\cite{Espinosa:2002cd} with the help of the so-called \emph{pinch technique}, 
\cite{Cornwall:1981zr,Cornwall:1989gv,Binosi:2009qm}, 
the mixed scalar self-energy~\eqref{mixselfNLG} obtained
in this way coincides with the gauge-invariant part of the 
\emph{pinched} result. While the results {proven} in 
Ref.~\cite{Espinosa:2002cd} are applied to 
the squark and Higgs sectors of the MSSM, the proof does not rely
on Supersymmetry and hence can be exported to the more general
case of a system of two gauge eigenstates which mix in the mass basis. In addition, self-energies computed
using the pinch technique are independent of the gauge-fixing scheme \cite{Binosi:2009qm}.
With these arguments in mind, we thus retain only 
the first line in Eq.~\eqref{mixselfNLG}  and define the mixed mass 
counterterm through
\begin{equation}
\label{eq:improvedOS}
 \delta m^2_{hH} = 
\mbox{Re}\,\Sigma_{hH}(\pstar^2)\big|_{\xi_{W}=\xi_Z=1,\tilde\delta_i=0} 
\quad \mbox{with}\quad\pstar^2 = \frac{m_h^2+m_H^2}{2}\, ,
\end{equation}\noindent
\noindent which must be therefore gauge-{independent} (as we again {confirm} numerically
in Section~\ref{sec:gauge-numerical}).
Finally, the non-diagonal field renormalization are once more fixed using OS conditions~\eqref{eq:onshellcond}
and fully determined in terms of $\delta m^2_{hH}$. 

\medskip{}
In Table~\ref{tab:scheme-summary} we provide a summarized overview of the different renormalization
schemes discussed in this section. Notice that they differ from each other in the
renormalization conditions used to fix the non-diagonal Higgs field renormalization $\delta Z_{hH,Hh}$
constants and the mixed mass counterterm $\delta m^2_{hH}$.

\begin{table}[htb!]
\begin{center}
\tiny{
\begin{tabular}{c|llll|} \hline
 & $\delta Z_{h,H}$ & $\delta Z_{hH,Hh}$ & $\delta m^2_{hH}$ \\ \hline \\
\multirow{2}{*}{Minimal field} & $\delta Z_{h} = -\retilde'_h(m_h^2)$ & 
$\delta\,Z_{hH} = \cfrac{1}{2}\,s_{2\alpha}\,
 \left[\delta Z_h + \delta Z_H\right]$
& $\retilde_{hH}(\mu_R^2) + \left[\mu_R^2-\cfrac{m_h^2+m_H^2}{2}\right]$ \\ 
 &   $\delta Z_{H} = -\retilde'_H(m_H^2)$ &   $\delta\,Z_{Hh} = \delta Z_{hH}$ & 
\\ \\ \hline \\
\multirow{2}{*}{OS} & $\delta Z_{h} = -\retilde'_h(m_h^2)$ 
& $\delta\,Z_{hH} = \cfrac{\retilde_{hH}(m_H^2) - \retilde_{hH}(m_h^2)}{m_h^2-m_H^2}$
& $\cfrac{\retilde_{hH}(m_h^2) + \retilde_{hH}(m_H^2)}{2}$ \\ 
 &   $\delta Z_{H} = -\retilde'_H(m_H^2)$ &   $\delta\,Z_{Hh} = \delta Z_{hH}$ & 
\\ \\ \hline \\
\multirow{2}{*}{Mixed $\msbar$/OS} & $\delta Z_{h} = -\retilde'_h(m_h^2)$ 
& $\delta\,Z_{hH} = \cfrac{2}{m_h^2-m_H^2}\,\left[\retilde_{hH}(m_H^2) - \delta m^2_{hH}\right]$
& Eq.~\eqref{eq:dMhHSloopS} \\ 
 &   $\delta Z_{H} = -\retilde'_H(m_H^2)$ &   $\delta\,Z_{Hh} = \cfrac{2}{m_H^2-m_h^2}\,\left[\retilde_{hH}(m_h^2) - \delta m^2_{hH}\right]$ & 
\\ \\ \hline \\
\multirow{2}{*}{Improved OS} & $\delta Z_{h} = -\retilde'_h(m_h^2)$ 
& $\delta\,Z_{hH} = \cfrac{2}{m_h^2-m_H^2}\,\left[\retilde_{hH}(m_H^2) - \delta m^2_{hH}\right]$
& $\retilde_{hH}(\pstar^2),\quad \pstar^2 = \cfrac{m_h^2+m_H^2}{2}$ \\ 
 &   $\delta Z_{H} = -\retilde'_H(m_H^2)$ &   $\delta\,Z_{Hh} = \cfrac{2}{m_H^2-m_h^2}\,\left[\retilde_{hH}(m_h^2) - \delta m^2_{hH}\right]$ &
\\ \hline
\end{tabular}}
\end{center}
\caption{Overview of the scheme-dependent counterterms in the different renormalization
setups considered in this paper.}
 \label{tab:scheme-summary}
\end{table}

\section{Heavy-to-light Higgs decay width}
\label{sec:decay}

\subsection{Leading-order contribution}

When kinematically accessible, the heavy-to-light Higgs decay mode $\Hzero \to \hzero\hzero $ proceeds at leading order (LO)
via the tree-level contact interaction $\lambda_{Hhh}$ with partial width~  \cite{Schabinger:2005ei,Bowen:2007ia}
\begin{alignat}{5}
 \Gamma^{\text{LO}}_{\Hhh} = \cfrac{\lambda^2_{\Hzero\hzero\hzero}}{32\,\pi\,m_{\Hzero}}\,\sqrt{1-\cfrac{4\mhd}{\mHHd}}
 \label{eq:lowidth},
\end{alignat}
\noindent where
\begin{\eqn}
\lambda_{\Hzero\hzero\hzero} \,=\, -\cfrac{i s_{2\alpha}}{v}\,\left[m^2_{\hzero}+\cfrac{m^2_{\Hzero}}{2}\right]\,(c_\alpha + s_\alpha\,t^{-1}_\beta)\; . 
 \label{eq:Hhhfirst} 
\end{\eqn}
{Notice that, owing to the structure
of the scalar self-coupling, the decay
width is not symmetric under a sign flip of the mixing angle $s_\alpha \to -s_\alpha$.}

As such, this decay mode constitutes a genuine new physics
contribution to the total heavy Higgs width - aside from the
global rescaling of its decay modes into SM particles. 
The opening of this novel channel is thus capable to alter
the Higgs boson lineshape significantly, as well as its decay pattern.
More specifically, the branching fractions of the heavy
Higgs boson of mass $m_H$ to SM fields $\phi$ are modified as 
\begin{alignat}{5}
 \br_{\Hzero \to \phi\phi} \lb m_H  \rb = \cfrac{s^2_\alpha\,\Gamma_{H\,\to\,\phi \phi}^{\text{SM}}\lb m_H  \rb }{s^2_\alpha\,\Gamma_{H_{\text{tot}}}^{\text{SM}}\lb m_H  \rb  + \GHhh\lb m_H  \rb }\, ,
 \label{eq:br-sm}
\end{alignat}
\noindent where $\Gamma^{\text{SM}}_H\lb m_H  \rb $ stands for the 
total width of a SM-like Higgs boson with mass $m_H$. 
For the lighter Higgs boson with mass $m_h$, the branching fractions are exactly
as for a SM-like Higgs with that mass.

\noindent Notice that for $\lambda_{Hhh}=0$, all partial decay widths are universally rescaled in terms of the Higgs mixing angle $\alpha$, 
leading to the same branching ratios that a Higgs boson of that mass would experience in the SM.

\begin{figure}[thb!]
\begin{center}
\includegraphics[width=0.35\textwidth,height=0.33\textwidth]{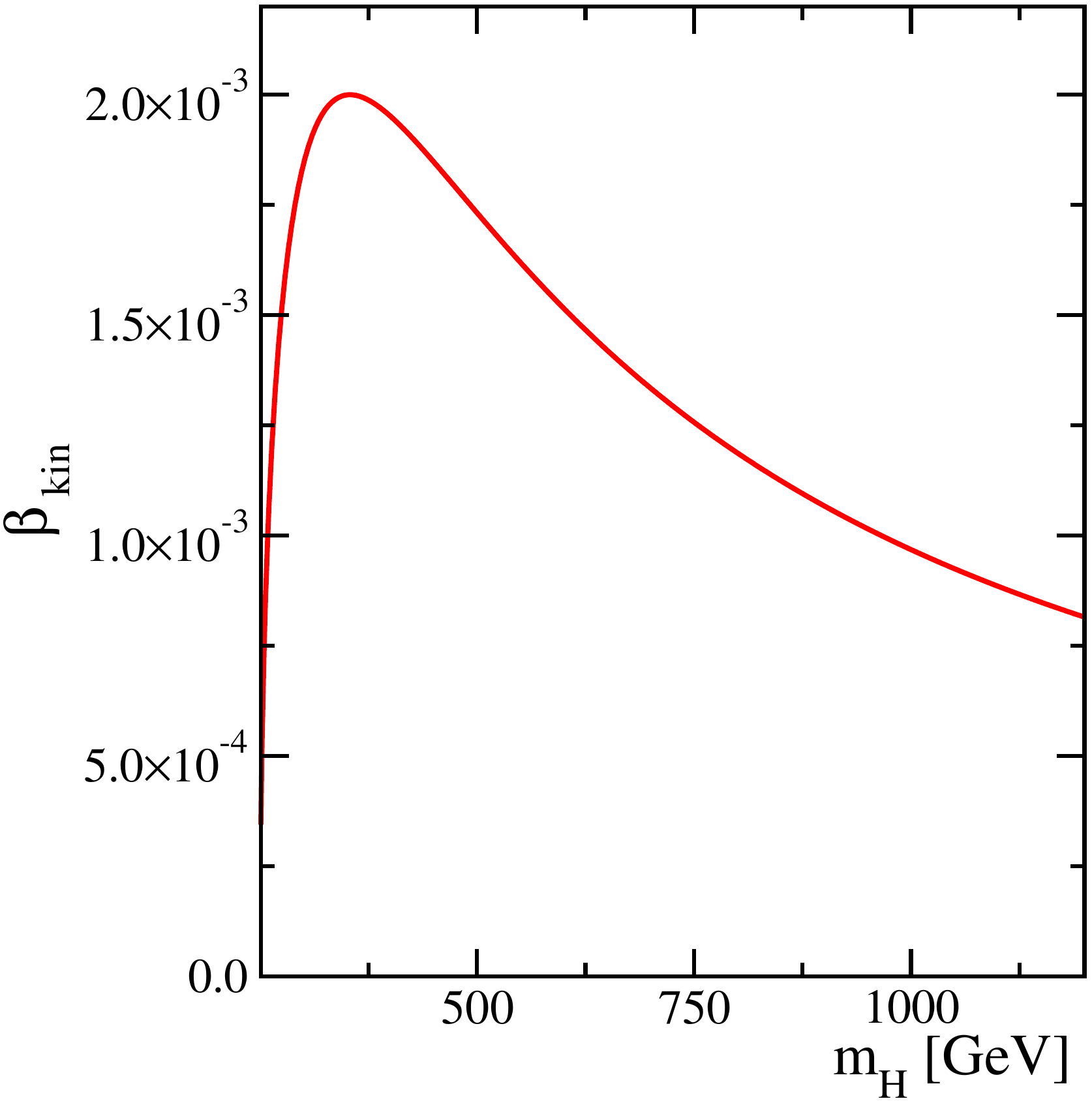} \hspace{1.7cm} 
\includegraphics[width=0.40\textwidth,height=0.35\textwidth]{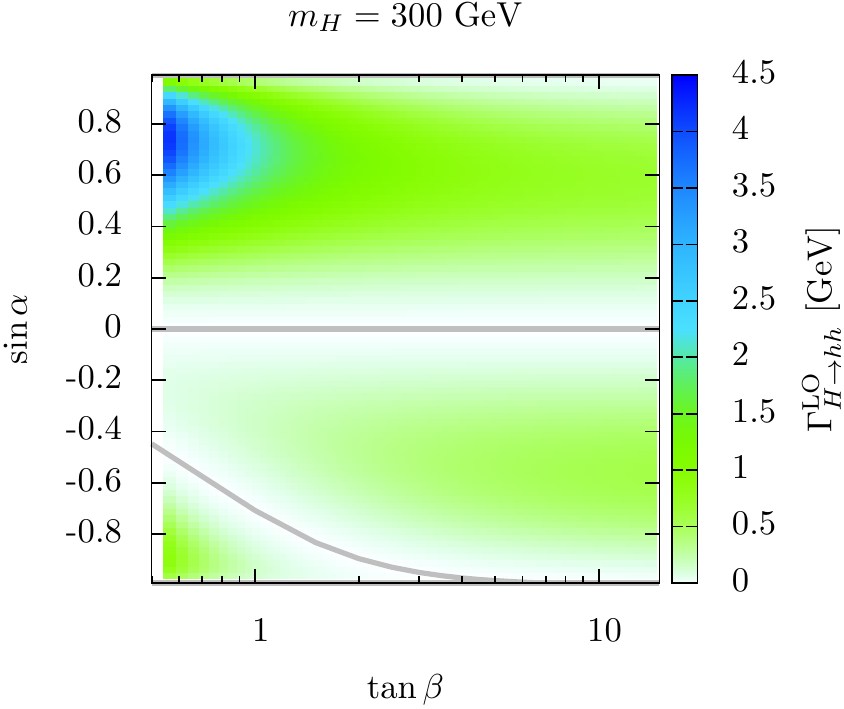} 
\end{center}
\caption{\underline{Left panel:} kinematical factor $\beta_{\text{kin}} = \sqrt{1-4m_h^2/m_H^2}$ as a function of 
the heavy Higgs mass, for $m_{\hzero} = 125.09$ GeV. \underline{Right panel:} leading-order heavy-to-light
Higgs decay width  $\Gamma^{\text{LO}}_{\Hhh}$ [in GeV] over the $\sin{\alpha} - \tan\beta$ plane for 
a fixed heavy Higgs mass of $m_H = 300$ GeV. 
The grey lines signal the configurations $\sin\alpha = 0$ and $\tan\beta = -\tan\alpha$ along
which $\Gamma^{\text{LO}}_{\Hhh}$ vanishes.}
\label{fig:kinematics}
\end{figure}

\medskip{}
Two competing mechanisms determine the overall size of $\GammaLO$.  On the one hand there is
the kinematic factor $\beta_{\text{kin}}/\mHHd = 1/\mHHd\sqrt{1-4m^2_{\hzero}/m^2_{\Hzero}}$, 
where $\beta_{\text{kin}}$ trades the light Higgs-pair velocity in the heavy Higgs boson rest frame.
Its dependence with respect to $\mHH$ is displayed in the left panel of Figure~\ref{fig:kinematics}, 
for a fixed light Higgs mass $m_{\hzero} = 125.09$ GeV.

The characteristic $\mathcal{O}(m_H^{-1})$ phase-space suppression 
is compensated by the trilinear Higgs coupling strength $\lambda_{Hhh}$,
which depends quadratically on both the light and the heavy Higgs masses.
On the other hand, there are $\cot\beta$-enhanced contributions
which can invigorate these Higgs self-interactions for $t_\beta < 1$, and push
the $\Hzero \to \hzero\hzero$ rates even higher. 
We illustrate these effects
in the right panel of Figure~\ref{fig:kinematics}, in which
we show the leading-order heavy-to-light Higgs decay width $\Gamma^{\text{LO}}_{\Hhh}$ in the $s_\alpha -t_\beta$ plane
for a heavy Higgs boson with mass $m_H = 300$ GeV. 
We can identify
three different configurations
in which the $\Hhh$ mode exactly vanishes {~\cite{Robens:2015gla}}:
i) the light Higgs decoupling limit, $s_\alpha = 0$; ii) the heavy Higgs
decoupling limit, $|s_\alpha| = 1$; and iii) the line $t_\beta = -t_\alpha$.
In cases i (resp. ii), all couplings of the heavy (resp. light) Higgs boson eigenstate
are identically zero. 
\medskip{}

\subsection{Electroweak one-loop corrections}

\begin{figure}[b!]
\begin{center}
\includegraphics[width=0.8\textwidth]{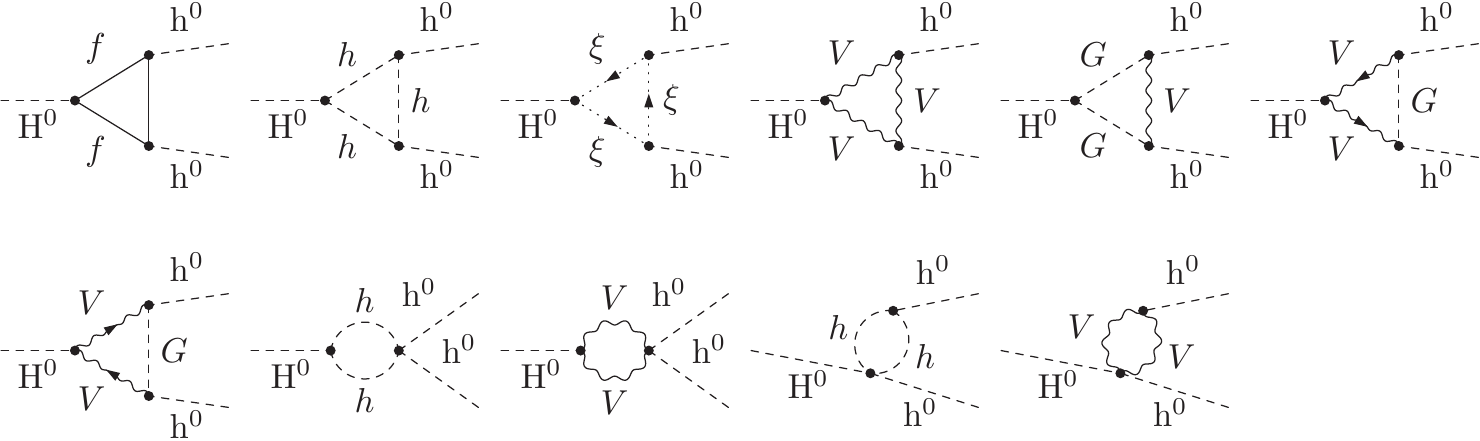}
\caption{\label{fig:oneloop} Representative Feynman diagrams 
for $\Hzero \to \hzero\hzero$ at one-loop electroweak accuracy in the ' t Hooft-Feynman gauge. 
The Feynman diagrams are generated using {\sc FeynArts.sty} \cite{Hahn:2000kx}.}
\end{center}
\end{figure}

Since all {external} particles involved in this process
are colorless and electrically neutral, the next-to-leading order (NLO)
corrections are given by purely weak one-loop effects. 
These $\mathcal{O}(\alpha_{\rm ew})$ corrections stem from the interference 
of the LO amplitude and different
subsets of one-loop graphs. On the one hand we have the genuine one-particle irreducible 
(1PI) vertex corrections. These include triangle and bubble-like three-point topologies
which involve the exchange of virtual heavy fermions, weak gauge bosons and Higgs bosons,
as generically illustrated in Figure~\ref{fig:oneloop}. The neutral Goldstone bosons and 
the $SU(2)_L$ Faddeev-Popov ghost contributions appear explicitly in the 't Hooft-Feynman 
gauge.
In addition to the genuine 1PI topologies, the one-loop  corrections
involve as well the $\Hzero\hzero\hzero$ vertex counterterm, which relies
on a combination of Higgs and gauge boson two-point functions,
as discussed beforehand 
in Section~\ref{sec:renormalization}. 
This contribution cancels the UV-divergent poles
of the 1PI amplitude and allows us to write the complete one-loop amplitude
in terms of physical (renormalized) parameters. Lastly, we must include
the finite wave-function
corrections to the external Higgs boson legs \eqref{eq:finitewf} in the minimal field scheme - while
for the on-shell schemes these are identically zero. 

\medskip{}
Combining all these pieces we may express the NLO heavy-to-light
Higgs decay width as
\begin{alignat}{5}
 \Gamma^{\text{NLO}}_{\Hhh} &= \cfrac{1}{32\pi\,m_H}\,\sqrt{1-\cfrac{4m_h^2}{m_H^2}}\,
 \left[
  \lambda_{Hhh}^2 + 2\,\text{Re}\,\lambda_{Hhh}\,
   \left(\delta\Gamma^{\bigtriangleup}_{Hhh} + 
  \delta\Gamma^{\text{WF}}_{Hhh} + \delta\lambda_{Hhh}\right)
  \right].
 \label{eq:nlowidth}
\end{alignat}
\noindent By $\delta\Gamma^{\bigtriangleup}_{Hhh}$ we denote the one-loop
contribution from the 1PI three-point vertex graphs.
The wave-function corrections yield 
\begin{alignat}{5}
 \delta\Gamma^{\text{WF}}_{Hhh} &= 2\,\hat{Z}_{hH}\,\lambda_{HHh}
 + \,\hat{Z}_{Hh}\,\lambda_{hhh} = {\cfrac{1}{m_h^2-m_H^2}}\,
  \left[\lambda_{hhh}\,\rself_{hH}(m_H^2) -2\,\lambda_{HHh}\,\rself_{hH}(m_h^2) \right]
 \label{eq:nlowf}\; ,
\end{alignat}
\noindent where we have introduced the finite field renormalization constants Eq.~\eqref{eq:finitewf}
and expanded them to first order in $\alpha_{\text{ew}}$. 
Finally, $\delta\lambda_{Hhh}$ stands for the counterterm of the trilinear scalar coupling. 
The latter is constructed from 
the tree-level expression \eqref{eq:Hhhfirst}, expanding all the bare quantities
as customary as $X^{0} \to X  + \delta X$. Doing so we find
\begin{alignat}{5}
 \delta\, \lambda_{\Hzero\hzero\hzero} &=  \lambda_{\Hzero\hzero\hzero}\,\left[\delta Z_{\hzero} + \cfrac{1}{2}\,\delta Z_{\Hzero} 
 + \cfrac{1}{2}\cfrac{\lambda_{\hzero\hzero\hzero}}{\lambda_{\Hzero\hzero\hzero}}\,\delta Z_{\hzero\Hzero} +
  \cfrac{\lambda_{\Hzero\Hzero\hzero}}{\lambda_{\Hzero\hzero\hzero}} \,\delta Z_{\hzero\Hzero}\right]   \notag \\
&\qquad + c_1^{Hhh}\,\delta m^2_{\hzero} + c_2^{Hhh}\,\delta m^2_{\Hzero} +  c_3^{Hhh}\,\delta\,\mhHd + c_4^{Hhh}\,\delta T_{\hzero} + c_5^{Hhh}\,\delta T_{\Hzero} 
+  c_6^{Hhh}\,\cfrac{\delta v}{v}
 \label{eq:ct-Hll},
\end{alignat}
\noindent where the coefficients $c_i$ are quoted separately in the Appendix.

\medskip{}

The relative size of the quantum effects is quantified through
the ratio
\begin{alignat}{5}
 \delta_{\alpha} &\equiv \cfrac{\Delta\Gamma_{\alpha}^{\text{1-loop}}}{\Gamma^{\text{LO}}_{\alpha}} =\cfrac{\Gamma^{\text{NLO}}_{\alpha} 
 - \Gamma_{\alpha}^{\text{LO}}}{\Gamma^{\text{LO}}_{\alpha}},
 \label{eq:rel-alpha}
\end{alignat}
\noindent where all quantities are given in the $\aem$-parametrization. 
The pure one-loop corrections $\Delta \Gamma^{\text{1-loop}}$ include all
terms stemming from the LO-NLO interference.

\section{Phenomenology}
\label{sec:numerical}

Hereafter we describe the phenomenology of heavy-to-light Higgs decays at NLO EW accuracy.
We begin in Section~\ref{sec:gauge-numerical}
by completing the discussion on the gauge dependence issues 
that were pointed out
qualitatively in Section~\ref{sec:renormalization}. We here revisit them on quantitative grounds
and justify the choice of the \emph{improved} on-shell scheme as our default
setup for the remainder of the analysis. Furthermore, we perform a dedicated numerical 
comparison of different schemes and show that these theoretical shortcomings
have arguably a negligible impact in practice.

\smallskip{}
We continue in Sections \ref{sec:highmass} and \ref{sec:lowmass} with
a detailed presentation of our phenomenological analysis. 
In line with Ref.~\cite{Robens:2015gla}, 
we separately consider two regions of interest, where
heavy-to-light Higgs decays are 
kinematically accessible.  

\begin{itemize}
 \item {\underline{High-mass region:} in which    
 the \textit{lighter} eigenstate is identified with the discovered SM-like Higgs {of (fixed) mass} $m_{h}$,
 while the heavier mass-eigenstate corresponds to
 an additional heavy Higgs companion with a variable mass $m_H$, such that
 $m_{H} > 2\,m_h$. 
 }
 \item {\underline{Low-mass region:}
 where one instead identifies the \emph{heavier} mass eigenstate with the SM-like Higgs {of (fixed) mass $m_H$},
 while
 $h$ represents now a light Higgs companion and $m_{H} > 2\,m_h$.}
\end{itemize}

Specific scenarios with maximal $H\to hh$ branching fractions in agreement
with all of the model constraints are analysed separately in Section~\ref{sec:maxbr}. 

\medskip{}
\subsection{Computational Setup}
\label{sec:setup}

{In the remainder of our numerical analysis, we
fix the SM Higgs boson mass to the best-fit value
based on the combined data
samples of the ATLAS and CMS experiments $m_{\hzero} = 125.09$ GeV \cite{Aad:2015zhl}.}
{Whenever needed,} we use in addition {the current best averages} of the top-quark mass 
$m_t = 173.07$ GeV; the (pole) bottom-quark
mass $m_b^{\text{pole}} = 4.78$ GeV; and the {weak gauge boson masses  $m_{\PW}
= 80.385$ GeV, $m_{\PZ} = 91.1875$ GeV \cite{Agashe:2014kda}}. 
{The singlet vev is linked to the \emph{physical} doublet vev 
through the 
input parameter $\tan\beta$ as $v_s = v_{\text{phys}}\,\tan\beta$,
with $v_{\text{phys}}\,\equiv v_{G_F} = (\sqrt{2}G_F)^{-1/2} = \,246.219$ GeV. This is in fact equivalent to
defining $\tan\beta$ in the $G_F$-parametrization. To perform our calculation in
the $\alpha_{\text{em}}$-parametrization, we must translate it accordingly
{through Eq.~\eqref{eq:deltar_def1}}
\begin{alignat}{5} {
\tan\beta\Big{]}_{\alpha_{\text{em}}} = \tan\beta\Big{]}_{G_F}\,\left(\frac{v_{\text{phys}}}{v_{\alpha_{\text{em}}}}\right) = 
\frac{\tan\beta\Big{]}_{G_F}}{\sqrt{1+\Delta r}}  
}
 \label{eq:tb-translation}\, ,
\end{alignat}
{\noindent where
\begin{alignat}{5}
 v^2_{\alpha_{\text{em}}} = \cfrac{\mw^2\left(1-\mw^2/\mz^2\right)}{\pi\,\alpha_{\text{em}}(0)}
 \qquad \text{and hence} \qquad v_{\alpha_{\text{em}}} = v_{G_F}\,\sqrt{1+\Delta r}
\label{eq:vev-translation}\,.
\end{alignat}
\noindent {Plugging the above relation along with Eq.~\eqref{eq:deltar_def1} into
the expression for the decay width~\eqref{eq:nlowidth}, 
\begin{\eqn}
\Gamma^\text{LO}_{G_F}\,=\,\Gamma^\text{LO}_{\aem}\,\lb 1+ \cfrac{\Delta\,r}{1+ t_\alpha/\tan\beta}  \rb, \label{eq:w-schemerel}
\end{\eqn}

\noindent {which relates the $\aem$ and $\gf$ parametrizations up to NLO EW accuracy through}
{\begin{alignat}{5}
  \delta_{G_F} &\equiv \cfrac{\Delta\Gamma^{\text{1-loop}}_{G_F}}{\Gamma^{\text{LO}}_{G_F}} 
  = \delta_{\alpha_{\text{em}}}\,\left( 1-\frac{\Delta\,r}{1+ t_\alpha/\tan\beta}  \right){ + \mathcal{O}(G_F^3)}
  \label{eq:rel-gf}.
 \end{alignat}}

\medskip{}
Feynman rules for the singlet model rely on two independent implementations. For one
of them we use {\sc LanHEP} \cite{Semenov:1996es,Semenov:2014rea} and {\sc
Sloops} \cite{Boudjema:2005hb,Baro:2007em,Baro:2008bg,Baro:2009gn}
and include a non-linear gauge fixing Lagrangian \eqref{gaugefixing}. For the second one
we generate {\sc UFO} ~\cite{Degrande:2011ua} {and {\sc FeynArts} \cite{Hahn:2000kx}} files using {\sc FeynRules} \cite{Alloul:2013bka},
while the counterterms are derived analytically and implemented by hand. Both 
implementations are in perfect agreement.

\medskip{}
The one-loop decay amplitude is generated with
{\sc FeynArts} and analytically processed via {\sc FormCalc} \cite{Hahn:2000kx}. 
The loop form factors are handled with dimensional regularization 
in the 't~Hooft--Veltman scheme, and written in terms of standard loop
integrals. These are further reduced via
Passarino--Veltman decomposition and evaluated with the help of {\sc LoopTools}
\cite{Hahn:1998yk}.

\begin{table}[htb!]
 \begin{center}
   \begin{tabular}{|l|c|c|c|}
   \cline{2-4}
  \multicolumn{1}{c}{}  & \multicolumn{3}{|c|}{$\delta\G^{\text{1-loop}}_{\Hhh}\, 
[\mbox{GeV}]$} \\
    \hline
  Scheme  &$\Delta = 0,\{\mbox{nlgs}\} = 0$ &$\Delta = 10^7,\{\mbox{nlgs}\} = 0$ &$\Delta = 
10^7,\{\mbox{nlgs}\} = 10$ \\
\hline 
   Minimal field &$+4.2807988{\bf 8}\times10^{-3}$ &$+4.2807988{\bf
2}\times10^{-3}$ & $-6.63340412\times10^{4}$\\
   OS &$+4.2633488{\bf 8}\times10^{-3}$ & $+4.2633488{\bf 6}\times10^{-3}$& 
$-5.27015844\times10^{3}$\\
   Mixed $\msbar$/OS & $+6.846750{\bf 6}\times10^{-3}$ &
$+6.846750{\bf 4}\times10^{-3}$ 
& $+6.846750{\bf 0}\times10^{-3}$\\
   Improved OS &$+3.939356{\bf 9}\times10^{-3}$ &$+3.939356{\bf 8}\times10^{-3}$ & 
$+3.93935{\bf 56}\times10^{-3}$\\
   \hline
   \end{tabular}
 \end{center}\caption{Checks on UV-finiteness and gauge independence of 
the one-loop correction to the heavy-to-light Higgs decay width
$\delta \Gamma^{\text{1-loop}}_{\Hhh}$ (in GeV)
within the different renormalization schemes
introduced in Section \ref{sec:nondiag}. The model parameters are fixed as in 
Eq.~\eqref{eq:bench}. For the (scale dependent) minimal field scheme
we set the renormalization scale at 
 $\mu_R^2 =(m_h^2+m_H^2)/2$. {Bold-faced numbers highlight the first
departing digits between the entries
 of the different columns in a given row.}\label{tab:checkwidth}}
\end{table}\noi 
\begin{table}[htb!]
 \begin{center}
   \begin{tabular}{|l|c|c||l|c|c|}
    \cline{1-3}\cline{4-6}
    $\delta m_{hH}^{2}|^\infty$ & $\{\mbox{nlgs}\} = 0$ & $\{\mbox{nlgs}\} = 10$ & $\delta 
m_{hH}^{2}|^{\rm fin}$ & $\{\mbox{nlgs}\}= 0$ & $\{\mbox{nlgs}\} = 10$  \\
\hline 
    Minimal field &$-5.80\times 10^{2}$ &$-9.44\times 10^{2}$ & Minimal field &
$+5.72\times 10^3$& $+8.48\times 10^{3}$ 
\\
    OS & $-5.80\times 10^{2}$ &$-9.44\times 10^{2}$  & OS& $+5.75\times 
10^{3}$ & $+8.80\times 10^{3}$  \\
    Mixed $\msbar$/OS & $-5.80\times 10^{2}$ &$-5.80\times 10^{2}$ & Mixed $\msbar$/OS 
&$-2.48\times 10^{2}$ & $-2.48\times 10^{2}$\\ 
    Improved OS & $-5.80\times 10^{2}$ & $-5.80\times 10^{2}$ &Improved OS & $+5.72\times 
10^{3}$ & $+5.72\times10^{3}$\\
\hline
   \end{tabular}
 \end{center}\caption{Dependence on the gauge-fixing parameters of the mixed mass counterterm $\delta 
m_{hH}^2$ (in GeV$^2$)
within the different renormalization schemes
introduced in Section \ref{sec:nondiag}.
The model parameters are fixed as in 
Eq.~\eqref{eq:bench}.  For the (scale dependent) minimal field scheme
we set the renormalization scale at 
 $\mu_R^2=(m_h^2+m_H^2)/2$.\label{tab:checkmixing}}
\end{table}

\subsection{Scheme choice and gauge invariance}
\label{sec:gauge-numerical}

Gauge-fixing parameters may appear explicitly
at intermediate stages in the calculation of $S$-matrix elements in gauge theories.
Taken separately, 
counterterms and unrenormalized loop amplitudes may in general
depend on the gauge-fixing parameters 
and are eventually also UV-divergent. 
We only expect these UV divergent contributions to cancel
once all
the different building blocks
are combined together into predictions
for physical observables. 
Nonetheless, depending on which renormalization conditions are chosen for a certain
input  parameter $X$, one may obtain loop amplitudes which, albeit finite, still
{depend on the gauge--fixing.}
These situations reflect that, for such a renormalization scheme,
the definition for $X$ is gauge-dependent. 

In the following 
we check the different
renormalization schemes introduced in Section~\ref{sec:nondiag} {in the light of gauge independence}. We compute
the one-loop correction to the heavy-to-light Higgs decay width
$\delta \Gamma^{\text{1-loop}}_{\Hhh}\equiv \Gamma^{\text{NLO}}_{\Hhh}- \Gamma^{\text{LO}}_{\Hhh}$
in the general non-linear gauge of Eq.~\eqref{gaugefixing}, 
where the quantities $\Gamma^{\text{LO}}_{\Hhh}$ and $\Gamma^{\text{NLO}}_{\Hhh}$ are given by Eqs.~\eqref{eq:lowidth} and ~\eqref{eq:nlowidth}
respectively. We resort to the {\sc SloopS} implementation of the singlet model Feynman
rules which includes
the general non-linear gauge-fixing Lagrangian of Eq.~\eqref{gaugefixing}, and vary the 
gauge-fixing parameters  
$\{\mbox{nlgs}\}=\{\anlg,\bnlg,\knlg,\dnlg_1,\dnlg_2 ,\enlg_1 ,\enlg_2\}$ within the fiducial range
$\{\mbox{nlgs}\}= 0 \dots 10$. Notice that the lower endpoint $\{\mbox{nlgs}=0\}$ 
reproduces the familiar 't~Hooft--Feynman linear gauge. 
As a sample parameter space point we take
\begin{equation}
 m_h = 125.09\,\mbox{GeV}, \quad m_H = 260\,\mbox{GeV},\quad  \sin{\alpha} = 0.3,\quad 
\tan 
\b = 5
\label{eq:bench}\, ,
\end{equation}\noi
which gives a leading-order width $\Gamma^{\text{LO}}_{\Hhh} = 0.137$ GeV.
In Table~\ref{tab:checkwidth} 
we compare the results for $\delta \Gamma^{\text{1-loop}}_{\Hhh}$ in the linear gauge ($\{\mbox{nlgs}=0\}$) and
one exemplary non-linear setup  ($\{\mbox{nlgs}=10\}$). Simultaneously,
we check the UV-finiteness of our results by sweeping the range $\Delta = 0 \dots 10^7$, where the parameter $\Delta$ trades
the UV-divergences of the one-loop amplitude as defined by ~Eq.~\eqref{eq:delta}. 
Gauge independence and UV-finiteness are verified if $\delta \Gamma^{\text{1-loop}}$ remains
unchanged (within numerical precision) under these varations  
\footnote{Using double precision we expect an agreement of 14 to 15 digits. 
Given the variation ranges $\Delta=0 \dots 10^7$ and $\{\mbox{nlgs}\} = 0 \dots 10$, we deem
the test as satisfactory if 6 to 8 common digits are achieved.}.
The fact that in the first two columns $\delta \Gamma^{\text{1-loop}}$ remain
constant confirms that all of the four schemes introduced
in Section~\ref{sec:nondiag} yield UV-finite results in the linear gauge. 
However,
only the mixed $\msbar$/OS and the improved OS setups produce UV-finite,
$\{\text{nlgs}\}$-independent results for the generalized non-linear gauge-fixing.
Instead, 
in the minimal field and the OS schemes 
we observe left-over 
$\delta \Gamma^{\text{1-loop}}$ dependencies
on the gauge-fixing
parameters. These $\{\text{nlgs}\}$-dependent remainders
affect both the finite parts and the UV-divergent contributions,
and are thus responsible 
for the incomplete cancellation of the UV-poles, cf. the last column of Table~\ref{tab:checkwidth}.
This breakdown can be ultimately 
traced back to the {renormalization condition that determines} the mixed mass counterterm $\delta m^2_{hH}$.
{Its definitions in the minimal field scheme~\eqref{eq:renorm-mixing} and 
the OS scheme \eqref{eq:mixing-scheme2} are not gauge-independent, and lead
to a $\{\text{nlgs}\}$-dependent decay width. Instead, we find no residual
$\{\text{nlgs}\}$-dependencies {in the mixed $\msbar$/OS and} the improved OS schemes, in 
which $\delta m^2_{hH}$ is fixed via 
the gauge-independent definitions {of Eq.~\eqref{eq:dMhHSloopS} and ~\eqref{eq:improvedOS} respectively}.}
We make these observations patent in Table~\ref{tab:checkmixing},
where we display 
the numerical value of the mixing 
counterterm $\delta m^2_{hH}$ corresponding to the four renormalization schemes under analysis. Since 
the counterterm is not UV finite,
we split it into a finite and singular part as (with $\Delta$ as defined in
\eqref{eq:delta})
\begin{equation}
 \delta m_{hH}^2 = \delta m_{hH}^{2} \Big{\lvert}^\infty \cdot \Delta  + \delta m_{hH}^{2} \Big{ \lvert }^{\text{fin}}\,.
\end{equation}\noi
\noindent Neither the coefficient of the UV pole $\delta m_{hH}^{2} \Big{\lvert}^\infty$
nor the finite remainder $\delta m_{hH}^{2} \Big{ \lvert
}^{\text{fin}}$
depend on the gauge-fixing parameters when we fix $\delta m_{hH}^{2}$ 
either in the mixed $\msbar$/OS or the improved OS conditions.
Instead, both terms 
are shifted  when we switch
from the linear $\{ \mbox{nlgs}\} = 0$ to the non-linear gauge-fixing choice $\{ \mbox{nlgs}\} = 10$,
when the calculation is performed using the minimal field or the OS schemes. 
{In view of the fact that
$\delta m^2_{hH}$ (along with the mixed
field renormalization $\delta Z_{hH}$, cf. Table~\ref{tab:scheme-summary}) are
the only different ingredients between these four schemes, they are ultimately
responsible for the finite $\{\text{nlgs}\}$-dependent remainders 
in  $\delta \Gamma^{\text{1-loop}}$
in the latter two schemes
-- and linked to them, of the uncancelled  
UV poles.} We emphasize as well that these concomitant UV divergences 
vanish for $\{\text{nlgs}\} = 0$ and hence 
do not feature in  
the customary 't Hooft--Feynman gauge,
where the results in all schemes are UV finite.}
{Finally, let us also notice that, given the relation between the mixed
mass counterterm and the mixing angle via Eq.~\eqref{eq:mixedmass-angle}, a gauge-independent
$\delta m^2_{hH}$ supports a more physical interpretation of the mixing angle, viz.
as value that could be extracted from e.g. a deviation in the LHC Higgs signal strengths or, alternatively, 
an excess which points to the direct production of the heavy scalar.\footnote{{Similar lines of argument are used in the context of the
renormalization of the $\tan\beta$ parameter in the MSSM, cf. e.g. 
Table~2 in Ref.~\cite{Baro:2008bg}.}} 

\medskip{} {For practical purposes, therefore,}
the proven robustness of the improved OS scheme 
{(giving in all cases UV-finite, $\{\text{nlgs}\}$-independent, {and numerically stable} 
renormalized {one-loop} amplitudes)
justifies its use as default scheme choice in our numerical analysis hereafter.} 
{Moreover, the
excellent agreement between the $\delta \Gamma^{\text{1-loop}}$
results for the different schemes {in} the linear 't~Hooft--Feynman gauge -- as explicitly shown further down -- 
give convincing arguments
that also the schemes where the mixed mass counterterm 
is gauge dependent render {reliable} results --
at least as long as the linear 't~Hooft--Feynman gauge is used {and, in the case of the minimal field scheme the renormalization
scale is chosen in the ballpark of the relevant physical scales.
This is again in line with analogue situations 
such as e.g. the squark sector of the MSSM ~\cite{Espinosa:2001xu,Espinosa:2002cd,Baro:2009gn}.} 

\subsection{High-mass region}
\label{sec:highmass}

\begin{table}[thb!]
\begin{center} \footnotesize{
\begin{tabular}{|c|c|ll|llll|} \hline
 $\mHH$ [GeV] & $\sin\alpha$ & & {$\Gamma_{\alpha}^{\text{LO}}(\Hhh)$ [GeV]} & & {$\Gamma_{\alpha}^{\text{NLO}}(\Hhh)$} [GeV] & $\delta_{\alpha}\,[\%]$ &  $\delta_{\gf}\,[\%]$ \\ \hline
\multicolumn{8}{|l|}{\qquad $\tan\beta = 5$} \\ \hline
\multirow{9}{*}{300} & \multirow{3}{*}{0.1} & & \multirow{3}{*}{4.374$\times 10^{-2}$} & OS & 4.516$\times 10^{-2}$ & 3.250 & 3.130 \\ 
 & & & & Improved OS & 4.509$\times 10^{-2}$ & 3.106 & 2.990 \\
  & & & & Minimal field & 4.544$\times 10^{-2}$ & 3.895 & 3.751 \\  \cline{2-8}
 & \multirow{3}{*}{0.2} & & \multirow{3}{*}{0.171} & OS & 0.177 & 3.371 & 3.248 \\ 
 & & & & Improved OS &  0.177& 3.218 & 3.100 \\
  & & & & Minimal field & 0.178  & 4.033 & 3.886 \\  \cline{2-8}
 & \multirow{3}{*}{0.3} & & \multirow{3}{*}{0.362} & OS & 0.375 & 3.583 & 3.455 \\ 
 & & & & Improved OS &  0.374& 3.400 & 3.278 \\
  & & & & Minimal field & 0.377  & 4.281 & 4.127 \\ \hline \hline
\multirow{9}{*}{500} & \multirow{3}{*}{0.1} & & \multirow{3}{*}{0.221} & OS & 0.234 & 5.667 & 5.456 \\ 
 & & & & Improved OS &  0.233& 5.438 & 5.236 \\
  & & & & Minimal field & 0.237  & 6.989 & 6.730 \\  \cline{2-8}
 & \multirow{3}{*}{0.2} & & \multirow{3}{*}{0.868} & OS & 0.920 & 5.980 & 5.761 \\ 
 & & & & Improved OS &  0.917 & 5.728 & 5.518 \\
  & & & & Minimal field & 0.932  & 7.441 & 7.168 \\  \cline{2-8}
 & \multirow{3}{*}{0.3} & & \multirow{3}{*}{1.831} & OS & 1.951 & 6.566 & 6.329 \\ 
 & & & & Improved OS &  1.945 & 6.237 & 6.012 \\
 & & & & Minimal field & 1.983  & 8.294 & 7.995 \\ \hline \hline
\multirow{9}{*}{700} & \multirow{3}{*}{0.1} & & \multirow{3}{*}{0.586} & OS & 0.597 & 1.948 & 1.876 \\ 
 & & & & Improved OS &  0.601& 2.569 & 2.473 \\
  & & & & Minimal field & 0.598  & 2.009 & 1.935 \\   \cline{2-8}
& \multirow{3}{*}{0.2} & & \multirow{3}{*}{2.296} & OS & 2.355 & 2.583 & 2.489 \\ 
 & & & & Improved OS & 2.369& 3.188 & 3.071 \\
 & & & & Minimal field & 2.366  &3.056 & 2.944 \\  \cline{2-8}
& \multirow{3}{*}{0.3} & & \multirow{3}{*}{4.845} & OS & 5.026 & 3.742 & 3.606 \\ 
 & & & & Improved OS & 5.056& 4.353 & 4.195 \\
  & & & & Minimal field & 5.082  &4.893 & 4.716 \\ \hline
\end{tabular}}
\caption{Heavy-to-light Higgs decay width $\GHhh$ at LO and NLO EW accuracy, for representative
parameter choices and different renormalization schemes, in the high-mass region. 
{The total decay widths are obtained in the $\alpha_{\text{em}}$-parametrization, as defined in Eqs.~\eqref{eq:rel-alpha},
while the relative
one-loop EW effects are quantified in both the $\alpha_{\text{em}}$-parametrization
and the $\gf$-parametrization, cf. Eq.~\eqref{eq:rel-gf}}.
For the (scale-dependent) minimal field scheme, the renormalization scale is fixed to the geometrical average mass $\mu^2_R = \pstar^2 = (m_h^2+m_H^2)/2$. 
{The input value for $\tan\beta$ is linked to the singlet vev through $v_s = (\sqrt{2}G_F)^{-1/2}\,\tan\beta$.} }
\label{tab:highmass}
\end{center}
\end{table}

In Table~\ref{tab:highmass} we evaluate $\GammaLO$ and $\GammaNLO$
for representative parameter choices and different renormalization schemes. 
The relative 
one-loop EW corrections are given in both the $\aem$-parametrization and the $\gf$-parametrization
introduced in Section~\ref{sec:decay}. Our results show decay rates that strongly vary
with the relevant parameters of the model. 
The heavy-to-light Higgs decay width significantly depends on the decaying Higgs mass $\mHH$,
changing by two orders of magnitude when sweeping the range $\mHH = 300 \dots 700$ GeV.
For heavy Higgs masses close to the di-Higgs threshold,
the partial Higgs widths lie in the ballpark of
$\mathcal{O}(0.01 - 0.1)$ GeV. These results depend as well on the mixing angle,
and change by roughly one order of magnitude from small (viz. $\sin{\alpha} \simeq 0.1$)
to moderate mixing angles (viz. $\sin{\alpha} \simeq 0.3$).
For larger heavy Higgs masses, the $\GammaNLO$ values may rise up to the few GeV level.
{The mild numerical discrepancies between the different schemes
are indicative of small theoretical uncertainties in the 't~Hooft--Feynman gauge.
For a more general gauge-fixing choice, though, 
the minimal field and on-shell schemes are no longer reliable, in view
of their proven gauge-dependent nature. {It is also worth noticing that 
the radiative corrections in the $G_F$-parametrization ($\delta_{G_F}$) are
generically smaller than in the $\alpha_{\text{em}}$-parametrization. The reason 
is twofold: i) part of the NLO EW corrections in the latter case ($\delta_\alpha$) 
are contained in the $\Delta r$ parameter, and hence already
embedded into the $G_F$-scheme LO calculation (cf. Eq.~\eqref{eq:w-schemerel}). Consequently,
the quantum effects encoded by
$\Delta r$ do not belong to $\delta_{G_F}$;
ii) for phenomenologically relevant scenarios, 
$\Delta r$ is dominated by purely SM effects, for which $\Delta^{\text{SM}} > 0$ \cite{Kennedy:1988sn,Hollik:1988ii,Hollik:1993cg,Langacker:1996qb,Hollik:2003cj,Hollik:2006hd}, and
thereby $\delta_{\alpha} > \delta_{G_F}$, given the relation between both~\eqref{eq:rel-gf}.
} 
}

 \begin{figure}[thb!]
 \begin{center}
\includegraphics[width=0.31\textwidth,height=0.302\textwidth]{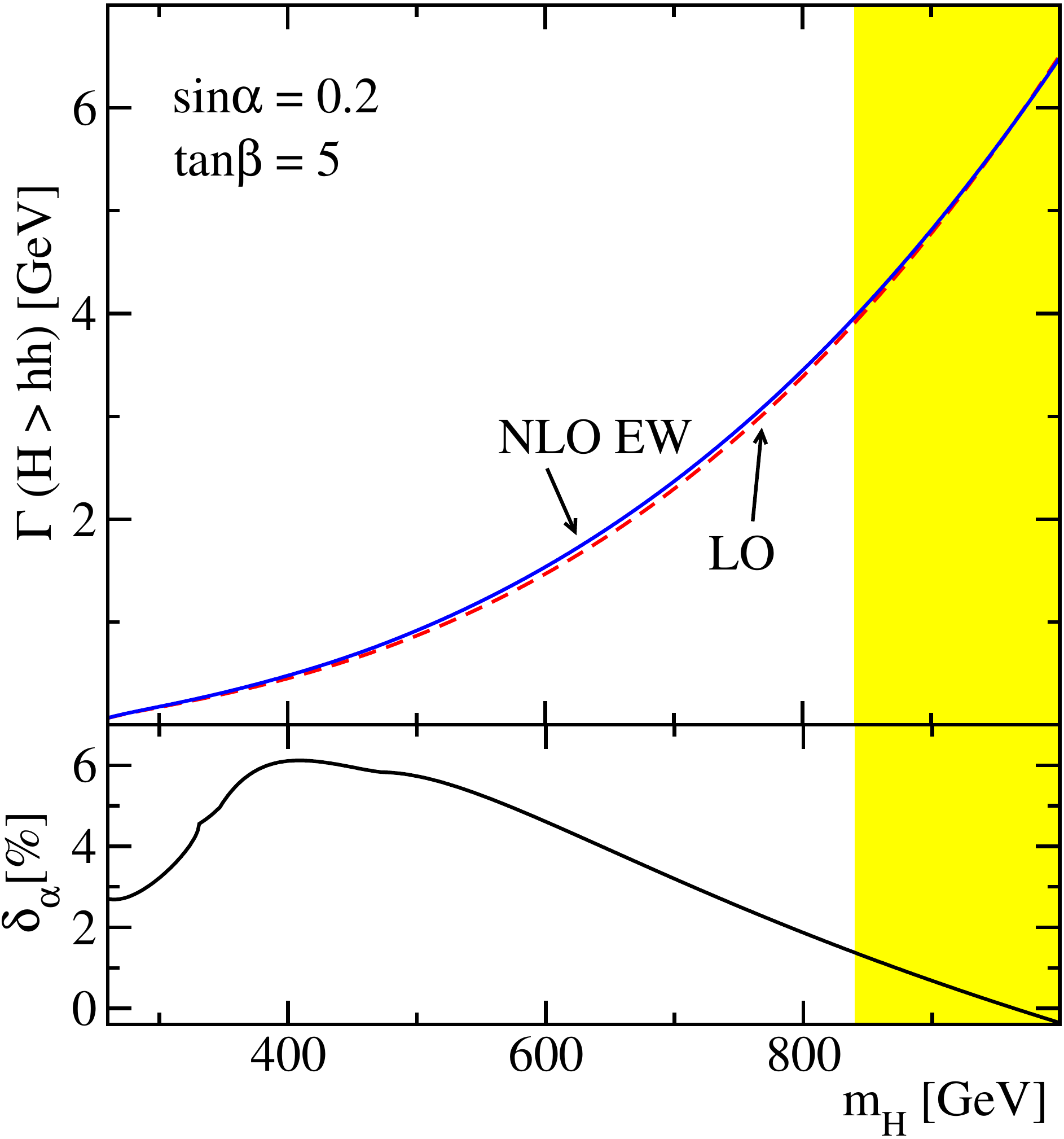}\hspace{0.02cm} 
\includegraphics[width=0.31\textwidth,height=0.3\textwidth]{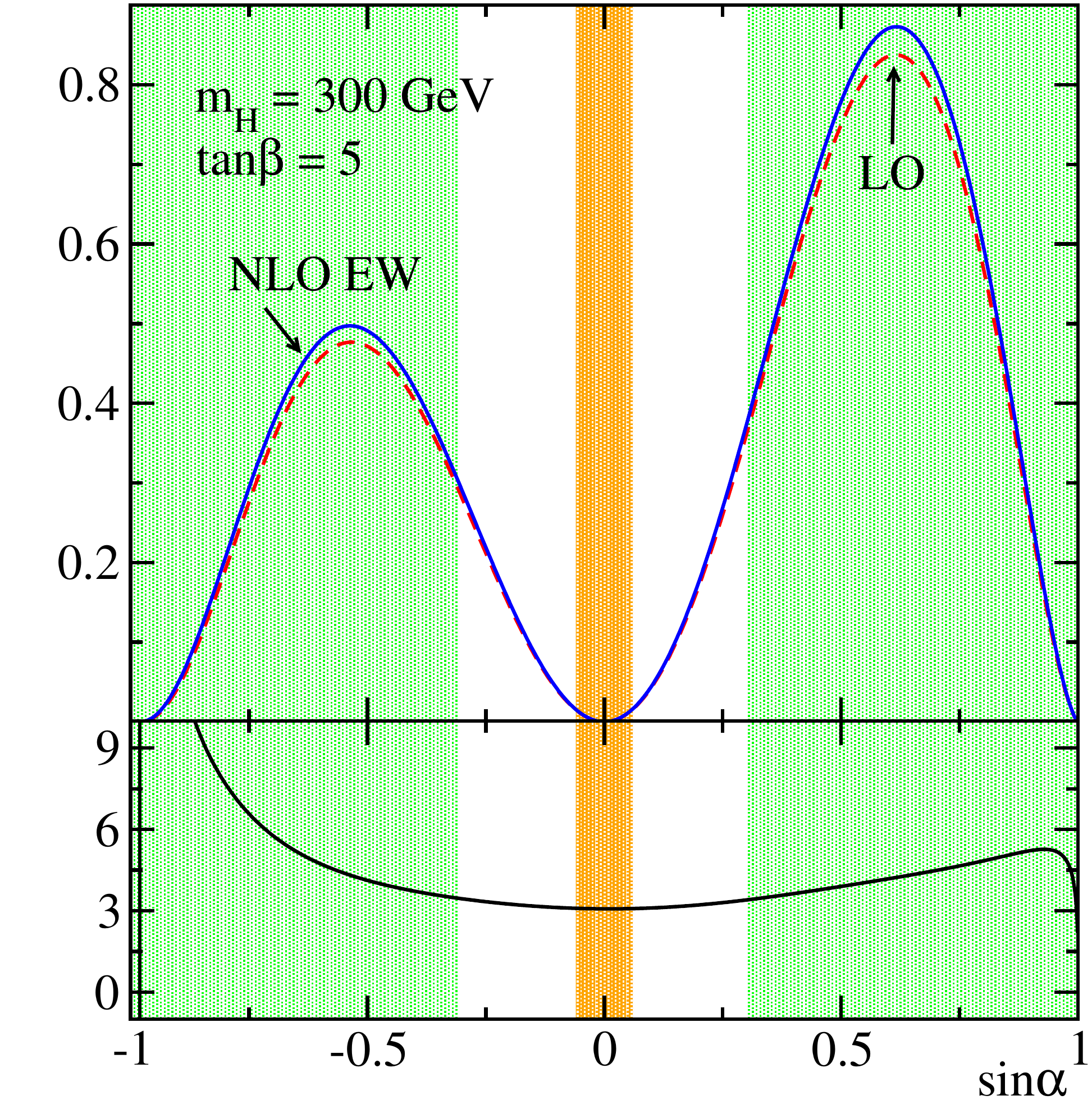} \hspace{0.02cm} 
\includegraphics[width=0.31\textwidth,height=0.3\textwidth]{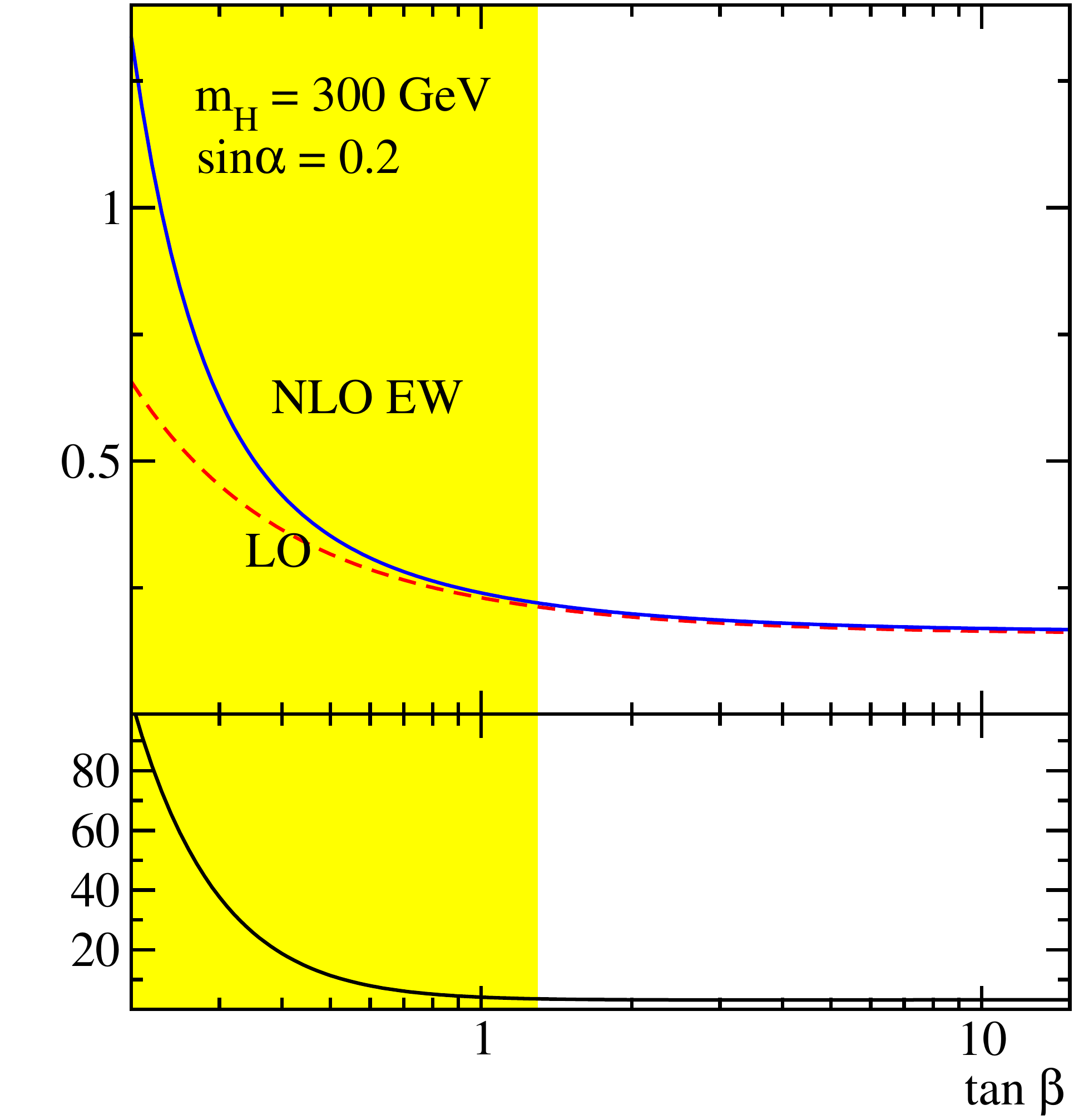} 
\caption{Heavy-to-light Higgs decay width $\GHhh$ in the high-mass region. 
The results are shown at LO (dashed, red) and NLO (full, blue) as 
a function of the relevant parameters of the model.
The lower subpannels show the relative
one-loop EW correction in the $\alpha_{\rm em}$-parametrization~\eqref{eq:rel-alpha}.
Renormalization is performed in the improved on-shell scheme. 
The shaded regions are excluded by constraints (see the text for more
details).}
\label{fig:highmass}
\end{center}
\end{figure}

\medskip{} The analysis is complemented in
Figure~\ref{fig:highmass} with a thorough
survey of the parameter space dependencies. The NLO results are 
calculated in the improved OS scheme. The shaded regions
are ruled out by different theoretical and experimental
constraints on the model: 
i) the ranges $m_H > 840$ GeV (left panel) and $\tan\beta < 1.27$ (right panel)
are excluded by perturbativity
ii) $|\sin{\alpha}|> 0.31$ (central panel, green shading) is incompatible with
electroweak constraints from the $m_W$ measurement; finally, 
the central range $|\sin{\alpha}| < 0.06$ (central panel, orange shading) is incompatible 
with vacuum
stability.

\medskip{}
Most features observed in Fig.~\ref{fig:highmass} can be readily traced back to the LO
dynamics which governs the decay process. The 
two key players, as alluded to above, are the trilinear Higgs self-coupling $\lambda_{Hhh}$ and the characteristic
$1\to 2$ kinematics. The former is responsible for the quadratic growth $\Gamma_{\Hhh} \sim \mathcal{O}(\mHH^2)$ 
(cf. Eq.~\eqref{eq:Hhh}), which
overcomes the phase space suppression at $m_H^2 \gg m_h^2$, and explains 
the power-like increase as a function of $\mHH$ (left panel in Figure~\ref{fig:highmass}). 
The NLO-corrected result with respect to the mixing angle mimics
the LO result, with the expected
nodes in the decoupling limits $|\sin{\alpha}|=0$ or 1 as well
as for $\tan\beta = -\tan\alpha$ (cf. the central panel of Fig.~\ref{fig:highmass}).

\medskip{}
Unlike the stark changes observed for the decay width, the relative
one-loop EW corrections are much more stable, positive,
and of the order of 
few percent.
Differences between the $\aem$-parametrization and the $\gf$-parametrization, 
as well as between the different
renormalization schemes, are mild and remain typically below the percent level.

\medskip{}
The slight kink in $\delta_{\alpha}$ for $\mHH \simeq 350$ GeV (left panel, Fig.~\ref{fig:highmass}) 
reflects the top-quark threshold. 
{The finite correction $\delta_{\alpha}\sim 3\%$ for $\sin{\alpha} \to  0$
(cf. the lower 
subpanel of Figure~\ref{fig:highmass}, center)
follows from the fact that both $\GammaLO$ and $\GammaNLO$ tend to zero
in this limit, while its ratio remains roughly constant.}
The unphysical
large effects at $\sin{\alpha} \lesssim -0.9$  are due to the LO node  
 in the
limit $\tan\alpha \to -\tan\beta$, for which $\lambda_{Hhh} = 0$.
The pronounced NLO slope at low $\tan\beta$
is ultimately due to the exchange
of virtual Higgs bosons, and constitutes a telltale
imprint of the singlet model dynamics at the one-loop level.
While the fermion and the gauge
boson-mediated contributions are all controlled by (globally rescaled)
gauge couplings, the size of the Higgs-mediated loops is
governed by the Higgs self-couplings. These are strongly enhanced for $\tan\beta\ll 1$,
specially the Higgs boson two-point graphs, which depend on them quadratically.
For low enough $\tan\beta$ values, {e.g. typically $\tan\beta \lesssim 0.3$ and for $m_H \gtrsim 300$ GeV}, 
the relative yield $\delta_{\alpha}$ exceeds $\sim 50\%$,  
indicating that the process becomes effectively loop-induced.
Such sizable
loop effects are nonetheless hampered in practice, owing to
the unitarity and perturbativity bounds which severely constrain the phenomenologically 
viable low-$\tan\beta$ range. The limit $\tan\beta \ll 1$
corresponds in fact to the onset of a strongly-coupled regime, in which {at least one of the
scalar self-couplings} becomes non-perturbative, cf. also the discussion in Sec.~\ref{sec:constraints}.
 \begin{figure}[t!]
 \begin{center}
\includegraphics[width=5cm, height=5cm]{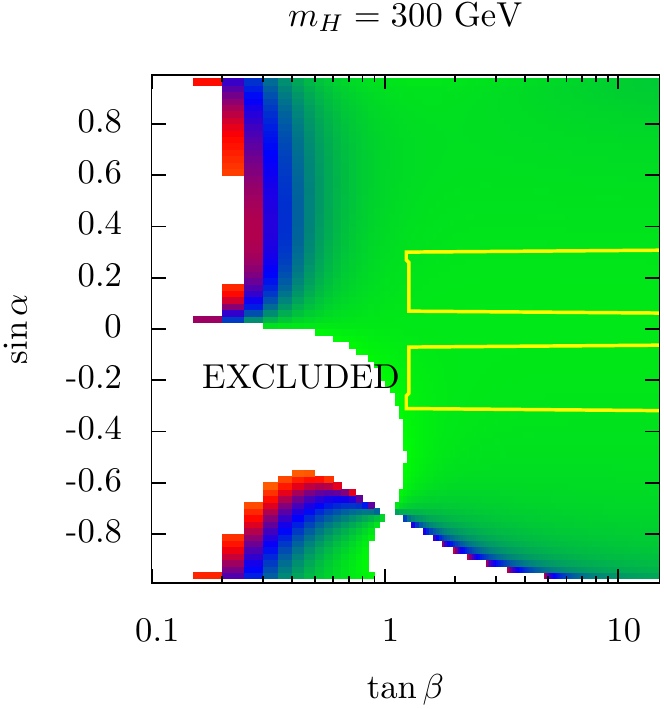} \hspace{0.2cm}
\includegraphics[width=5cm, height=5cm]{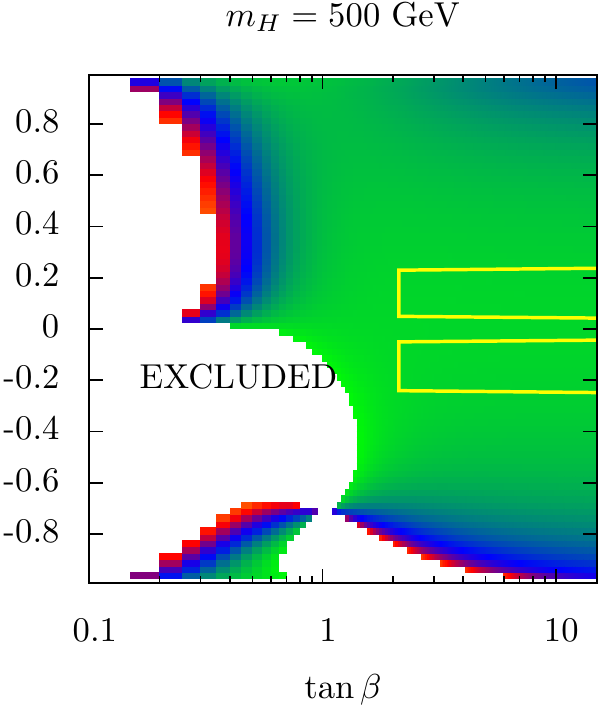} \hspace{0.2cm} 
\includegraphics[width=6cm, height=5cm]{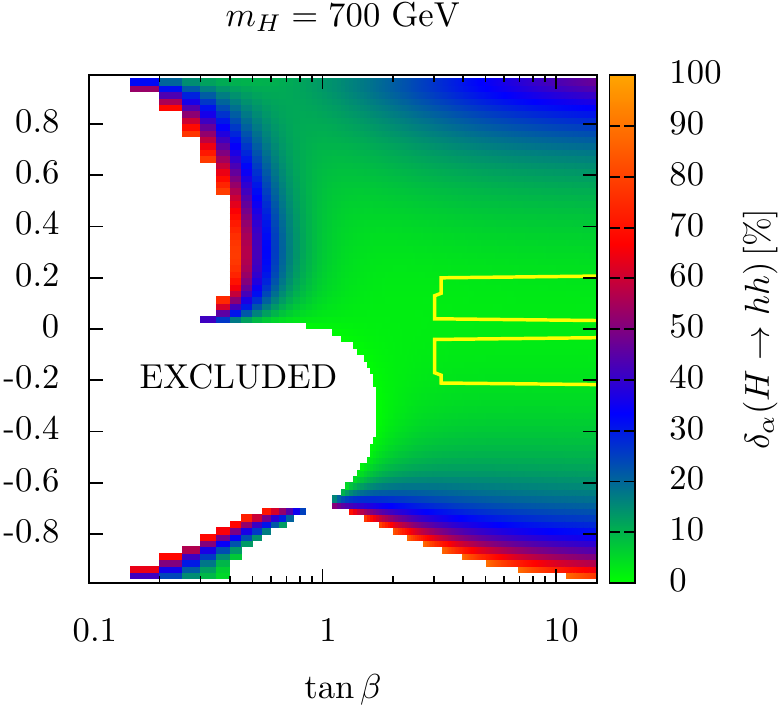} 
\caption{Relative one-loop EW corrections 
in the $\alpha_{\rm em}$-parametrization~\eqref{eq:rel-alpha},
projected on the $\sin{\alpha} - \tan\beta$ plane for 
exemplary heavy Higgs masses in the 
high-mass
region. The white voids
correspond to regions with $\delta_{\alpha} \gtrsim 100$ \%. 
Renormalization is performed in the improved on-shell scheme.
The yellow contour separates the allowed and excluded regions in the parameter space.}
\label{fig:scan}
\end{center}
\end{figure}

\medskip{} Complementary vistas to the $\Hhh$ landscape
are displayed in Figure~\ref{fig:scan}. Here we show
the relative one-loop effects $\delta_{\alpha}$ \eqref{eq:rel-alpha}
as density contours in the $\sin{\alpha} - \tan\beta$ plane. 
The yellow contour separates the allowed and excluded regions in the parameter space. Only
the horizontal fringes enclosed by the contour are compatible with all constraints on the model.
The white voids stand for values of
$\delta_\alpha \gtrsim 100 \%$
and correspond to regions where 
$\delta_\alpha$ is no longer a meaningful measure
of the relative quantum effects, while it instead indicates that
the decay process becomes loop-induced.
We find this situation: i) along 
the {strip $\tan\alpha \simeq -\tan\beta$}, 
due to the suppressed tree-level couplings;
and ii) for $\tan\beta < 1$, due to the $\cot\beta$-enhanced
Higgs-mediated loops.

\begin{figure}[t!]
\begin{center}
\includegraphics[width=0.32\textwidth,height=0.32\textwidth]{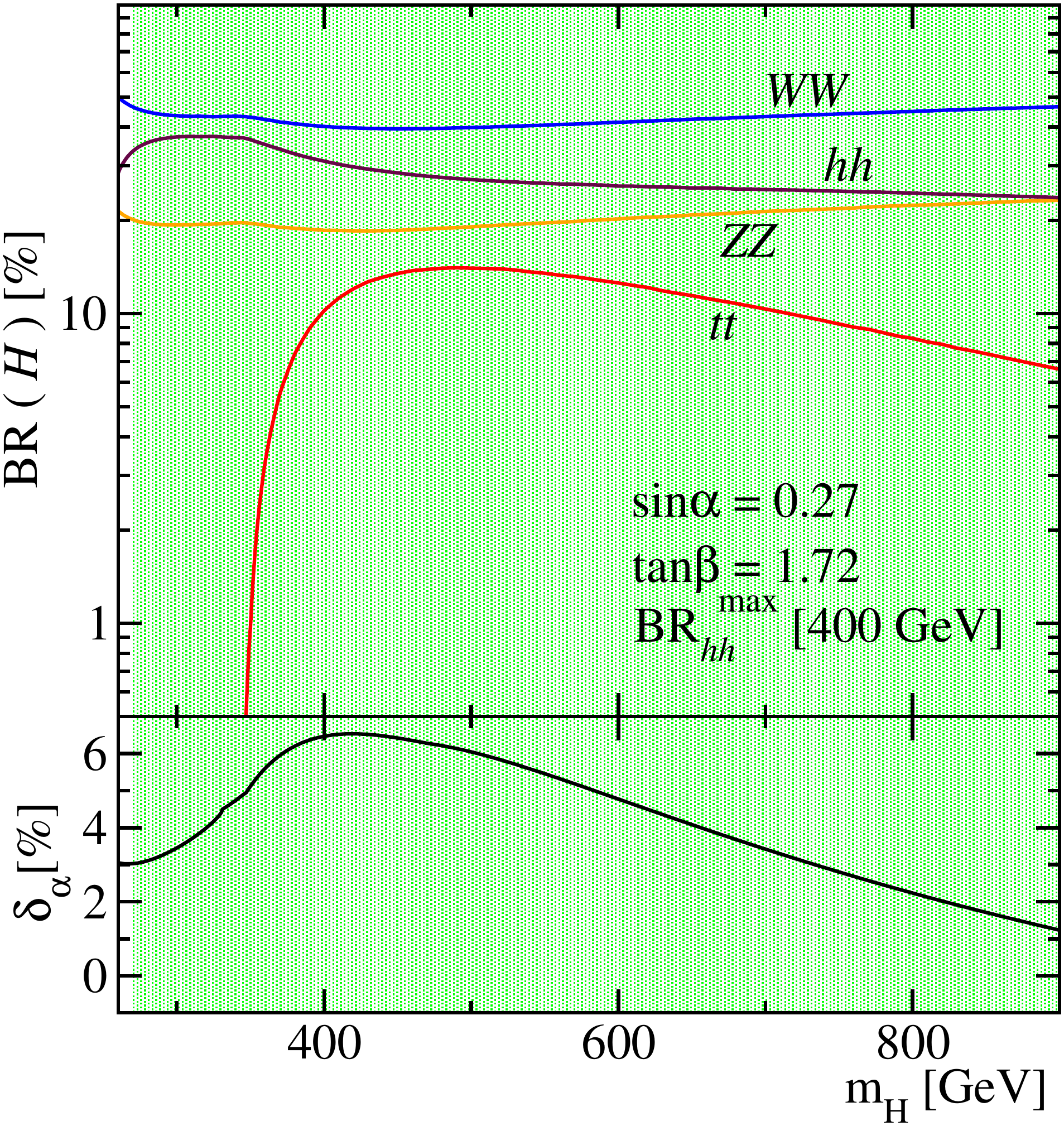} \hspace{0.1cm}
\includegraphics[width=0.32\textwidth,height=0.32\textwidth]{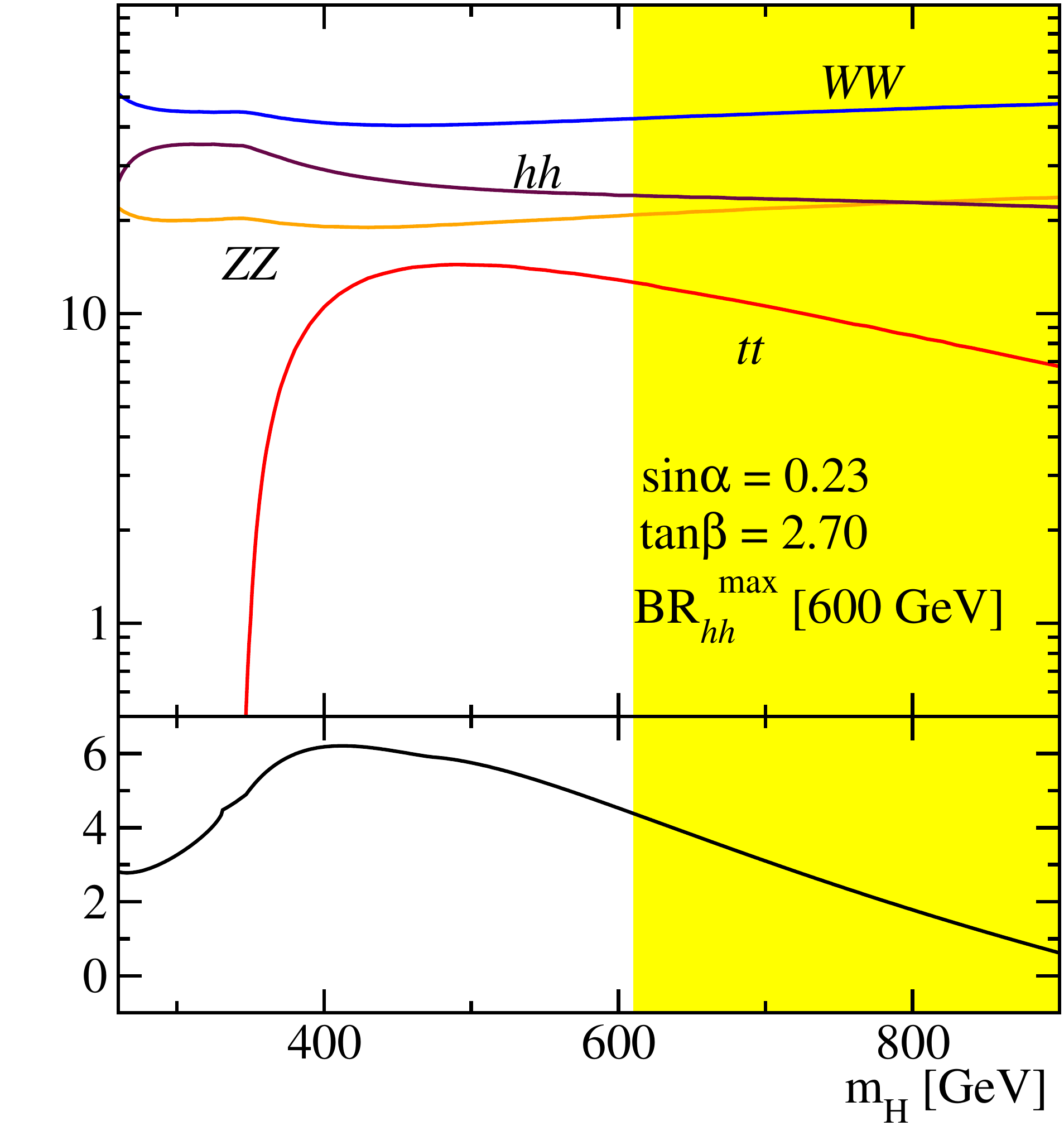} \hspace{0.1cm}
\includegraphics[width=0.32\textwidth,height=0.32\textwidth]{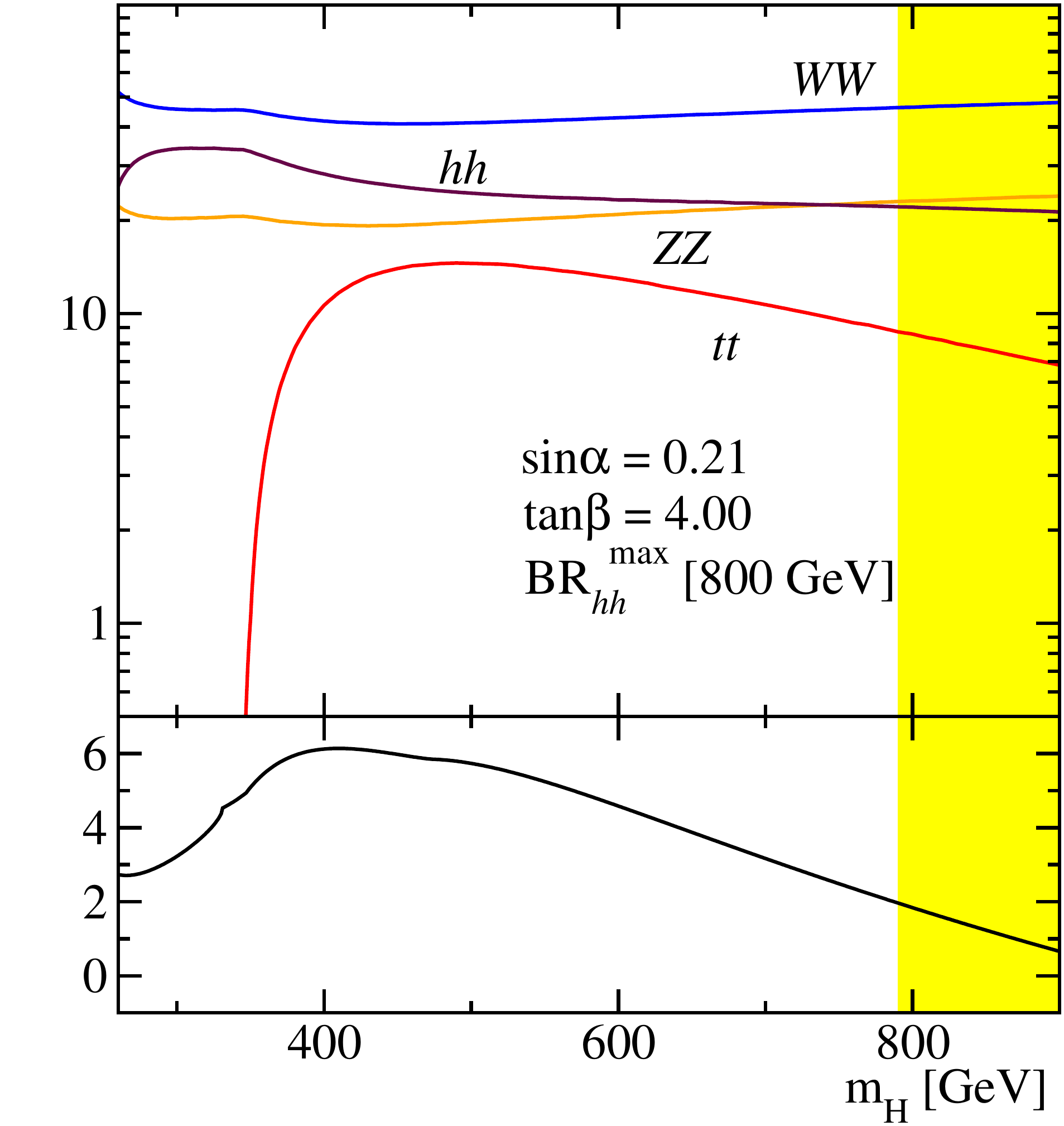} 
\caption{Heavy Higgs branching ratios (in $\%$) 
as a function of the 
heavy Higgs mass in the high-mass region. The mixing angle and $\tan\beta$ values
are fixed in each panel such that they maximize the $\Hhh$ branching ratio at LO 
for a given heavy Higgs mass of 400 (left), 600 (center) and 800 GeV (right) \cite{sbm}.
Decay modes into light fermions and loop-induced decays into gauge bosons 
lie below $\mathcal{O}(0.1)\%$ and are not shown. 
The lower subpanels show the relative
one-loop EW correction ($\alpha_{\rm em}$ parametrization~\eqref{eq:rel-alpha}) to $\GammaLO$
for the same parameter variations.
{The shaded (green) area in the left plot is ruled out by the W-mass measurement.
Excluded regions in the central and right panels (in yellow) are incompatible with perturbativity} and $m_W$.}
\label{fig:br-highmass}
\end{center}
\end{figure}

\bigskip{}
The impact of heavy-to-light Higgs decays 
{on the decay pattern of the heavy Higgs state} is portrayed in Figure~\ref{fig:br-highmass}. 
The branching ratios 
for the leading decay channels are represented
as a function of the heavy Higgs mass.
The {mixing angle and $\tan\beta$ values are} fixed in each panel such that they maximize the $\Hhh$ branching ratio
for a given heavy Higgs mass \cite{sbm}, as explicitly indicated in the figure. 
In this plot, we show the partial decay widths to SM fields by
rescaling the SM predictions \cite{Heinemeyer:2013tqa}, while for the
$\Hhh$ we use the LO result
\footnote{A global study including all Higgs decays in the singlet 
model to state-of-the-art accuracy 
lies beyond
the scope of the present study and will be discussed in a forthcoming publication~\cite{future-rsinglet}.}.
As well known, bosonic modes dominate the Higgs boson 
decays at high masses \cite{Spira:1997qz,Djouadi:2005gi,Plehn:2009nd}.
We find a rather featureless profile, with $\PW\PW$ being {the leading mode}
and with roughly no changes {over the whole mass range}.
Only the decays into top-quark pairs are also competitive, and attain up
to $\br \sim \mathcal{O}(10)\%$. 
The remaining fermionic channels, as well as the loop-induced
$\gamma\gamma, \gamma\PZ$ and $gg$ modes, stagnate at the $\mathcal{O}(0.1)\%$ 
level or below and are not shown. In the lower
subpanels we show the relative one-loop EW correction
to the heavy-to-light Higgs decay width for the same parameter variation. 

 \begin{figure}[t!]
 \begin{center}
\begin{tabular}{p{0.31\textwidth}p{0.31\textwidth}p{0.31\textwidth}} 
\vspace{0pt} 
\includegraphics[width=0.318\textwidth]{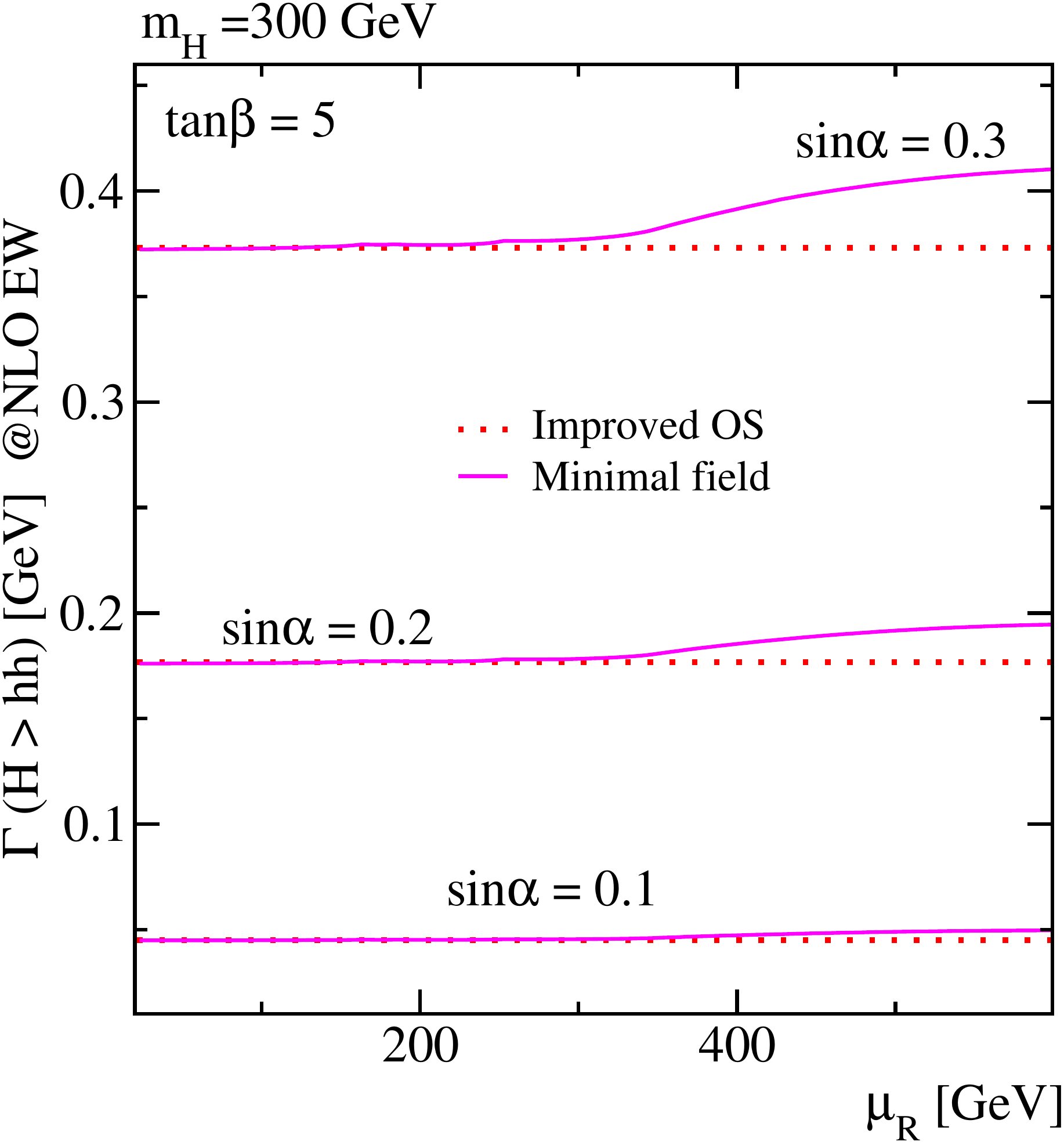} & \vspace{0pt} 
\includegraphics[width=0.3\textwidth]{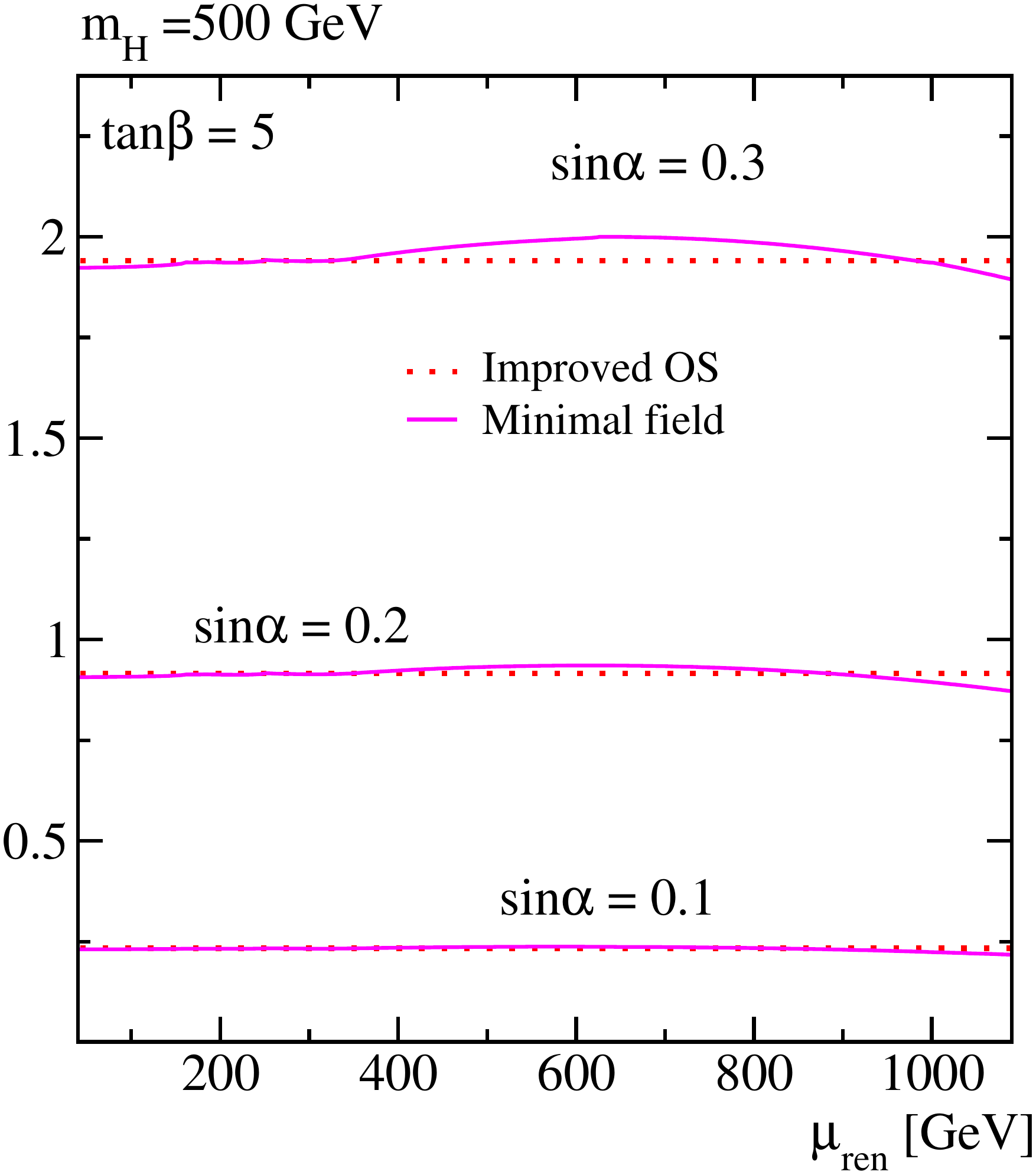} & \vspace{0pt} 
\includegraphics[width=0.295\textwidth]{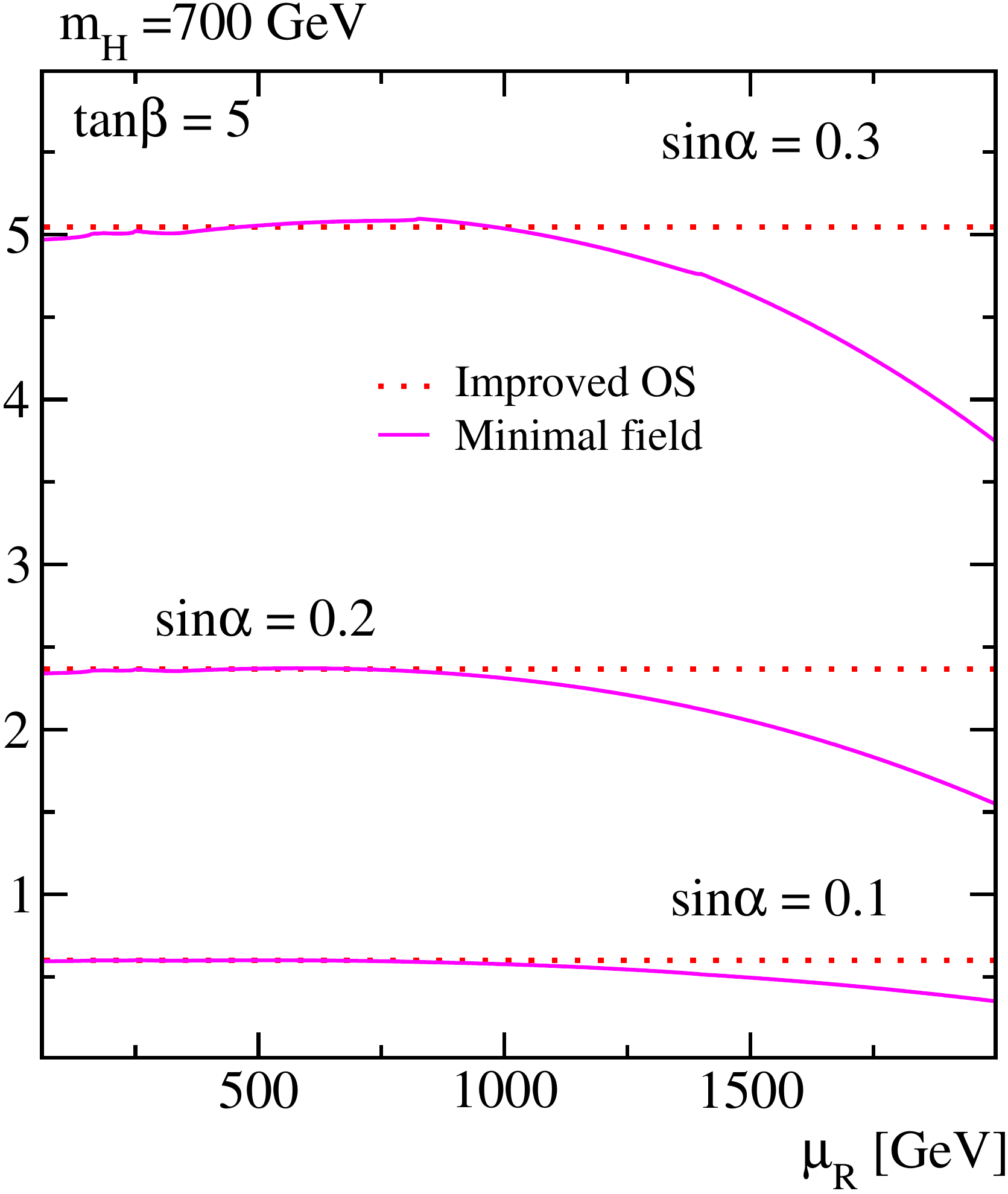} \\
\vspace{0pt} 
\includegraphics[width=0.32\textwidth]{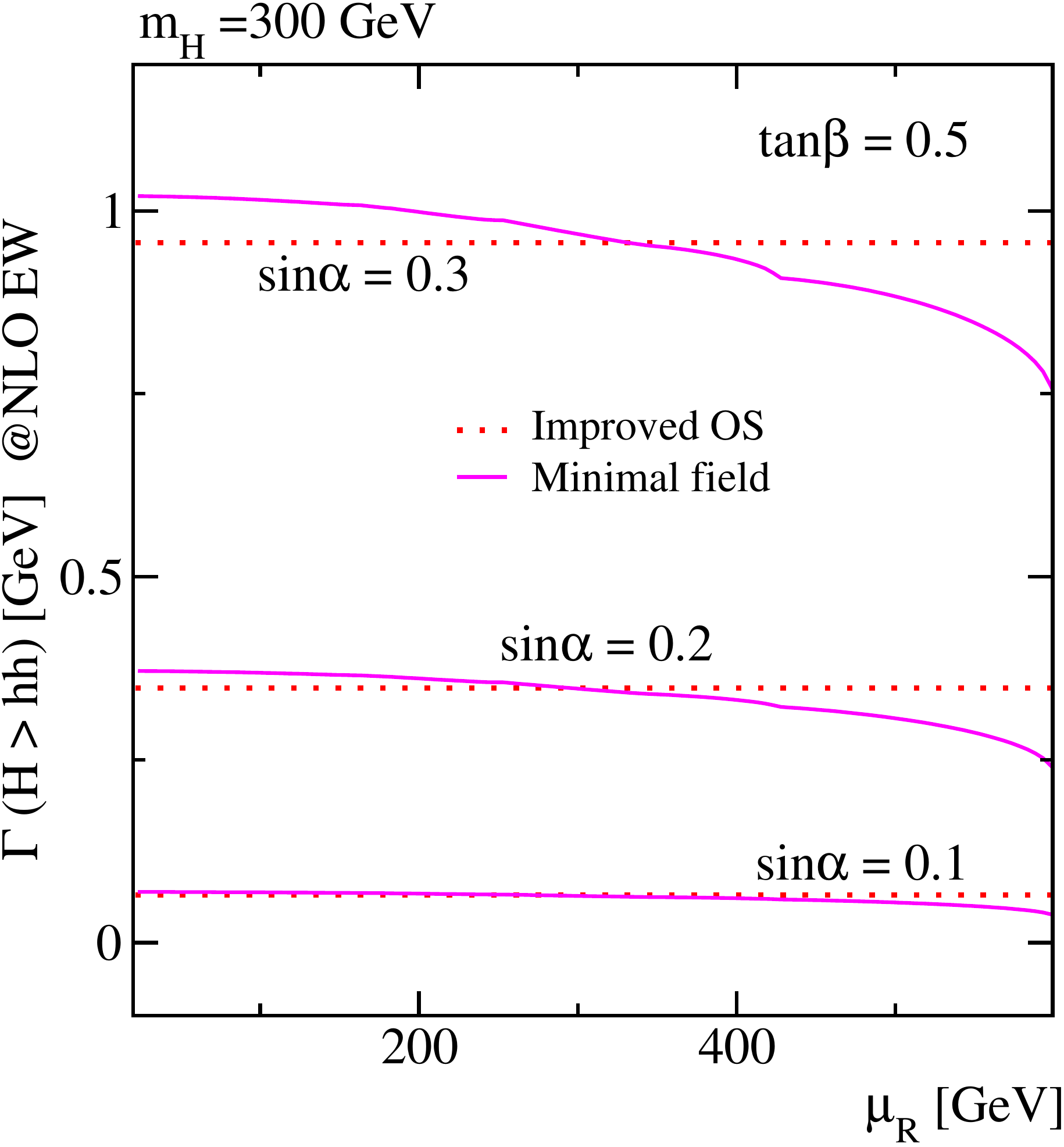} & \vspace{0pt} 
\includegraphics[width=0.3\textwidth]{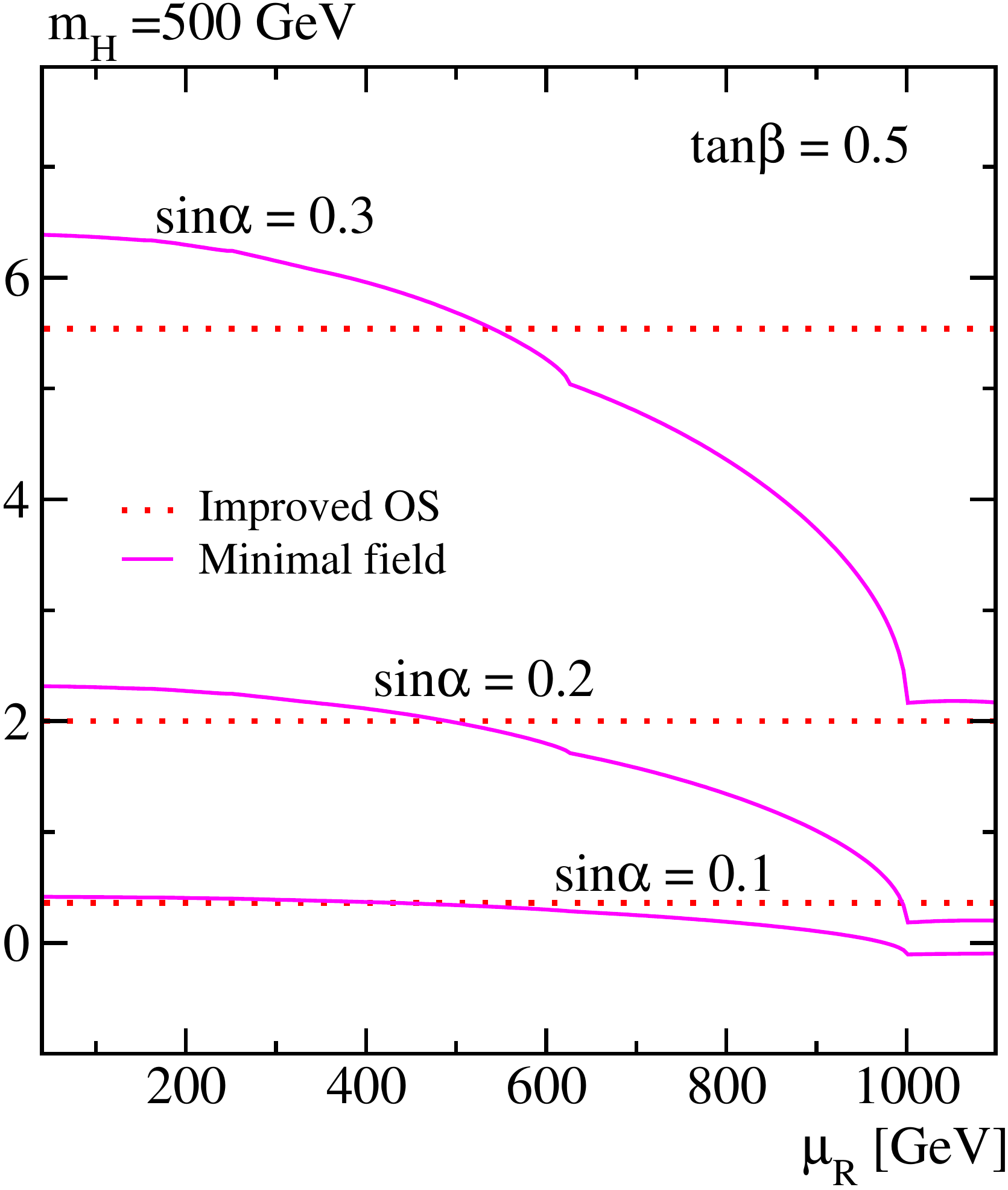} & \vspace{0pt} 
\includegraphics[width=0.31\textwidth]{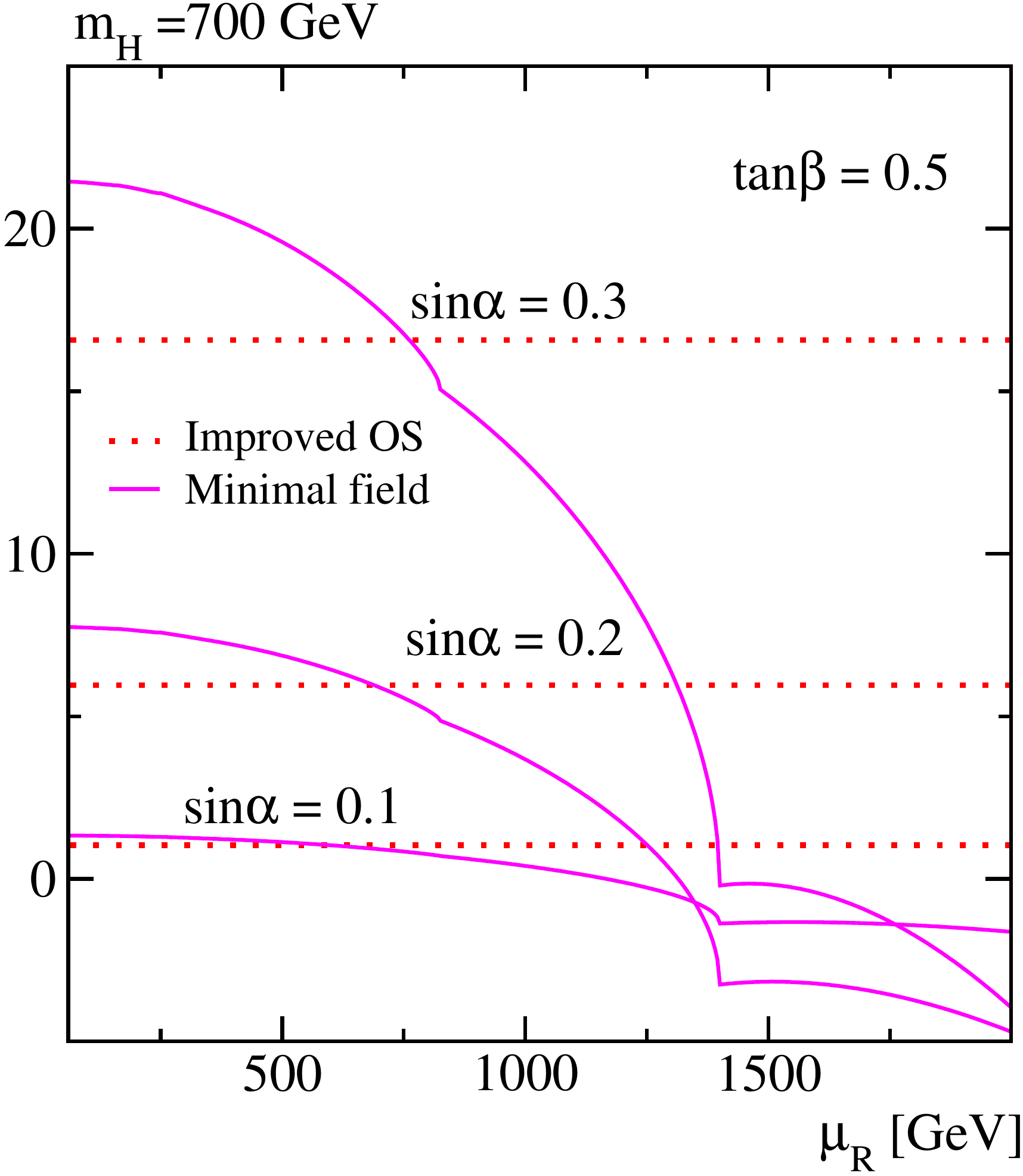} \\
\end{tabular}
\caption{NLO decay width $\GammaNLO$ as a function of the renormalization scale in the high-mass region, for exemplary heavy
Higgs masses, mixing angles, and $\tan\beta$ choices.
The scale-dependent predictions for the minimal field scheme
are represented by the solid (magenta) lines. The scale-independent
reference value (dotted, red lines) we obtain in the improved OS scheme. Parameter space constraints are  not shown. }
\label{fig:highmass-scale}
\end{center}
\end{figure}

\medskip{}
Finally, in Figure~\ref{fig:highmass-scale}
we analyse how $\GammaNLO$ varies with the
renormalization scale {in the (scale-dependent) minimal field scheme, as} introduced in Eq.~(\ref{eq:renorm-mixing}). We compare  
the \emph{minimal field}  to the (scale-independent) \emph{improved} OS scheme, which we show
as reference value. 
For $\tan\beta = 5$, $\GammaNLO$
flattens 
not far from the geometrical average mass scale 
$\mu_R^2 \simeq \pstar^2 = (m^2_h+m^2_H)/2$. 
Precisely around this value, both the
 \emph{minimal field} and the \emph{improved} OS predictions
tend to converge,  
suggesting that $\mu_R^2 = \pstar^2$ in Eq.~\eqref{eq:renorm-mixing} is indeed a convenient scale choice for the former.
Moreover, the very stable NLO predictions around this scale, added up
to the mild changes with the different renormalization schemes shown in
Table~\ref{tab:highmass}, indicate a small theoretical uncertainty.

\smallskip{}
Much steeper scale variations arise instead in the $\tan\beta < 1$ region.
Here, the $\GammaNLO$ predictions become {unstable},
especially
for heavy Higgs masses. {Such instability} may be once more traced back to
the Higgs-mediated scalar two-point graphs: these become overly  
large owing to the enhanced Higgs self-couplings, and artificially dominate
the scale dependence in these regions. 
Such a stark scale dependence is simply the reflect
of the poor perturbative behavior of the model in the vicinity
of {a strongly-coupled regime $\lambda \sim \mathcal{O}(4\pi)$}, which obviously 
translates into a huge
theoretical uncertainty. 

\bigskip{}
\subsection{Low-mass region}
\label{sec:lowmass}
Assuming now $m_{\Hzero} = 125.09$ GeV
and a free light Higgs mass $m_{\hzero}$,
direct LEP and LHC mass bounds, and most remarkably
the measured LHC Higgs signal strengths, narrow the viable $\sin{\alpha}$ region
down to a slim fringe $|\sin{\alpha}| \lesssim 1$.
Constraints become particularly tight in the region of interest $\mHH > 2\mh$,  given the limited tolerable room
for deviations in the total SM-like Higgs width when additional decay modes feature. 
State-of-the-art LHC constraints on the total Higgs width
place an upper limit of 
$\Gamma_{\hzero}\,\leq\,22\,\MeV$ \cite{Khachatryan:2014iha,Aad:2015xua}.
For definiteness, we hereafter adopt the fiducial choice $\sin{\alpha} = 0.998$ 
\cite{Robens:2015gla}. 

 \begin{figure}[thb!]
 \begin{center}
\begin{tabular}{c} \vspace{0pt} 
\includegraphics[width=0.3\textwidth,height=0.3\textwidth]{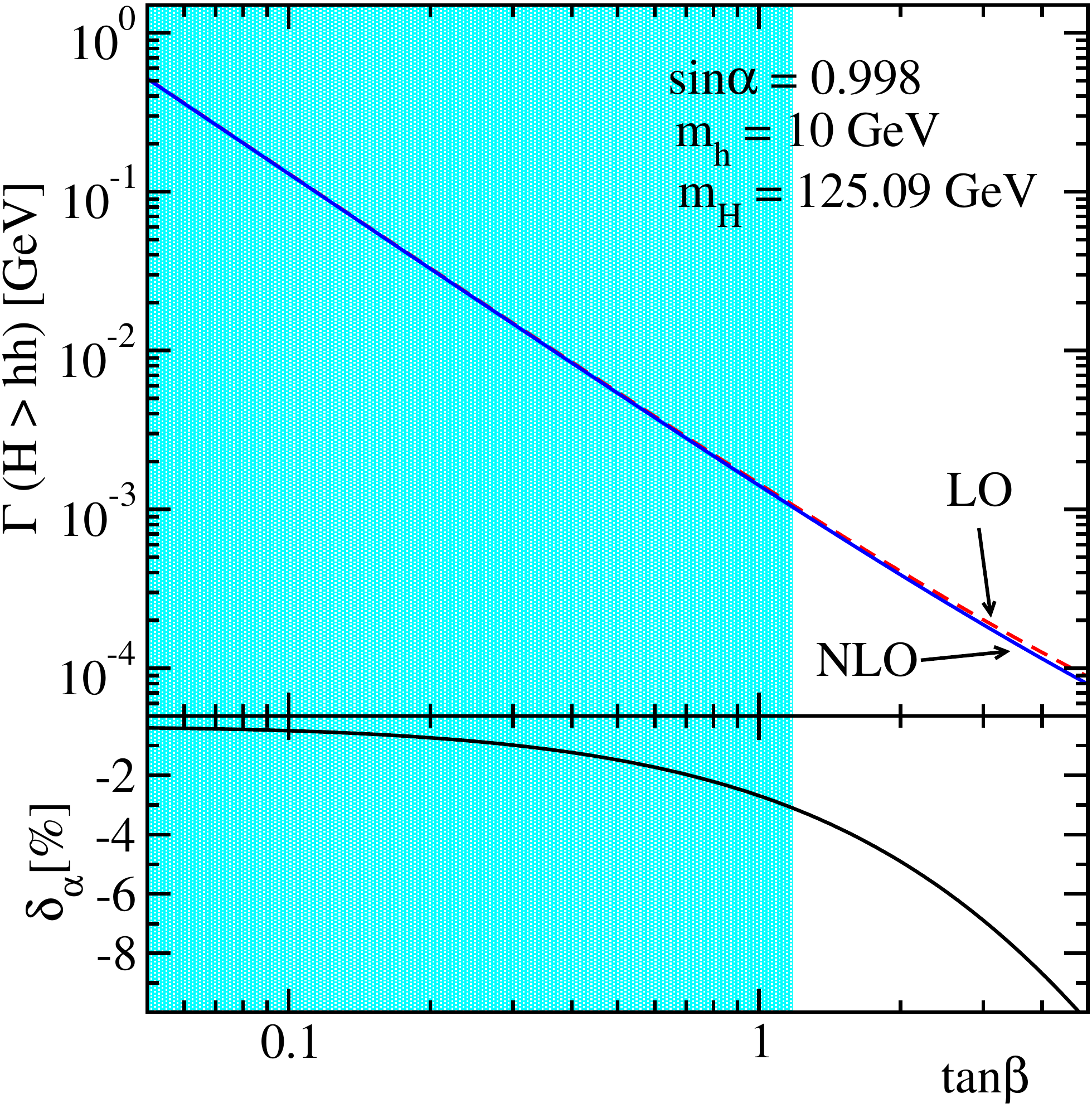}  
\hspace{0.05cm}
\includegraphics[width=0.3\textwidth,height=0.3\textwidth]{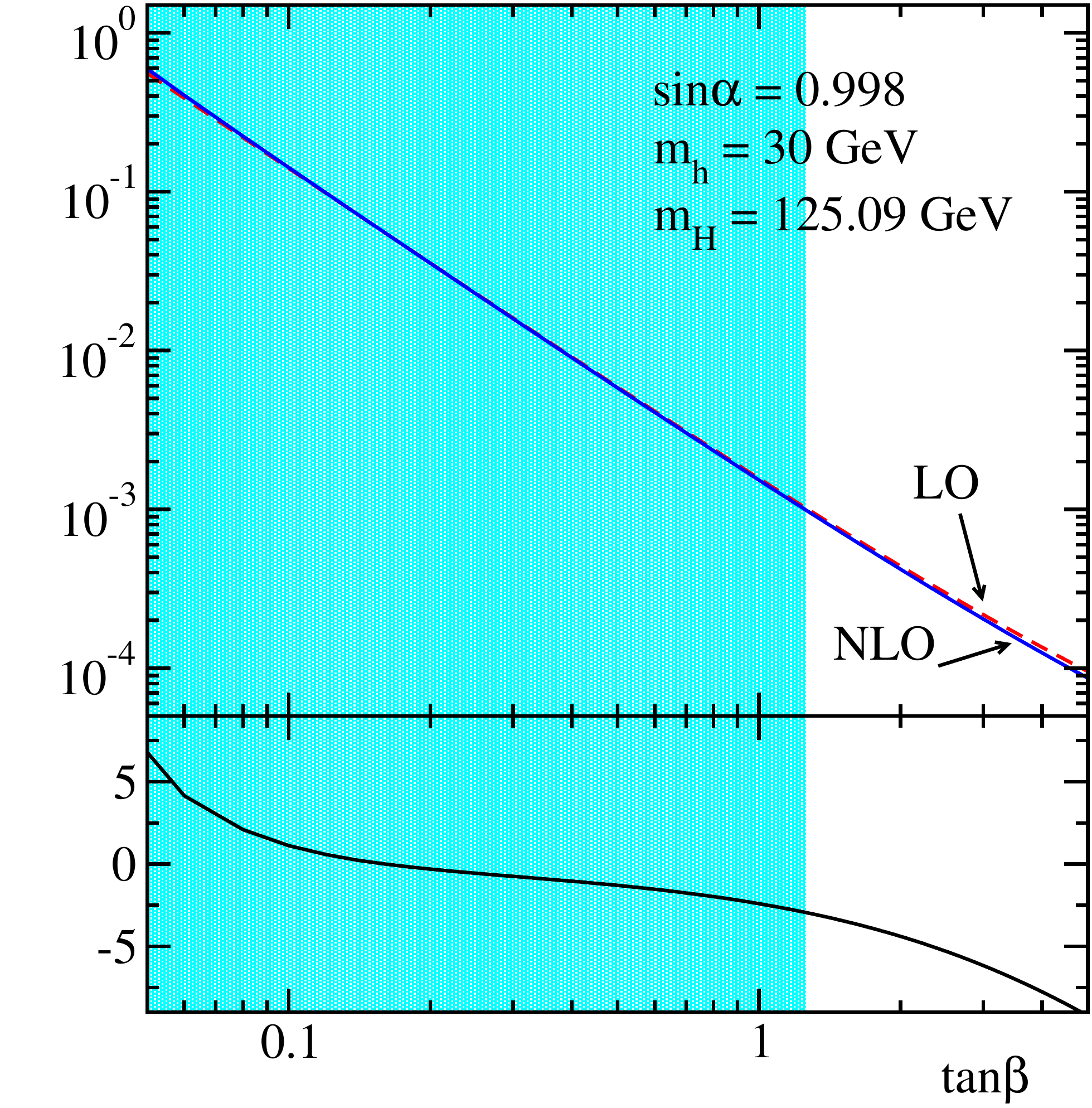}  
\hspace{0.05cm}
\includegraphics[width=0.3\textwidth,height=0.3\textwidth]{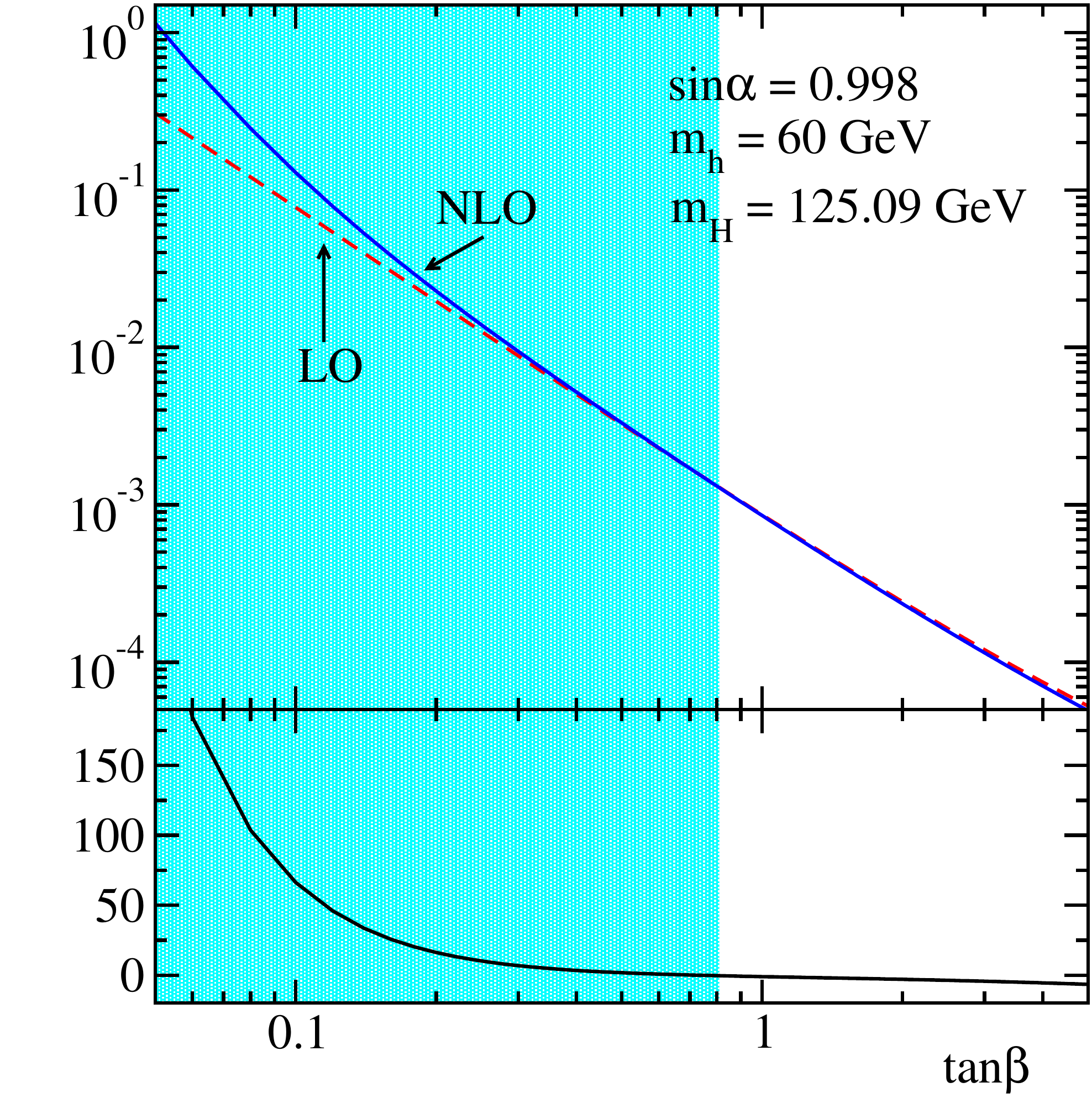}  
\\ \vspace{0.2cm}\\
\includegraphics[width=0.31\textwidth,height=0.3\textwidth]{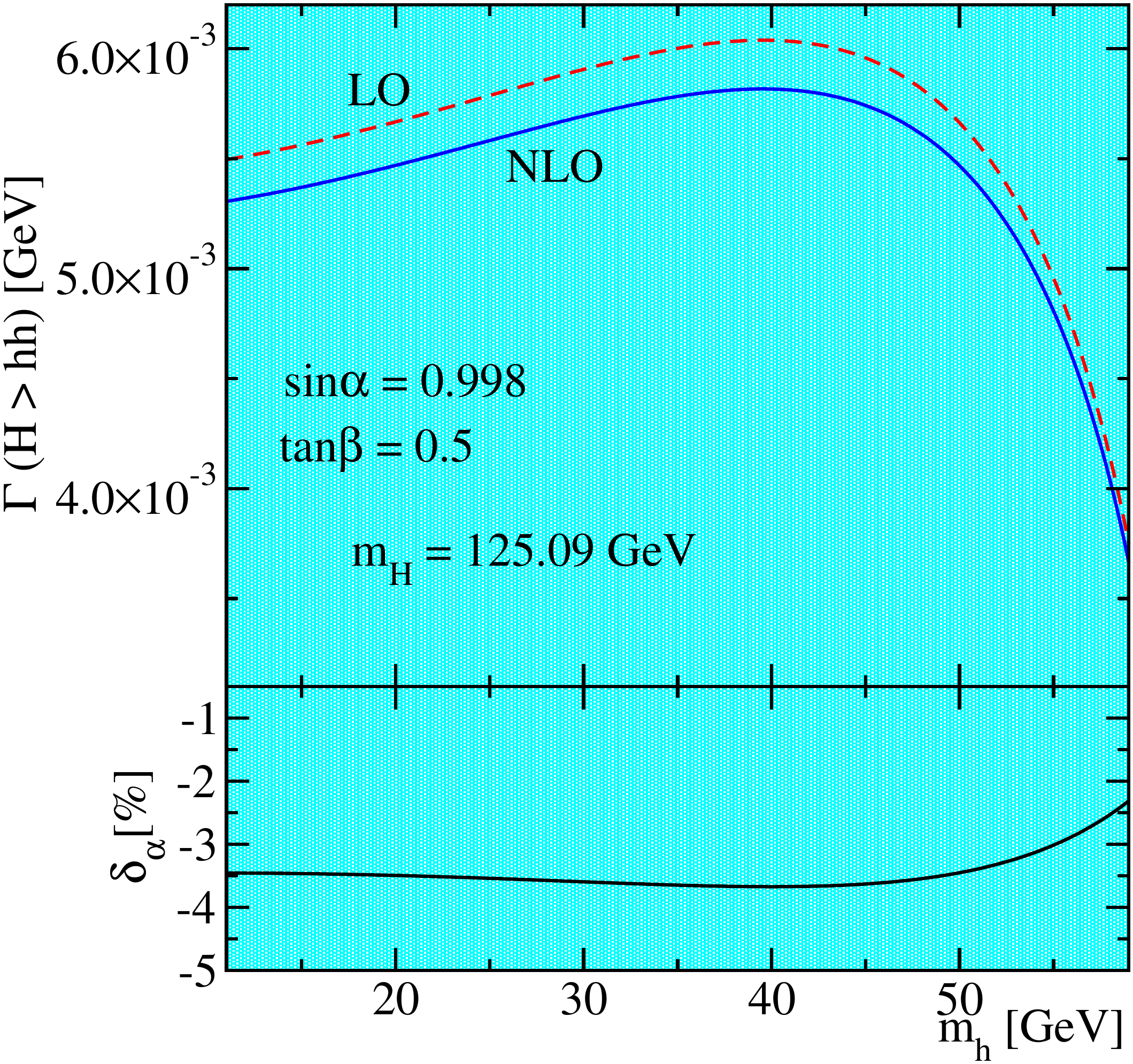}  
\hspace{0.05cm}
\includegraphics[width=0.31\textwidth,height=0.3\textwidth]{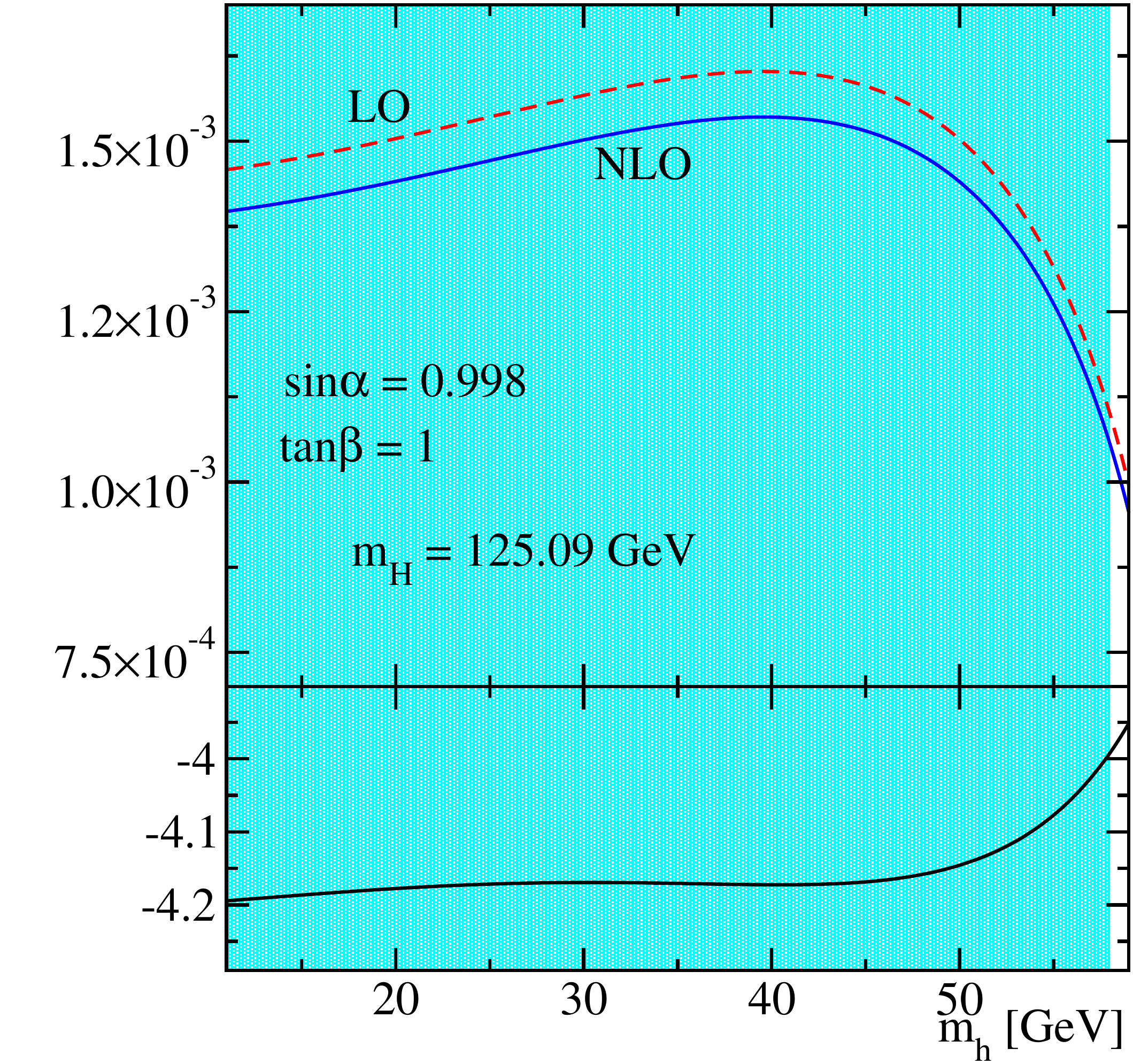}  
\hspace{0.05cm}
\includegraphics[width=0.31\textwidth,height=0.3\textwidth]{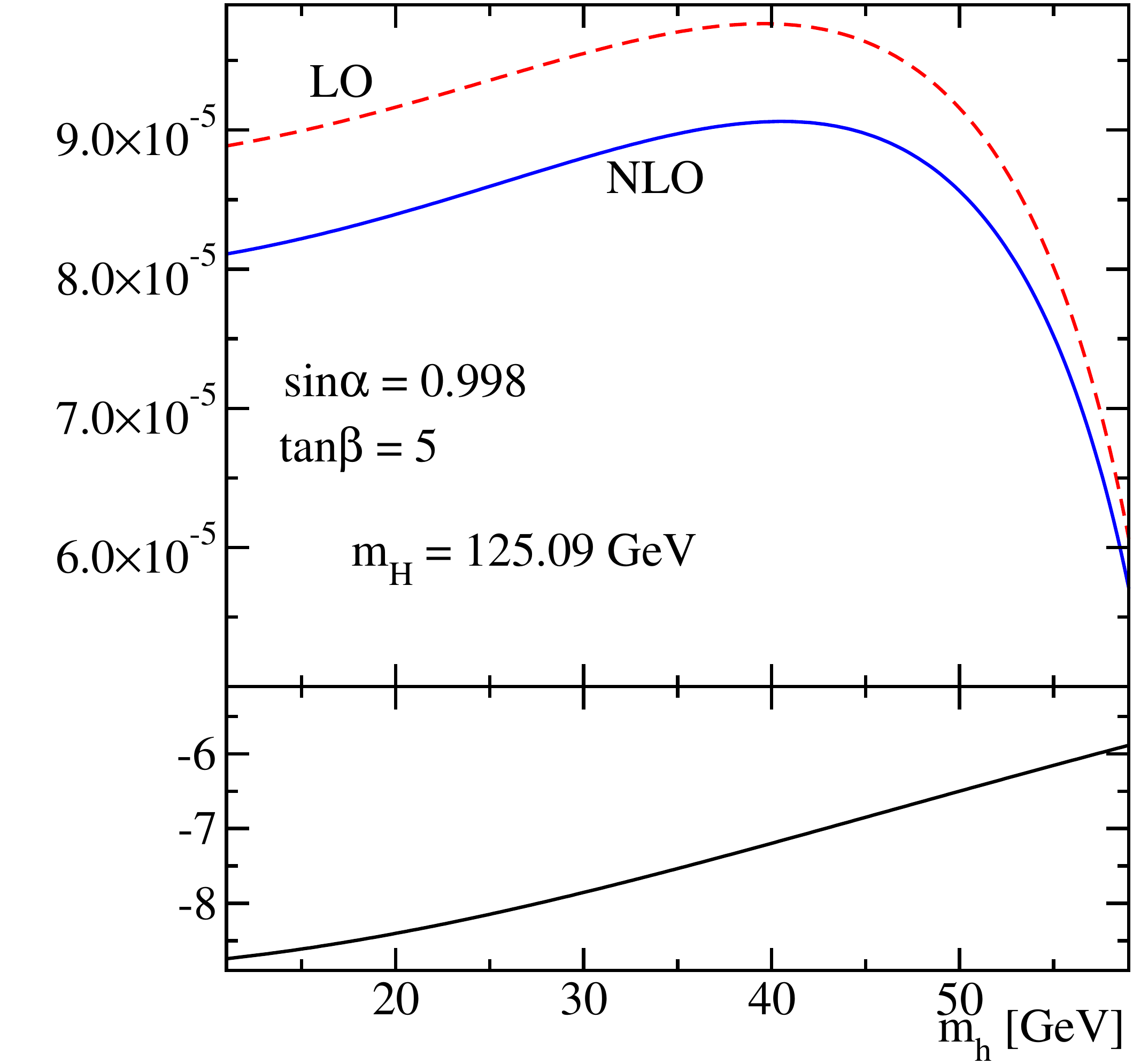}  
\end{tabular}
\caption{Heavy-to-light Higgs boson width $\GHhh$ in the low-mass region.
The results are shown at LO (dashed, red) and NLO (full, blue) as a function of the
relevant parameters of the model. The mixing angle is fixed to $\sin{\alpha} = 0.998$ in 
all cases.
The lower subpanels show the relative EW
one-loop correction in the $\alpha_{\rm em}$-parametrization ~\eqref{eq:rel-alpha}.
Renormalization is performed in the improved on-shell scheme. The shaded  areas in the low $\tan\beta$ and $m_h$ ranges are ruled out by the
LHC Higgs signal strength measurements.}
\label{fig:over-lowmass}
\end{center}
\end{figure}

In Figure~\ref{fig:over-lowmass} we examine the parameter space
dependence of $\Gamma_{\Hhh}$ in this scenario. 
Complementarily,
in Table~\ref{tab:lowmass} we provide precise
predictions for specific parameter space points, while comparing again the different renormalization schemes. 
In
Figure~\ref{fig:scanlight} we analyse the $\mh - \tan\beta$ interplay 
by showing the total NLO amplitude $\GammaNLO$~\eqref{eq:nlowidth} and 
the relative NLO correction $\delta_{\alpha}$~\eqref{eq:rel-alpha} 
in the form of 
two-dimensional density maps. 

\smallskip{}
The obtained $\GammaNLO$ values span two orders of magnitude, 
ranging
from $\mathcal{O}(10^{-3})$ down to $\mathcal{O}(10^{-5})$ GeV
as we navigate throughout the different parameter space regions.
This sharp variation is again connected to the behavior of the leading-order coupling $\lambda_{Hhh}$:
whilst it tends to zero in the limit $\sin{\alpha} \to 1$, it can
yet contribute if the $\cot\beta$-enhanced terms are large enough. Either way, 
let us once more recall that a significant patch of the low-$\tan\beta$ range is in practice
precluded by the different constraints on the model (see {\it e.g.} the
top panels of Fig.~\ref{fig:over-lowmass} and Fig.~\ref{fig:scanlight}). In
particular, the shaded 
regions at small $\tan\beta$ and low Higgs masses 
are incompatible with the LHC Higgs signal strength measurements.
Another salient feature is the steep rise of the quantum corrections at low
$\tan\beta$ (see the top panels of Fig.~\ref{fig:over-lowmass}):
these are positive, tend to increase with the light Higgs mass, and may surmount
the $\mathcal{O}(50\%)$ level.
Aside from the discussed Higgs-mediated loop enhancements,
additional mechanisms reinforce this behavior in this case:
i) the suppressed tree-level decay amplitude, due to the lesser
phase space available, the closer we move to the kinematical threshold; 
ii) the vicinity of the di-Higgs loop threshold, which
invigorates the light Higgs-mediated loops even further. 
For $\tan\beta > 1$ the corrections are instead 
moderate and negative, becoming even more so 
for very light $m_{h}$ values. The latter effect
may be traced back to the fermionic (viz. the top-mediated ) three-point loops,
which are in this case the dominant source of quantum corrections and present 
a trademark logarithmic dependence  
$\sim \log(m_t^2/m_h^2)$.

\medskip{}
The scale dependence of the NLO results in this low-mass region 
is analysed in Figure~\ref{fig:lowmass-scale}. 
The $\GammaNLO$ predictions obtained in the improved OS and the minimal field schemes
schemes agree very well in the ballpark of the geometric average mass $\mu_R^2 = \pstar^2 = (m_h^2+m_H^2)/2$,
and the latter barely varies with the scale. The very stable slope
even in the  $\tan\beta < 1$ region, which is in contrast to the strong
scale dependence in the high-mass region, is explained by the much lower scales 
$\mu_R^2 \simeq \pstar^2$ involved in this case, for which the finite
Higgs-mediated contributions to the Higgs field two-point functions are much smaller.

 \begin{figure}[t!]
 \begin{center}
\includegraphics[width=0.47\textwidth,height=0.39\textwidth]{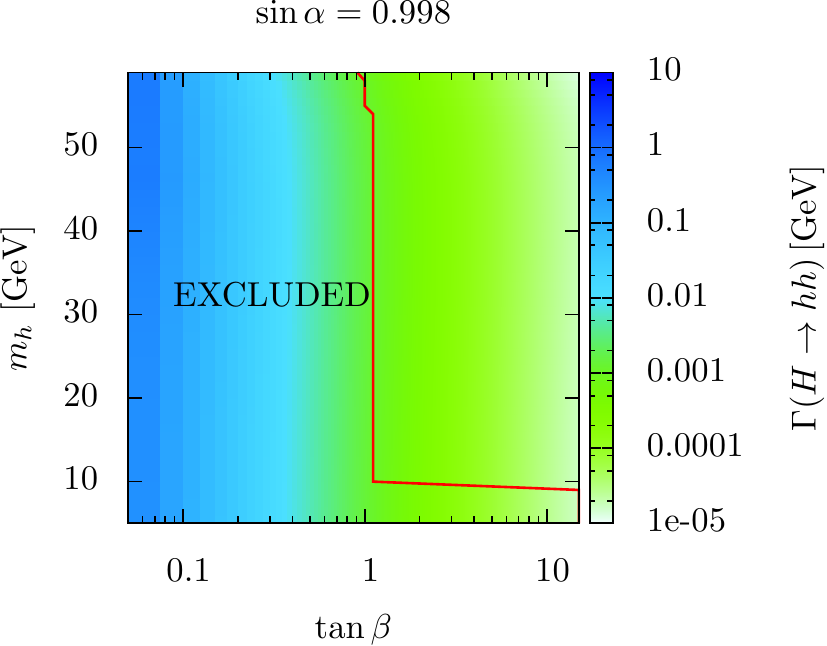} \hspace{0.7cm}
\includegraphics[width=0.44\textwidth,height=0.39\textwidth]{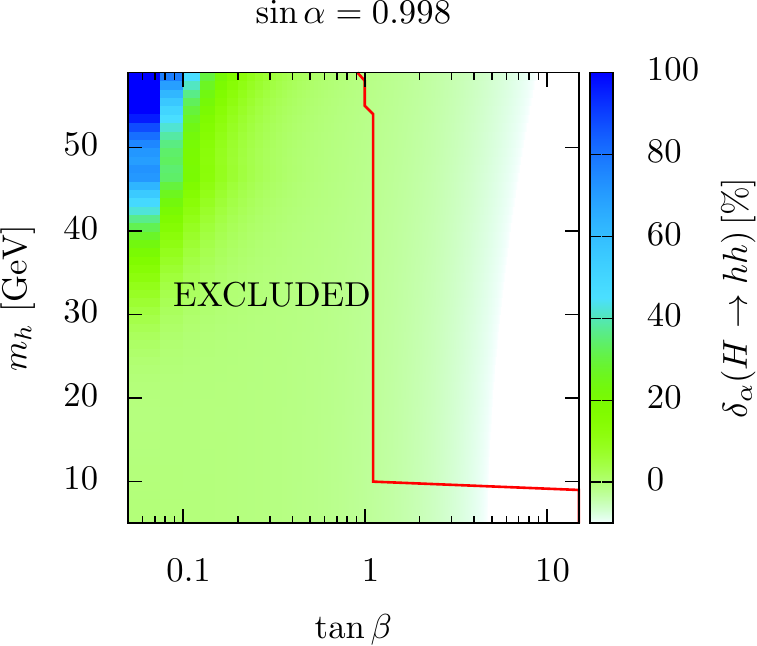} 
\caption{Loop-corrected partial $\Hhh$ width (left panel) and
relative one-loop EW corrections $\delta_{\alpha}$
in the $\alpha_{\rm em}$-parametrization ~\eqref{eq:rel-alpha},
projected on the $\mh-\tan\beta$ plane in the low-mass region. The mixing angle 
is fixed to the fiducial choice $\sin{\alpha} = 0.998$. 
{Renormalization is performed in the improved on-shell scheme.}
The regions on the  left and below 
the 
red contour are excluded by constraints. The strong exclusion in the low mass region follows from direct LEP searches \cite{Schael:2006cr}.}
\label{fig:scanlight}
\end{center}
\end{figure}

\begin{table}[thb!]
\begin{center}
\footnotesize{
\begin{tabular}{|c|ll|llll|} \hline
 $\mHH$ [GeV] &  & {$\Gamma_{\alpha}^{\text{LO}}(\Hhh)$ [GeV]} & &{$\Gamma_{\alpha}^{\text{NLO}}(\Hhh)$ [GeV]} & $\delta_{\alpha}\,[\%]$ &  $\delta_{\gf}\,[\%]$ \\ \hline
\multicolumn{7}{|l|}{\qquad $\tan\beta = 0.5$, \qquad $\sin\alpha = 0.998$} \\ \hline
\multirow{3}{*}{10} & & \multirow{3}{*}{5.496$\times 10^{-3}$} & OS & 5.416$\times 10^{-3}$ & -1.458 & -1.456 \\ 
 &  & & Improved OS & 5.415$\times 10^{-3}$ & -1.480 & -1.479 \\
   & & & Minimal field  & 5.414$\times 10^{-3}$ & -1.489 & -1.488 \\ 
   \cline{1-7}
\multirow{3}{*}{30} & & \multirow{3}{*}{5.920$\times 10^{-3}$} & OS & 5.841$\times 10^{-3}$ & -1.345 & -1.344 \\ 
 &  & & Improved OS & 5.844$\times 10^{-3}$ & -1.289 & -1.288 \\
   & & & Minimal field   & 5.844$\times 10^{-3}$ & -1.281 & -1.280 \\ 
   \cline{1-7}
\multirow{3}{*}{60} & & \multirow{3}{*}{3.267$\times 10^{-3}$} & OS & 3.333$\times 10^{-3}$ & 2.019 & 2.017 \\ 
 &  & & Improved OS & 3.323$\times 10^{-3}$ & 1.733 & 1.731 \\
   & & & Minimal field  & 3.331$\times 10^{-3}$ & 1.952 & 1.949 \\ 
   \hline    
\multicolumn{7}{|l|}{\qquad $\tan\beta = 5$, \qquad $\sin\alpha = 0.998$} \\ \hline
\multirow{3}{*}{10} & & \multirow{3}{*}{8.881$\times 10^{-5}$} & OS & 7.966$\times 10^{-5}$ & -10.310 & -10.216 \\ 
 &  & & Improved OS & 7.967$\times 10^{-5}$ & -10.296 & -10.202 \\
   & & & Minimal field   & 7.958$\times 10^{-5}$ & -10.394 & -10.300 \\ 
   \cline{1-7}    
\multirow{3}{*}{30} & & \multirow{3}{*}{9.567$\times 10^{-5}$} & OS & 8.686$\times 10^{-5}$ & -9.212 & -9.128 \\ 
 &  & & Improved OS & 8.687$\times 10^{-5}$ & -9.201 & -9.118 \\
   & & & Minimal field   & 8.680$\times 10^{-5}$ & -9.279 & -9.195 \\ 
   \cline{1-7}
\multirow{3}{*}{60} & & \multirow{3}{*}{5.279$\times 10^{-5}$} & OS & 4.931$\times 10^{-5}$ & -6.589 & -6.529 \\ 
 &  & & Improved OS & 4.932$\times 10^{-5}$ & -6.584 & -6.525 \\
   & & & Minimal field   & 4.929$\times 10^{-5}$ & -6.627 & -6.567 \\ 
   \hline\hline
   \end{tabular}
\caption{\label{tab:lowmass}
 Heavy-to-light Higgs decay width $\GHhh$ at LO and NLO EW accuracy for representative
parameter choices and renormalization schemes in the low-mass region. 
{The total decay widths are obtained in the $\alpha_{\text{em}}$-parametrization, as defined in Eqs.~\eqref{eq:rel-alpha},
while the relative
one-loop EW effects are quantified in both the $\alpha$-parametrization
and the $\gf$-parametrization, cf. Eq.~\eqref{eq:rel-gf}}.
For the (scale-dependent) minimal field scheme, the renormalization scale is fixed to the geometrical average mass $\mu_R^2 = \pstar^2 = (m_h^2+m_H^2)/2$.
{The input value for $\tan\beta$ is linked to the singlet vev through $v_s = (\sqrt{2}G_F)^{-1/2}\,\tan\beta$.}}
}
\end{center}
\end{table}

 \begin{figure}[t!]
 \begin{center}
\includegraphics[width=0.3\textwidth,height=0.3\textwidth]{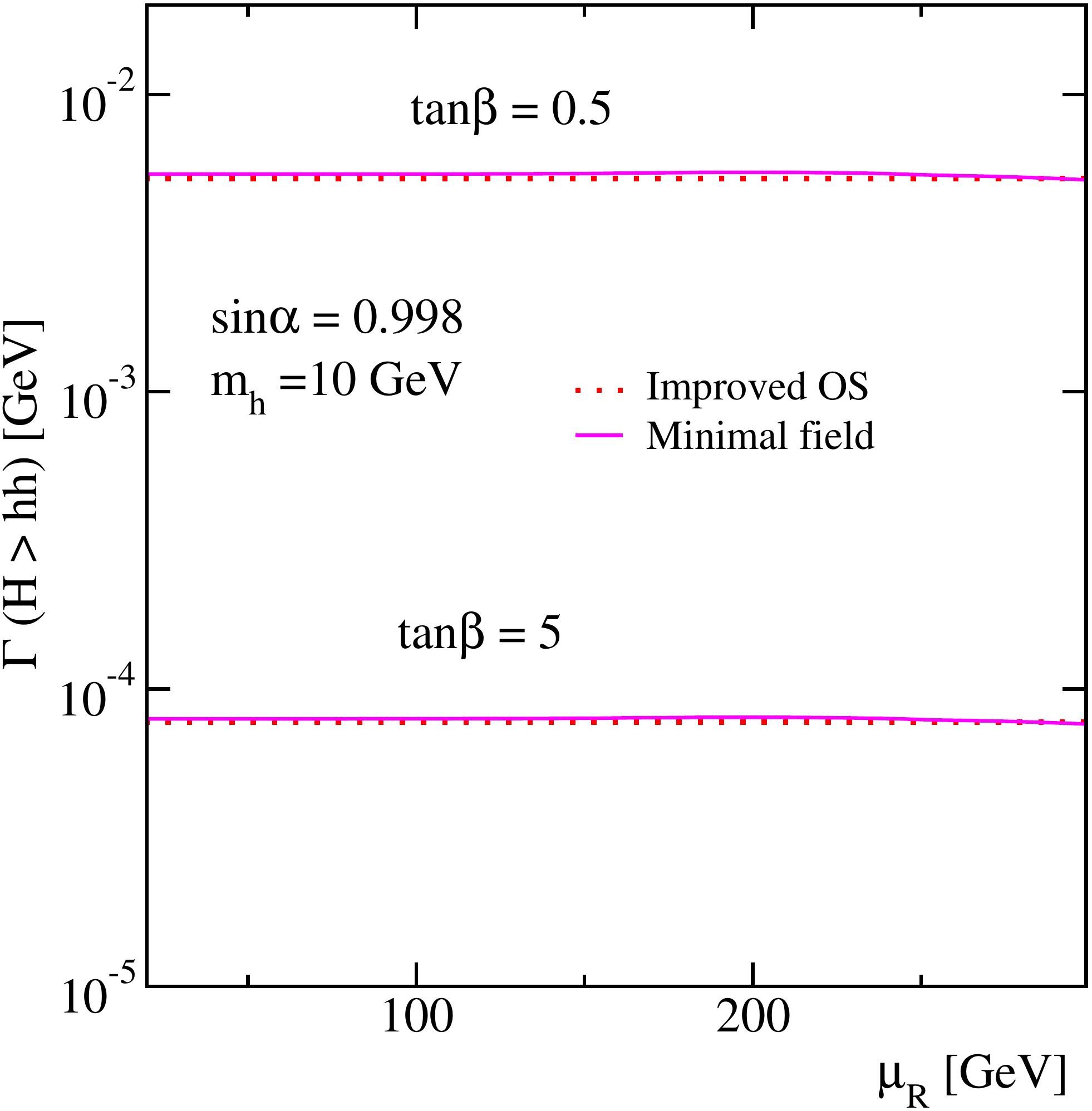} \hspace{0.2cm} 
\includegraphics[width=0.3\textwidth,height=0.3\textwidth]{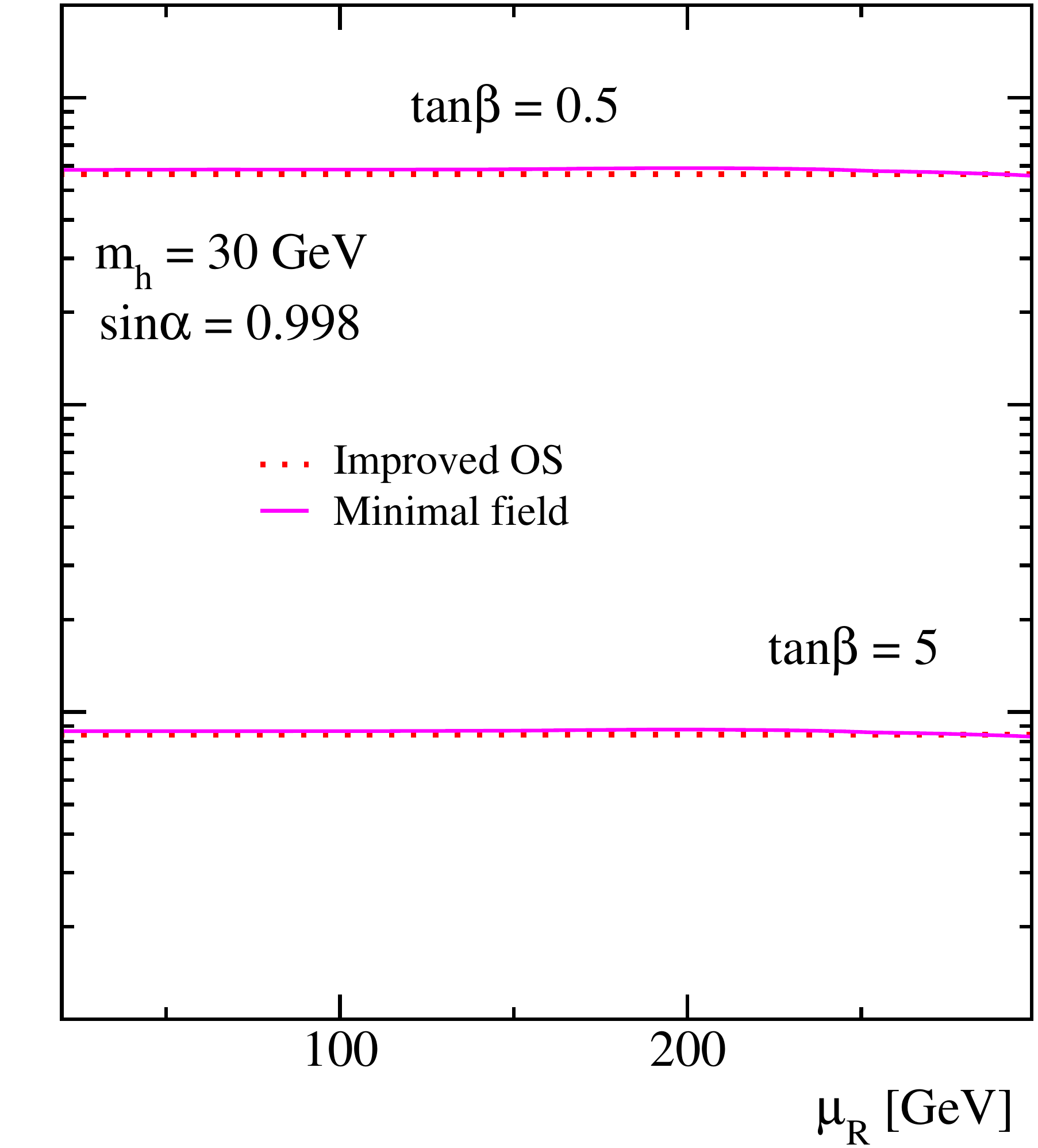} \hspace{0.2cm} 
\includegraphics[width=0.3\textwidth,height=0.3\textwidth]{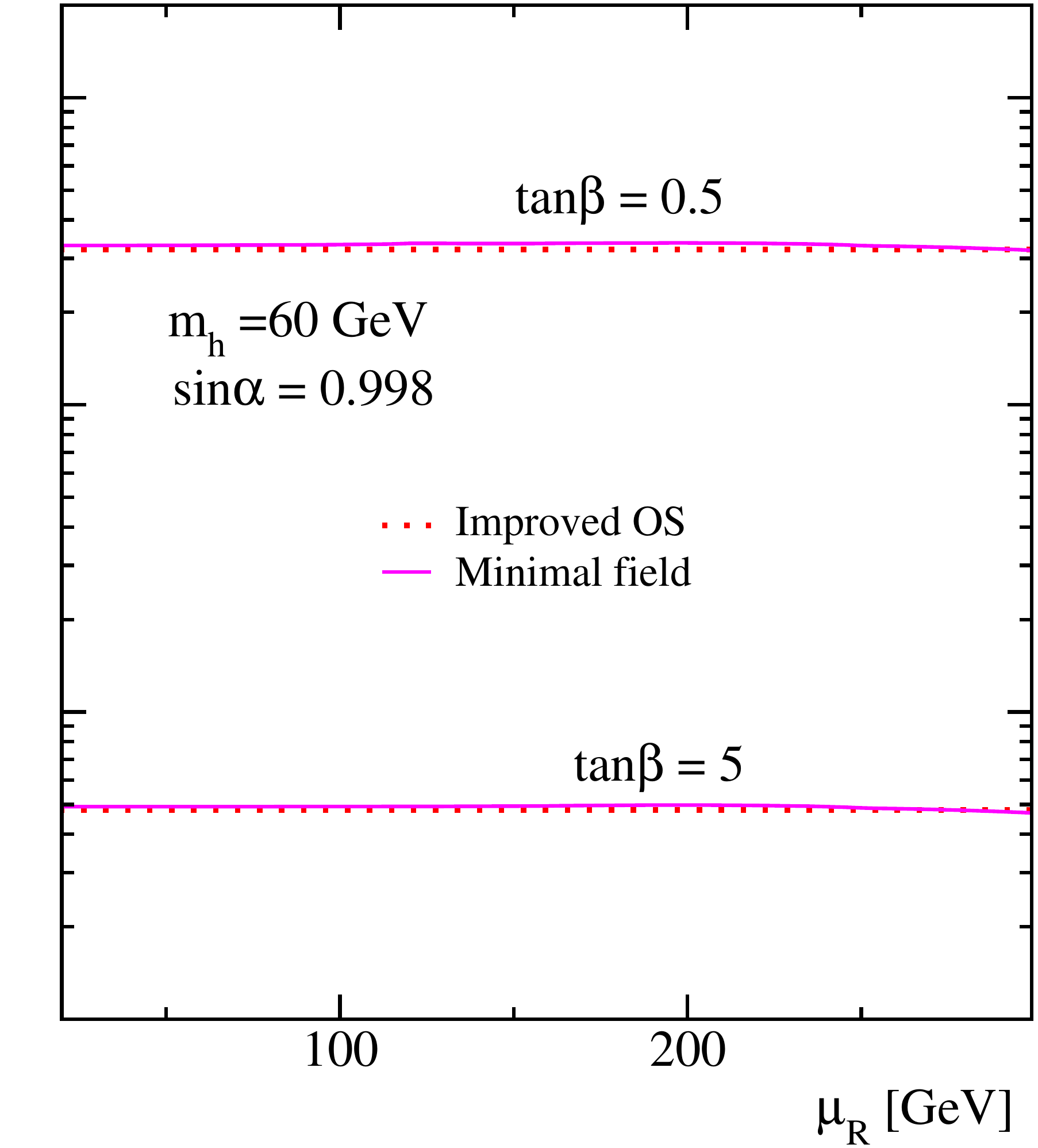}  
\caption{NLO decay width $\GammaNLO$ as a function of the renormalization scale in the low-mass region, 
for exemplary heavy
Higgs masses and $\tan\beta$ values. The mixing angle is fixed
to $\sin{\alpha}=0.998$.
The scale-dependent predictions for the minimal field scheme
are represented by the solid (magenta) lines. The scale-independent
reference value (dotted, red lines) we obtain in the improved OS scheme. Parameter space constraints are not shown.}
\label{fig:lowmass-scale}
\end{center}
\end{figure}

\medskip{}
In Figure~\ref{fig:br-lowmass} we recast the above analysis in terms of 
the heavy Higgs branching ratios. We track down their 
behavior as a function
$\tan\beta$ for exemplary light Higgs masses and fiducial mixing $\sin{\alpha} = 0.998$.
From values of $\tan\beta \lesssim 1$ onwards,
the obtained decay pattern approaches that of a purely SM-like Higgs boson.
The dominant mode is $b\bar{b}$,  while the di-Higgs final state 
is hampered due to the tiny tree-level coupling $\lambda_{Hhh} \sim c_{\alpha}$. 
In this case, the $\Hhh$ mode
carries not more than a few percent
of the total budget - on equal footing with
the loop-induced decay $H \to gg$. If we instead move towards lower $\tan\beta$ values, 
the $\cot\beta$-enhanced terms 
overcome in part the $\sin{\alpha} \to 1$ suppression and 
promote $\Hhh$ again to a chief role. 

 \begin{figure}[thb!]
 \begin{center}
\includegraphics[width=0.31\textwidth,height=0.31\textwidth]{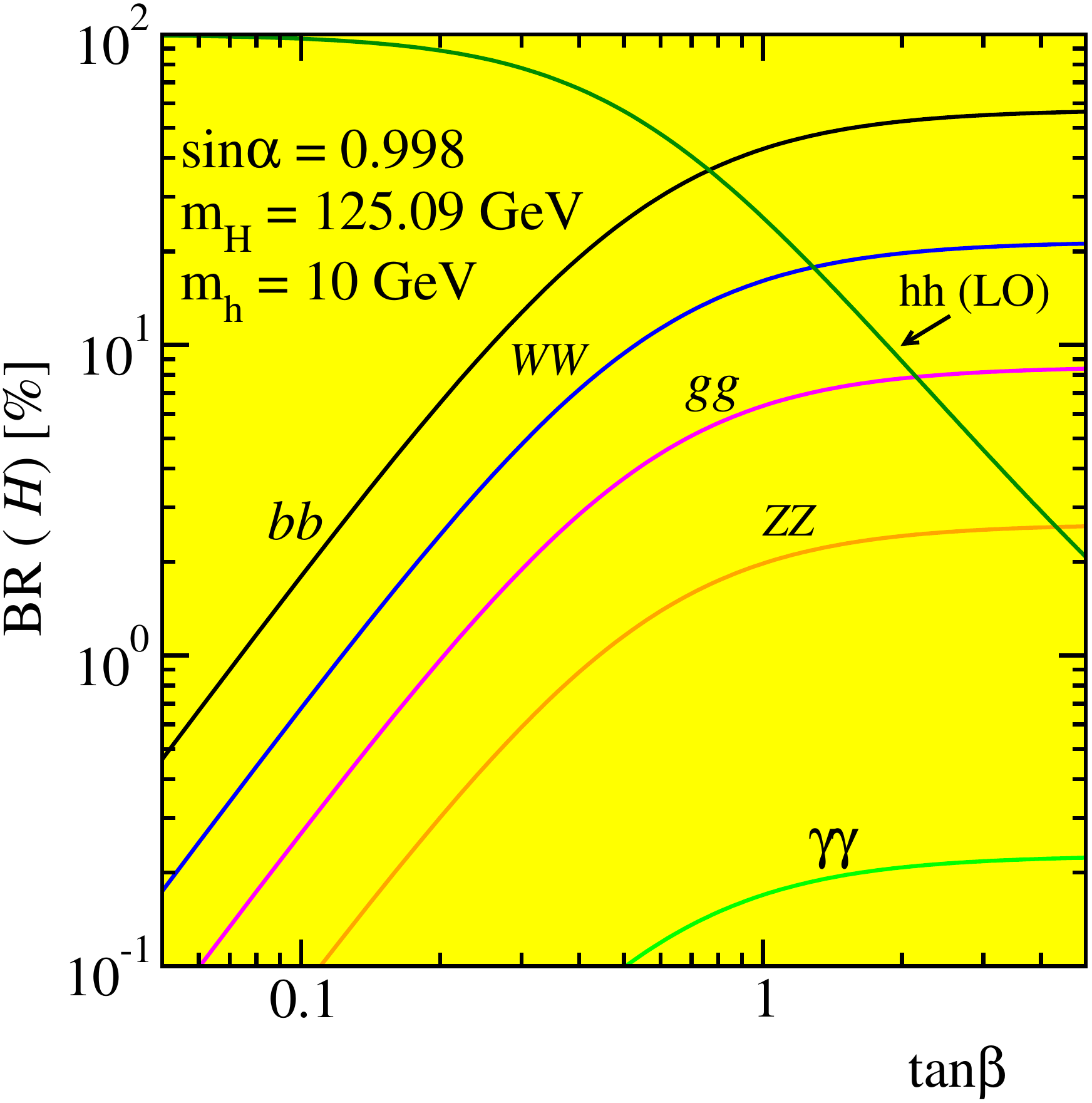} \hspace{0.01cm}
\includegraphics[width=0.31\textwidth,height=0.31\textwidth]{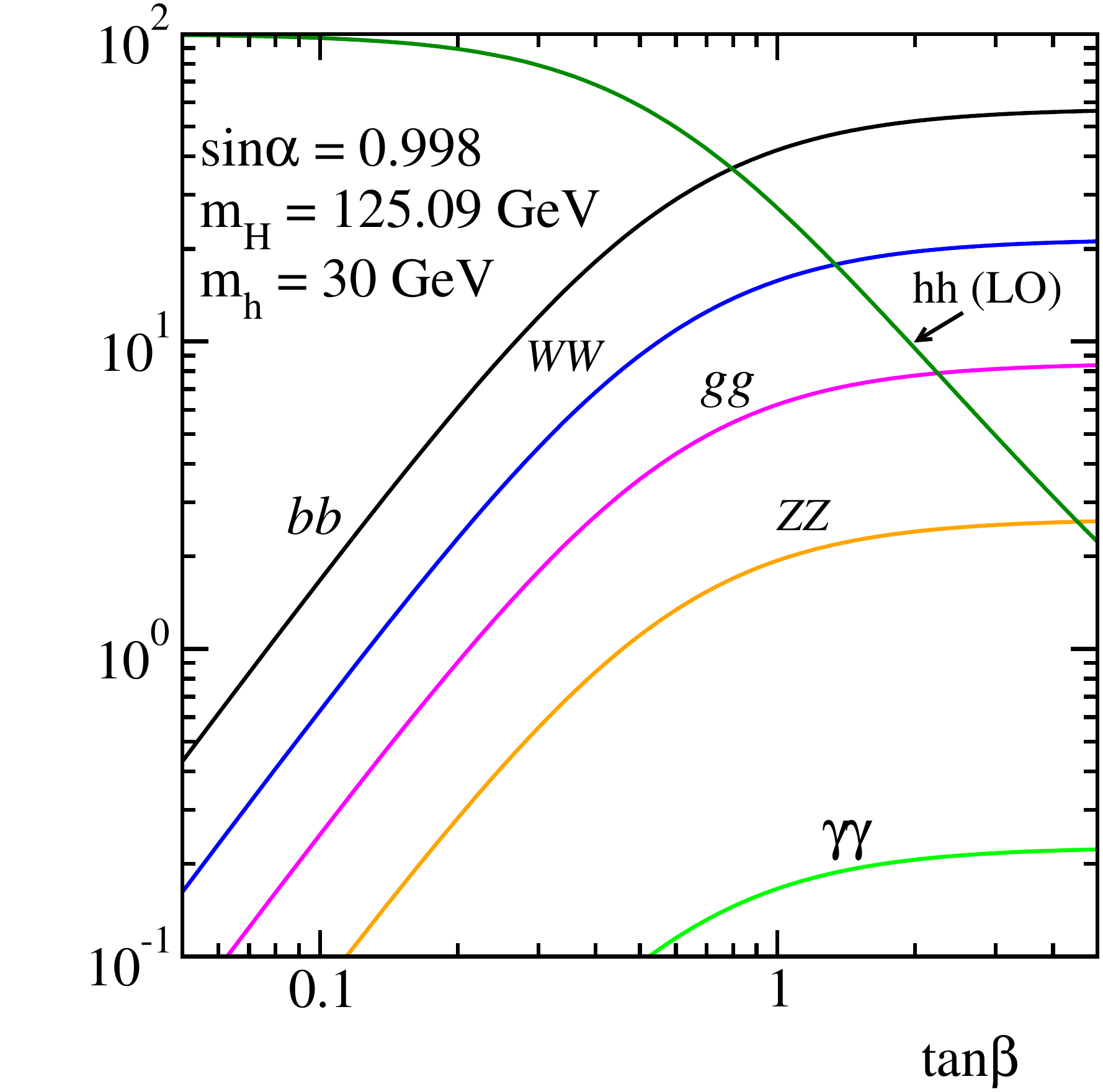} \hspace{0.01cm}
\includegraphics[width=0.31\textwidth,height=0.31\textwidth]{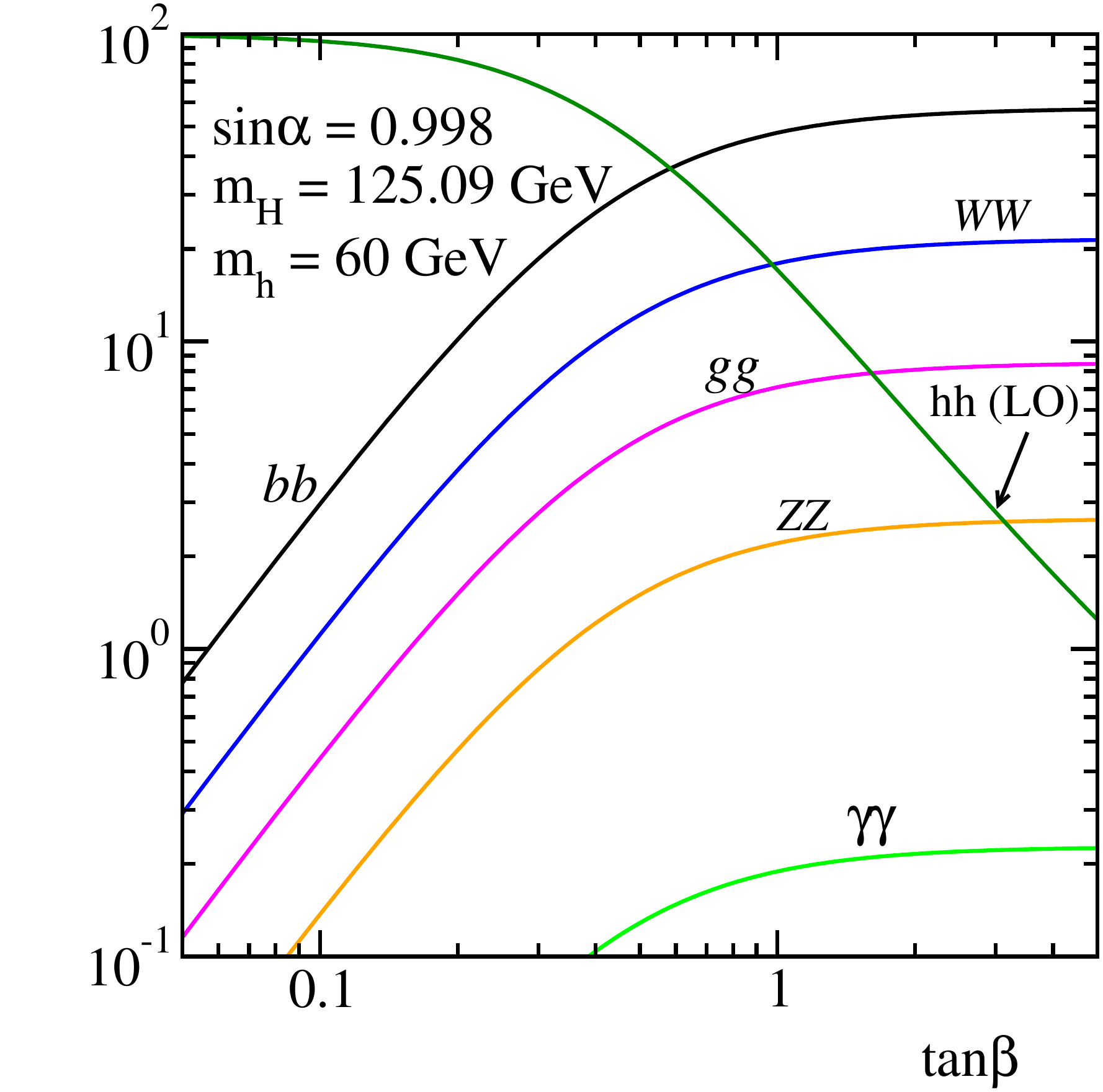} 
\caption{Heavy Higgs branching ratios (in $\%$) 
as a function of $\tan\beta$ in the low-mass region. 
The results are shown 
for representative light Higgs mass values, with fiducial mixing
angle $\sin{\alpha} = 0.998$. The whole parameter space region shown
in the left panel is excluded by perturbative unitarity.}
\label{fig:br-lowmass}
\end{center}
\end{figure}

\subsection{Maximal branching ratios}
\label{sec:maxbr}

So far we have discussed the general behavior of the NLO {EW} corrections to the 
heavy-to-light Higgs decay width 
along the relevant parameter space directions
$\sin{\alpha},\;\tan\be,\;m_{h/H}$. Before closing, we {focus on} the series of
benchmarks with maximal tree-level heavy-to-light Higgs branching ratio {proposed in \cite{sbm}}.
{These are defined}
as a function
of the heavy Higgs through the parameter choices quoted in Tables~\ref{tab:brmax},
for the high and the low mass regions respectively. In these regimes, 
the decays of the heavy Higgs state provide a particularly interesting phenomenological 
 ground for studying
finite width effects and lineshape modifications in the production of a heavy scalar resonance, cf. e.g. \cite{Franzosi:2012nk,Bonvini:2013jha,Kauer:2015hia,Kauer:2015dma,Maina:2015ela,Ballestrero:2015jca}\footnote{{See also e.g.
Ref.~\cite{Dawson:2015haa} in the context of Higgs pair production.}}.

Numerical predictions for $\Gamma_{\Hhh}^{\text{LO}}$ and $\Gamma_{\Hhh}^{\text{NLO}}$,
together with the relative one-loop correction in the 
two parametrizations $\delta_{\alpha_{\text{em}}}$ and $\delta_{\text{GF}}$ \eqref{eq:rel-gf}
are provided in Tables~\ref{tab:brmaxhigh} and ~\ref{tab:brmaxlow}, for the high and low mass
regions respectively. Complementarily, we list down the corresponding
branching fractions for the additional decay modes (barring those channels below $0.01\%$).
Let us recall that for the latter we use the rescaled partial widths from ~\cite{Heinemeyer:2013tqa}, while
for {$H\to hh$ we quote the LO result in the $\alpha_{\text{em}}$-parametrization, in line with Figures~\ref{fig:br-highmass} and \ref{fig:br-lowmass}.}

%

\begin{table}[thb!] 
\begin{center}
\footnotesize{
\begin{tabular}{|lccccc|lcccc|} \hline
\multicolumn{5}{|c}{high mass region} & \qquad &\multicolumn{5}{c|}{low mass region} \\ \hline
 & $m_H [\GeV]$&$|\sin\alpha|_\text{max}$&$BR^{H\rightarrow\,h\,h}_\text{max}$&$\tan\be$ & \qquad&  &$m_h [\GeV]$&$|\sin\alpha|_{\text{min}}$&$BR^{H\rightarrow\,h\,h}_\text{max}$&$\tan\be$ \\ \hline
BHM1 & 300&0.31&0.34&3.71 & \qquad & BLM1 & 60&0.9997 &0.26&0.29\\
BHM2 & 400&0.27&0.32&1.72 &\qquad& BLM2 &50&0.9998&0.26&0.31\\

BHM3 &500&0.24 &0.27&2.17&\qquad&  BLM3& 40&0.9998&0.26&0.32\\
BHM4 &600&0.23&0.25&2.70&\qquad& BLM4& 30&0.9998&0.26&0.32\\
BHM5 &700&0.21&0.24&3.23&\qquad& BLM5&20&0.9998&0.26&0.31\\
BHM6 &800&0.21&0.23&4.00&\qquad& BLM6&10&0.9998&0.26&0.30\\ \hline
\end{tabular}}
\end{center}
\caption{
Maximal branching ratios for the heavy-to-light Higgs decay mode $\Hhh$
in the high-mass (left) {and low-mass} regions {(right)} as proposed in Refs. \htb{\cite{sbm,Robens:2016xkb}}; 
{the results quoted here are obtained {in} the setup {of the mentioned references,} evaluating $\Gamma(H \to hh)$ 
at LO in the $G_F$-parametrization.}
Note that the maximal branching ratios are determined for a maximal mixing, 
to ensure a large production rate. In this case, the lower limit of $\tan\be$ is mainly determined by the requirement 
of perturbativity for $\lam_2$, cf. the extensive discussion in \cite{Pruna:2013bma}. 
The same strategy was followed for the low-mass region, where again for fixed $\sin\al$ values the minimal value of $\tan\be$ is determined. {Here, the lower limit on $\tan\be$ stems from the signal strength fit.}}
\label{tab:brmax}
\end{table}

\begin{table}[htb!]
\begin{center}
\footnotesize{
\begin{tabular}{|l|llll|cllllll|r|} \hline
 & $\Gamma_{\Hhh}^{\text{LO}}$  & $\Gamma_{\Hhh}^{\text{NLO}}$  & $\delta_{\alpha}\,[\%]$ &  $\delta_{\gf}\,[\%]$ & & $b\bar{b}$ & $t\bar{t}$ &  WW & ZZ &  $gg$& $hh$ & $\Gamma_H$  \\ \hline \hline
 BHM1 & 0.399 & 0.413 & 3.411 & 3.291 & & 0.04 & $< 0.01$ & 46.35 & 20.56 & 0.04 & 33.02& 1.210 \\ 
 BHM2 & 0.963 & 1.026 & 6.485 & 6.272 & & 0.01 & $10.19$  & 40.07 & 18.52 & 0.06 & 31.15& 3.092 \\
 BHM3 & 1.383 & 1.463 & 5.803 & 5.604 & & $0.01$ & 14.19 & 40.36 & 19.29 & 0.04 & 26.09& 5.299 \\ 
 BHM4 & 2.067 & 2.161 & 4.520 & 4.361 & & $0.01$ & 12.82 & 42.35 & 20.64 & 0.03 & 24.11& 8.574 \\ 
 BHM5 & 2.637 & 2.717 & 3.027 & 2.918 & & $<0.01$ & 10.61 & 44.37 & 21.91 & 0.02 & 23.11& 11.413 \\ 
 BHM6 & 3.798 & 3.867 & 1.826 & 1.759 & & $<0.01$ & 8.57  & 46.29 & 23.07 & 0.02 & 22.07& 17.204 \\ \hline 
\end{tabular}
 }
\end{center} 
\caption{Heavy-to-light Higgs decay width $\GHhh$ at LO and NLO EW accuracy for 
the maximal branching fraction scenarios in the high-mass region given in Table~\ref{tab:brmax}. 
The relative
one-loop EW effects are quantified in both the $\alpha_{\text{em}}$-parametrization
and the $\gf$-parametrization, as defined in Eqs.~\eqref{eq:rel-alpha}-\eqref{eq:rel-gf}. 
{Renormalization is performed in the improved on-shell scheme.}
In the 
right columns we document the branching
ratios (in $\%$) for the leading Higgs decay channels and the total Higgs width. 
{Like in Figures ~\ref{fig:br-highmass} and \ref{fig:br-lowmass}, 
all partial decay widths to SM fields are evaluated by
rescaling the SM predictions \cite{Heinemeyer:2013tqa}, while for
$\Hhh$ we use the LO result evaluated in the $\alpha_{\text{em}}$-parametrization.}
All partial widths are given in GeV.  \label{tab:brmaxhigh}} \end{table}

\begin{table}[htb!]
\begin{center}
\footnotesize{
\begin{tabular}{|l|llll|clllll|l|r|} \hline
 & $\Gamma_{\Hhh}^{\text{LO}}$ & $\Gamma_{\Hhh}^{\text{NLO}}$ & $\delta_{\alpha}\,[\%]$ 
 &  $\delta_{\gf}\,[\%]$ & $b\bar{b}$ & $\gamma\gamma$ &  WW & ZZ &  $gg$& $hh$ & $\Gamma_H$ \\ \hline 
BLM1 & $1.426$&$1.536$& 7.765 & 7.763 & 42.65  & 0.17 & 16.04 & 1.97 & 6.34 & 25.90 & $5.506$ \\ 
BLM2 & $1.439$&$1.472$& 2.305 & 2.304 & 42.55  & 0.17 & 16.00 & 1.97 & 6.33 & 26.07 & $5.520$ \\
BLM3 & $1.423$&$1.432$& 0.586 & 0.586 & 42.67  & 0.17 & 16.05 & 1.97 & 6.35 & 25.86 & $5.504$ \\ 
BLM4 & $1.419$&$1.415$& -0.272 & -0.272 & 42.71 &  0.17 & 16.06 & 1.97 & 6.35 & 25.80 & $5.500$ \\ 
BLM5 & $1.431$&$1.425$& -0.445 & -0.445 & 42.61 &  0.17 & 16.02 & 1.97 & 6.34 & 25.96 & $5.512$ \\ 
BLM6 & $1.427$&$1.421$& -0.438 & -0.438 & 42.64 &  0.17 & 16.04 & 1.97 & 6.34 & 25.91 & $5.508$ \\ \hline
\end{tabular}
 }
\end{center} 
\caption{ \label{tab:brmaxlow}
 As in Table~\ref{tab:brmaxhigh} for the low-mass region. Notice that in this case all partial widths are given in MeV.}
\end{table}

\section{Summary}
\label{sec:conclusions}

Heavy-to-light Higgs decays $\Hzero \to \hzero\hzero $ are of undisputed relevance
in the phenomenological characterization of extended Higgs sectors. When kinematically accessible, 
these may contribute to,
and in some scenarios even dominate, the heavy Higgs lineshape, while at the same
time they significantly modify
its decay pattern with respect to the SM picture. 
On the other hand, both the tree-level and the  
leading one-loop contributions to this process are governed by the scalar self-interactions,
which makes this decay a unique handle on the architecture of the scalar potential. 

\smallskip{}
While electroweak corrections to the Higgs
self-couplings have been the subject of dedicated analyses 
in the 2HDM ~\cite{Kanemura:2004mg,Kanemura:2004ch,Kanemura:2015mxa}, the NMSSM\cite{Muhlleitner:2015dua} or the Inert Doublet model \cite{Arhrib:2015hoa},
a corresponding study for the singlet extension was lacking.
Extending upon previous work~\cite{Lopez-Val:2014jva},
we have presented herewith a detailed analysis of the heavy-to-light Higgs
decays at NLO electroweak accuracy. 
To renormalize the singlet-extended Higgs sector we have proposed four renormalization
schemes: {i) a minimal field setup; ii) 
a traditional on-shell prescription; iii) a mixed $\msbar$/on-shell scheme; 
and iv) an \emph{improved} on-shell scheme.}
{Using the general non-linear gauge-fixing of {\sc Sloops},
we have discussed in detail the
gauge} {independence of the different} renormalization setups. 
{We have found that, while the minimal field and on-shell approaches} {still lead to a residual dependence on the {non-linear} gauge fixing {parameters},}
the mixed $\msbar$/OS and improved OS schemes render {gauge independent one-loop predictions} {for physical observables}.
Furthermore, the improved OS scheme {is numerically stable} 
in all regions of the parameter space.} {We therefore advocate 
for the use of this scheme to investigate the phenomenology of singlet extensions of 
the SM at higher orders.}

We have applied the above schemes to compute
the corresponding heavy-to-light Higgs decay widths $\Gamma_{\Hhh}$ including one-loop electroweak corrections.
We have performed a comprehensive phenomenological analysis, with a separate
study of 
two
possible realizations of the model:  
a high-mass and a low-mass region, in which the additional scalar field
corresponds to a heavy (resp. a light) companion of the SM-like Higgs boson.

The phenomenological implications of our study can be summarized as follows:

\begin{itemize}
\item The heavy-to-light Higgs decay width at LO is governed by two competing
mechanisms: i) the Higgs self-coupling strength $\lambda_{Hhh}$; ii) the one-to-two body
decay kinematics. We pinpoint a strongly varying width 
with the relevant model parameters.  
Overall, $\Gamma_{\Hhh}$ may attain
up to $\mathcal{O}(1)$ GeV for heavy Higgs masses above $m_H \sim 500$ GeV. 
\item Aside from the $\tan\beta < 1$ region, the relative one-loop effects are mild and
show tempered variations over the parameter space. 
In the high-mass region, electroweak corrections are positive, loosely variable,
and stagnate in the ballpark of few percent. In the low-mass region, mainly
for $\tan\beta > 1$ and small $m_{h}$ values, these may be 
pulled down to $\delta_\alpha \sim -10\%$. 

\item For certain parameter choices, the $\Hhh$ decay  
becomes effectively loop-induced: i) along the tree-level nodes
where
the LO contribution vanishes; ii) at low $\tan\beta$,
where the $\cot\beta$-enhancements
lead to increased 
scalar self-couplings, and thereby 
to large Higgs-mediated loop graphs. In practice, though, 
these sizable quantum effects are precluded once the constraints on the model are {taken into account}.

\item Let alone extreme parameter space corners, the NLO predictions are robust 
under changes of renormalization schemes
and renormalization scale choices, as indicative of a small theoretical uncertainty. 

\end{itemize}

Having constructed 
a complete renormalization scheme for the Higgs sector, the path ahead is clear 
for further analyses on the topic. On the one hand, it will be interesting to further explore
the role of the quantum effects on the Higgs self-couplings themselves, and whether 
                 these may have relevant implications e.g. for collider searches or
                 in electroweak baryogenesis. 
                 On the other hand, the  complete renormalization of the Higgs sector 
                 paves the way towards characterizing  
      the singlet model phenomenology at one-loop electroweak accuracy, including 
      all Higgs production modes and decay channels, and exploiting the rich possibilities
      of off-shell effects. Work in this direction is underway \cite{future-rsinglet}.

\section*{Acknowledgements}
We are grateful to Fawzi Boudjema, Sven Heinemeyer, Dominik St\"ockinger
and Hyejung St\"ockinger-Kim
for enlightening discussions.  
DLV   
acknowledges the support of the   
F.R.S.-FNRS ``Fonds de la Recherche Scientifique'' (Belgium). The work of GC
is supported by the Theory- LHC-France initiative of CNRS/IN2P3. \\

\begin{appendix}
\section{Appendix: Analytical details}
\subsection{Feynman rules}

For a sake of completeness, we collect the relevant
Feynman rules for the three--point and four--point scalar
field self--interactions
in the singlet model. The complete set of Feynman rules has been arranged
in the form of a {\sc FeynArts} model file \cite{Hahn:2000kx}, which 
we have obtained via two independent implementations
using {\sc FeynRules} \cite{Alloul:2013bka} and {\sc Sloops}
\cite{Boudjema:2005hb,Baro:2007em,Baro:2008bg,Baro:2009gn}.
Here we give the results in the ' t Hooft--Feynman gauge.
The shorthand notation $s_{\theta}\equiv \sin(\theta), c_{\theta}\equiv \cos(\theta)$
$t_{\theta}\equiv \tan(\theta)$ is employed throughout. 

\begin{itemize}
\item {\underline{Trilinear self--couplings at tree--level}:
\begin{alignat}{5}
 \lambda_{\hzero\hzero\hzero} &= -3i\left(2 c^3_{\alpha}\,\lambda_1\,v +  c_{\alpha} s^2_{\alpha}\,\lambda_3\,v -  c^2_{\alpha}\, s_{\alpha}\lambda_3\,v_s-2 s_{\alpha}^3\,\lambda_2\,v_s\right) =
-\cfrac{3\,i\,m^2_{\hzero}}{v}\,( c^3_{\alpha} -  s_{\alpha}^3 t_{\beta}^{-1}) \label{eq:hhh} \\
 \notag \\
 \lambda_{\Hzero\hzero\hzero} &= -i\left( 6 c^2_{\alpha} s_{\alpha} \lambda_1\,v - 2 c^2_{\alpha} s_{\alpha}\lambda_3\,v +  s_{\alpha}^3 \lambda_3 v  
 +  c^3_{\alpha} \lambda_3\,v_s - 2 c_{\alpha} s^2_{\alpha}\,\lambda_3 v_s + 6 c_{\alpha} s^2_{\alpha}\,\lambda_2\,v_s \right) \notag \\
 & = -\cfrac{is_{2\alpha}}{v}\,\left[m^2_{\hzero}+\cfrac{m^2_{\Hzero}}{2}\right]\,( c_{\alpha} +  s_{\alpha}\, t_{\beta}^{-1}) 
 \label{eq:Hhh} \\ \notag \\
  \lambda_{\Hzero\Hzero\hzero} &= -i\left( 6 c_{\alpha} s^2_{\alpha} \lambda_1\,v - 2 c_{\alpha} s^2_{\alpha}\lambda_3\,v +  c^3_{\alpha} \lambda_3 v  
 -  s_{\alpha}^3 \lambda_3\,v_s + 2 c^2_{\alpha} s_{\alpha}\,\lambda_3 v_s - 6 c^2_{\alpha}  s_{\alpha}\,\lambda_2\,v_s \right) \notag \\
 & = \cfrac{is_{2\alpha}}{v}\,\left[\cfrac{m^2_{\hzero}}{2}+m^2_{\Hzero}\right]\,(- s_{\alpha} +  c_{\alpha}\, t_{\beta}^{-1}) 
 \label{eq:HHh} \\
   \lambda_{\Hzero\Hzero\Hzero} &=  -\cfrac{3\,i\,m^2_{\Hzero}}{v}\,( c^3_{\alpha} t_{\beta}^{-1} +  s_{\alpha}^3) 
 \label{eq:HHH}.  
\end{alignat}
}
 \item{ \underline{Higgs -- Goldstone boson three--point couplings}:
\begin{alignat}{5}
 \lambda_{\hzero\gzero\gzero} &=  \lambda_{\Hzero\gp\gm} = (-i)\,\mhd c_{\alpha}/v; \qquad   \lambda_{\Hzero\gzero\gzero} =  \lambda_{\Hzero\gp\gm} = (-i)\,\mHHd s_{\alpha}/v. 
\label{eq:coup-sgg} 
\end{alignat}
}
\item {\underline{Higgs quartic couplings}:
\begin{alignat}{5}
 \lambda_{\hzero\hzero\hzero\hzero} &= -\cfrac{3i}{v^2}\,[\mhd  c^6_{\alpha}  + \mHHd  c^4_{\alpha}\,s^2_{\alpha} - 2\,(\mhd-\mHHd)\, c^3_{\alpha}  t_{\beta}^{-1} s_{\alpha}^3 
 + \mHHd\, c^2_{\alpha}\,s^4_{\alpha}\,t_{\beta}^{-2} + \mhd\,t_{\beta}^{-2} s^6_{\alpha}] \\ 
 \lambda_{\Hzero\hzero\hzero\hzero} &= \cfrac{3i}{8v^2s^2_{\beta}\, } s_{2\alpha}\,s_{\alpha+\beta} [(3 \mhd + \mHHd) s_{\alpha-\beta} + (\mHHd-\mhd) s_{3\alpha+\beta}] \\
\lambda_{\Hzero\Hzero\hzero\hzero} &= -\cfrac{i}{16v^2\, s^2_{\beta}}\, s_{2\alpha}\, [6\,(\mhd + \mHHd)\,s_{2\alpha} - (\mhd - \mHHd) (s_{2\beta} + 3 s_{4\alpha+2\beta})] \\
\lambda_{\Hzero\Hzero\Hzero\hzero} &= \cfrac{i}{8v^2s^2_{\beta}\, }\, s_{2\alpha}\,c_{\alpha+\beta}\,[(\mhd+3\mHHd) c_{\alpha-\beta} + (\mHHd-\mhd)\,c_{3\alpha + \beta}] \\
\lambda_{\Hzero\Hzero\Hzero\Hzero} &= 
 -\cfrac{3i}{v^2}\,[\mHHd  c^6_{\alpha}\,t_\beta^{-2} + \mhd\,c_{\alpha}^4  s_{\alpha}^2 t_\beta^{-2} - 2(\mhd - \mHHd)\,c_{\alpha}^3 t_\beta^{-1} s_{\alpha}^3 + \mhd c^2_{\alpha} s^4_{\alpha}
 + \mHHd s_{\alpha}^6]
 \label{eq:higgs-quartics}
\end{alignat}}
\item {\underline{Higgs -- Goldstone four--point couplings}:
\begin{alignat}{5}
  \lambda_{\hzero\hzero\gzero\gzero} &= \lambda_{\hzero\hzero\gp\gm} = \cfrac{-i}{v^2}\, c_{\alpha}\,[\mhd c^3_{\alpha} + \mHHd\,s^2_{\alpha} c_{\alpha} + (\mHHd-\mhd)\, s_{\alpha}^3 t_{\beta}^{-1}]  \\
  \lambda_{\Hzero\Hzero\gzero\gzero} &=  \lambda_{\Hzero\Hzero\gp\gm} = \cfrac{-i}{v^2}\, s_{\alpha}\,[\mhd c^2_{\alpha} s_{\alpha} + \mHHd\, s_{\alpha}^3 + (\mHHd-\mhd)\, c^3_{\alpha} t_{\beta}^{-1}]  \\
 \label{eq:coup-ssgg}
\end{alignat}}
\end{itemize}

\subsection{Trilinear Higgs counterterm}

The coefficients entering the trilinear Higgs coupling counterterm $\delta\lambda_{\Hhh}$ in Eq.~\eqref{eq:ct-Hll} yield
\begin{alignat}{5}
c_1^{Hhh} & = \cfrac{s_{2\alpha}\,s_{\alpha+\beta}}{v\, s_{\beta}}; \qquad c_2^{Hhh} = \cfrac{s_{2\alpha}\,s_{\alpha+\beta}}{2v\, s_{\beta}};
\qquad c_3^{Hhh} = -\cfrac{s_{\beta+3\alpha} - 5 s_{\beta-\alpha}}{4v s_{\beta}}\; \notag \\
c_4^{Hhh} &= \cfrac{3s_{2\alpha}\,( c^2_{\alpha} - s^2_{\alpha}\,t^{-2}_{\beta})}{2v^2}\; \qquad c_5^{Hhh} = \cfrac{3\,c^2_{\alpha}\, s^2_{\alpha}}{ s^2_{\beta}\,v^2};\; 
\qquad c_6^{Hhh} = -\cfrac{s_{2\alpha}}{v}\,\left[m^2_{\hzero}+\cfrac{m^2_{\Hzero}}{2}\right]\, c_{\alpha}
\label{eq:coefHll}.
\end{alignat}
\end{appendix}
\bibliography{hnlo}
\end{document}